\documentclass[a4paper]{lipics-v2021}
\usepackage[utf8]{inputenc}

\usepackage{amsmath}
\usepackage{amsfonts}
\usepackage{amsthm}
\usepackage{amssymb}
\usepackage{mathtools}
\mathtoolsset{centercolon}
\usepackage{thmtools, thm-restate}
\usepackage{comment}

\usepackage{lmodern}
\usepackage{ifthen}
\usepackage{tikz}
\usepackage{url}
\usepackage{stmaryrd}
\usepackage{xparse}
\usepackage{bbm}
\usepackage{xcolor}
\usepackage{etoolbox}
\usepackage[subrefformat=parens]{subcaption}
\usepackage{float}

\usepackage[most]{tcolorbox}
\usepackage[sort&compress,numbers]{natbib}

\usetikzlibrary{decorations.pathmorphing}
\usetikzlibrary{decorations.pathreplacing}
\usetikzlibrary{calc}
\usetikzlibrary{positioning}
\usetikzlibrary{patterns}
\usetikzlibrary{patterns.meta}

\pgfdeclarelayer{background}
\pgfdeclarelayer{foreground}
\pgfsetlayers{background,main,foreground}

\definecolor{lightblue}{rgb}{0.5,0.5,1.0}
\definecolor{darkred}{rgb}{0.5,0,0}
\definecolor{darkgreen}{rgb}{0,0.5,0}
\definecolor{darkblue}{rgb}{0,0,0.5}

\hypersetup{
	breaklinks=true,
	colorlinks=true,
	citecolor=darkblue,
	linkcolor=darkred,
	urlcolor=darkred,
	filecolor=darkgreen,
	pdflang={en},
}

\renewcommand{\phi}{\varphi}
\renewcommand{\epsilon}{\varepsilon}
\newcommand{\nat}{\mathbb{N}}
\newcommand{\ZZ}{\mathbb{Z}}
\newcommand{\FF}{\mathbb{F}}

\newcommand{\defining}[1]{\textbf{#1}}

\DeclarePairedDelimiter\set{\lbrace}{\rbrace}

\DeclarePairedDelimiterX\setcond[2]{\{}{\}}{\mathchoice{\,}{}{}{}#1 \;\delimsize\vert\; #2\mathchoice{\,}{}{}{}}

\newcommand{\msetopen}[0]{\{\hspace{-3pt}\{}
\newcommand{\msetclose}[0]{\}\hspace{-3pt}\}}
\newcommand{\mset}[1]{\msetopen #1 \msetclose}
\newcommand{\msetcondition}[2]{\msetopen #1 \mid #2 \msetclose}

\newcommand{\tup}[1]{\bar{#1}}

\DeclareMathSymbol{\shortminus}{\mathbin}{AMSa}{"39}
\newcommand{\inv}[1]{#1^{\shortminus 1}}

\newcommand{\hiddenEqq}{\phantom{{}:={}}}

\newcommand{\CFIsym}{\mathsf{CFI}}
\newcommand{\CFI}[1]{\CFIsym(#1)}

\newcommand{\compEquiv}{\equiv}

\newcommand{\clo}[1]{\operatorname*{cl}(#1)}

\definecolor{colA}{RGB}{0,80,250}
\definecolor{colB}{RGB}{60,200,230}
\definecolor{colC}{RGB}{60,180,75}
\definecolor{colD}{RGB}{230,25,75}
\definecolor{colE}{RGB}{245,130,48}
\definecolor{colF}{RGB}{145,30,180}
\definecolor{colG}{RGB}{240,50,230}
\definecolor{colH}{RGB}{170,110,40}

\newcommand{\restrict}[2]{#1|_{#2}}
\newcommand{\Hom}[2]{\mathrm{Hom}(#1,#2)}

\newcommand{\vars}[1]{\mathcal{V}(#1)}

\newcommand{\bag}{\beta}
\newcommand{\bIso}[2]{\text{bIso}(#1,#2)}

\newcommand{\Surj}[2]{\text{Surj}(#1,#2)}
\newcommand{\bExt}[2]{\text{bExt}(#1,#2)}
\newcommand{\Aut}[2]{\text{Aut}(#1,#2)}

\newcommand{\cl}[1]{\operatorname*{cl}(#1)}

\newcommand{\comucgrid}{\mathcal{C}/_{\equiv}}

\newcommand{\comcwgrid}[2]{\mathcal{C}_{#2}^{#1}/_{\equiv}}

\newcommand{\ucgrid}{\mathcal{C}}
\newcommand{\cgrid}[1]{\mathcal{C}_{#1}}
\newcommand{\cwgrid}[2]{\mathcal{C}_{#2}^{#1}}

\newcommand{\eclass}[1]{#1/_{\equiv}}
\newcommand{\eclassw}[1]{#1/_{\sim}}

\newcommand{\opvvariable}[4]{x^{#1}_{#2,#3,#4}}
\newcommand{\invref}[1]{(I\ref{#1})}

\newcommand{\opgridequation}[4]{e^{#1}_{#2,#3,#4}}

\newcommand{\sheight}[1]{\operatorname*{\mathsf{sep-height}}(#1)}

\newcommand{\homsys}[1]{#1^0}
\newcommand{\sols}[1]{\operatorname*{Sol}(#1)}

\newcommand{\exts}[2]{#1 \triangleright #2}
\newcommand{\forgt}[2]{#1^{\setminus #2}}
\newcommand{\eqTA}{\tup{f}}
\newcommand{\eqTAm}{f}
\newcommand{\eqTB}{\tup{e}}
\newcommand{\eqTBm}{e}

\newcommand{\thin}[1]{\operatorname*{thin}(#1)}
\newcommand{\cld}[1]{\operatorname*{col}(#1)}

\title{An Almost-tight Size Lower Bound for Bounded Treewidth Homomorphism-Distinguishability}
\titlerunning{A Size Lower Bound for Bounded Treewidth Homomorphism-Distinguishability}

\author{Moritz Lichter}{RWTH Aachen University, Germany}{lichter@lics.rwth-aachen.de}{https://orcid.org/0000-0001-5437-8074}{}
\author{Harry Vinall-Smeeth}{RWTH Aachen University, Germany}{vinall-smeeth@lics.rwth-aachen.de}{https://orcid.org/0000-0003-2422-9435}{}
\authorrunning{M. Lichter and H. Vinall-Smeeth}

\Copyright{Moritz Lichter and Harry Vinall-Smeeth}

\ccsdesc[500]{Theory of computation~Finite Model Theory}
\ccsdesc[300]{Mathematics of computing~Graph theory}

\keywords{Homomorphism distinguishability, treewidth, Weisfeiler-Leman, lower bound}

\funding{Funded by the European Union (ERC, SymSim, 101054974). Views and opinions expressed are those of the author(s) and do not necessarily reflect those of the European Union or the ERC. Neither the European Union nor the granting authority can be held responsible for them.}

\nolinenumbers

\hideLIPIcs

\begin{document}
	\maketitle

	\begin{abstract}
		One way to compare two graphs $G_1, G_2$ is to check whether there is some graph $H$ in a graph class $\mathcal{H}$, such that the number of homomorphisms from $H$ to $G_1$ is not equal to that from $H$ to $G_2$.
		This is the field of homomorphism distinguishability which has gained a lot of traction in recent years. In this paper, we study the case where $\mathcal{H}$ is the class of graphs of treewidth $k$. This has connections to important concepts in logic, machine learning, combinatorial optimization and proof complexity. In particular,  Dvořák showed that $G_1$ and $G_2$ can be distinguished by a homomorphism count from a graph of treewidth $k$ iff $G_1$ and $G_2$ are distinguishable by the $k$-dimensional Weisfeiler-Leman algorithm.
		While the relation between the iteration number of this algorithm and the depth of the tree-decomposition of a distinguishing graph is understood, our understanding of the \emph{size} of $H$ is limited. We show a $2^{\Omega(n^{k/2})}$ lower bound on the size of a graph of treewidth $k$ required to distinguish two graphs of order $n$. This bound is proved via a new combinatorial game along with a construction which uses a recent compression technique for CFI graphs. It is almost tight to the known $2^{O(n^{k+1})}$ upper bound and is an exponential improvement on the previous best known lower bound.
	\end{abstract}

	\subparagraph{AI Declaration}
	AI tools were used in the preparation of this manuscript for proof reading and to check the correctness of proofs towards the final stages of the project. We did not use AI in any other way in the project.

\section{Introduction}

Suppose that there exists a graph $H$ and two other graphs $G_1$ and $G_2$ such that $\Hom{H}{G_1} \neq \Hom{H}{G_2}$, where $\Hom{H}{G_i}$ denotes the number of homomorphisms from $H$ to $G_i$. Then clearly $G_1$ is not isomorphic to $G_2$. A beautiful theorem of Lovász gives a converse to this observation: if no such $H$ exists, then $G_1$ and $G_2$ are isomorphic \cite{lovasz1967operations}. This leads to the question: can relaxations of the isomorphism relation be similarly characterized?
The answer is a resounding yes, with a plethora of recent results (see~\cite{FluckSeppeltSpitzer2024,DBLP:journals/jct/AtseriasMRSSV19,DBLP:conf/lics/Grohe21,DBLP:conf/icalp/GroheRS22,DBLP:conf/aaai/0001RFHLRG19,DBLP:conf/lics/DawarJR21,DBLP:conf/focs/MancinskaR20}) of the following form: $G_1$ and $G_2$ satisfy some relation if and only if there is a graph $H \in \mathcal{H}$ such that $\Hom{H}{G_1} \neq \Hom{H}{G_2}$, where $\mathcal{H}$ is a graph class depending on the relation.

This work focuses on the class of graphs of treewidth at most $k$, for some fixed integer $k$. We say $G_1, G_2$ are \defining{treewidth-$k$-distinguishable} if $\Hom{H}{G_1} \neq \Hom{H}{G_2}$ for some graph $H$ of treewidth at most $k$.
From a computational perspective, these are natural classes to focus on since the problem of computing $\Hom{H}{G}$, for $H \in \mathcal{H}$ and any graph $G$ is in polynomial time if and only if $\mathcal{H}$ has bounded treewidth (under a common complexity-theoretic assumption)~\cite{DBLP:journals/tcs/DalmauJ04}. Moreover,
 $G_1$ and $G_2$ are treewidth-$k$-distinguishable if and only if they are distinguishable by the $k$-dimensional Weisfeiler-Leman ($k$-WL) algorithm \cite{DBLP:journals/jgt/Dvorak10, DellGroheRattan2018}, a simple yet powerful graph isomorphism heuristic \cite{DBLP:journals/corr/abs-2201-07083}.
In fact, treewidth-$k$-distinguishability has many other characterizations in terms of important concepts from
of
logic~\cite{CaiFuererImmerman1992,immerman1990describing}, 
machine learning~\cite{DBLP:journals/jmlr/0001LMRKGFB23,DBLP:conf/aaai/0001RFHLRG19,DBLP:conf/icdm/0001KM17,DBLP:conf/iclr/XuHLJ19}, 
combinatorial optimization~\cite{DBLP:journals/siamcomp/AtseriasM13,DBLP:journals/jsyml/GroheO15,DBLP:journals/disopt/Malkin14}, 
and proof complexity~\cite{BerkholzGrohe2015,DBLP:journals/siamcomp/AtseriasF23}. 
For example, two graphs are treewidth-$k$-distinguishable if and only if
 they are not equivalent in  $(k+1)$-variable first-order logic with counting quantifiers.

It is therefore natural to seek a finer-grained understanding of treewidth-$k$-distinguishability. We know that for any two graphs $G_1, G_2$ distinguishable by $k$-WL there exists a graph $H$ of treewidth at most $k$ with $\Hom{H}{G_1} \neq \Hom{H}{G_2}$. But what else can we say about $H$? For example, we might be interested in the depth of the tree-decomposition of $H$. This is connected to the iteration number of $k$-WL and hence is well understood~\cite{DBLP:journals/jacm/GroheLNS25,GroheLichterNeuen2025,FluckSeppeltSpitzer2024}.

This work studies another natural complexity measure of $H$: its number of vertices $|H|$, which we call the \defining{order of $H$}.
The first question is whether this quantity is even bounded (in terms of $n:= |G_i|$). This has an affirmative answer, due to Seppelt,
who showed that $|H| = 2^{O(n^{k+1})}$~\cite[Theorem 13]{DBLP:conf/mfcs/Seppelt24}.
However, to the best of our knowledge, the best lower bound on $|H|$  in the literature is an $\Omega(n^{k/2})$ bound obtained from work on the iteration number of $k$-WL~\cite{DBLP:journals/jacm/GroheLNS25}.
We almost completely close the gap between the upper and lower bounds by improving the lower bound.

\begin{theorem} \label{thm:main-conf}
	For every $k \ge 3$ and every sufficiently large $n$, there are two treewidth-$k$-distinguishable graphs $G_1$ and $G_2$ of order $|G_i|= \Theta(n)$, such that every graph $H$ of treewidth at most $k$
	with
	$\Hom{H}{G_1} \neq \Hom{H}{G_2}$ is of order $|H| = 2^{\Omega(n^{k/2})}$.
\end{theorem}

The lower bound is tight up to a multiplicative factor of just over two in the second exponent. Thus, the graph $H$ witnessing treewidth-$k$-distinguishability can be extremely large in general.

We obtain a weaker but still super-polynomial $2^{\Omega(n^\epsilon)}$ lower bound (see Section~\ref{ss:for-free-lower-bound})
by combining a robust iteration number lower bound for Weisfeiler-Leman~\cite{GroheLichterNeuen2025}
with a result for tree-decompositions~\cite{Bodlaender1988}.
But the available robust bound is too weak to go further. For that we require new ideas.

This robust lower bound, our new lower bounds, and many other lower bounds in this area~\cite{BerkholzNordstroem23, DBLP:conf/mfcs/BerkholzLV25, DBLP:conf/icalp/Furer01, DBLP:journals/jacm/GroheLNS25},
are obtained from a Cai-Fürer-Immerman-like encoding~\cite{CaiFuererImmerman1992,BerkholzGrohe2015} of systems of linear equations over the two-element field.
Our work depends on a recent compression technique for CFI graphs used in the super-linear $\Omega(n^{k/2})$ iteration number lower bound for $k$-WL~\cite{DBLP:journals/jacm/GroheLNS25} and its linear-system-based formulation~\cite{RezendeFJN025}, which 
compress CFI graphs/Tseitin-systems
 obtained from $k \times \Theta(n)$ grid graphs.
However, the compressed CFI graphs obtained for this bound are homomorphism-distinguishable by the $k\times\Theta(n)$ grids themselves~\cite{DBLP:journals/jctb/Roberson26},
which are of treewidth $k$ and order $\Theta(n^{k/2})$.

To lift the bound from $\Omega(n^{k/2})$ to $2^{\Omega(n^{k/2})}$,
we obtain an inconsistent system from not one but two $k \times \Theta(n)$ grids, 
which are connected by ``crossings''.
 At each crossing, the inconsistency can be moved between the grids, so the tree-decomposition of a distinguishing graph has to branch at each crossing, 
 leading to an exponential size blow up.
We then show that the compression technique of~\cite{DBLP:journals/jacm/GroheLNS25} can be applied in this setting while maintaining the lower bound.

Proving this bound requires new techniques for two reasons.
Firstly, 
existing techniques
 build on the Cops and Robber game played on the grid.
However, the correspondence with these games 
only holds for (compressed) CFI encodings of Tseitin-systems and 
so they are not applicable to our system.
Secondly, other known game-based methods such as the bijective pebble game~\cite{Hella1996} or other games played on equation systems~\cite{GroheLichterNeuen2025} only give insights into the depth of a tree-decomposition of a distinguishing graph, not its size.
Therefore, we develop new games to prove size lower bounds, which 
turn out to be more complicated than those for depth.
Such difficulties also occur when proving succinctness results in logic for formula size instead of quantifier depth~\cite{DBLP:journals/tocl/AdlerI03, DBLP:conf/lics/GroheS04,DBLP:conf/lics/Vinall-Smeeth24, DBLP:conf/csl/Pfau26, DBLP:journals/lmcs/FaginLRV25}.

\subparagraph{Structure of the paper}
The paper is structured as follows. After giving preliminaries in Section~\ref{sec:prelims-full}, in Section~\ref{sec:CFI-graphs} we introduce the necessary background and some preliminary results on CFI-like graphs obtained from $\FF_2$ equation systems.
  We should note that the content of Section~\ref{sec:CFI-graphs} is standard in flavor but is necessary to lay the groundwork for the rest of the paper; we encourage the reader to not get too bogged down in the details here on a first reading. In Section~\ref{sec:homo-bounded-treewidth-full}, we turn to the main topic of this paper: showing equal homomorphism count from bounded treewidth graphs. In Section~\ref{sec:ind-game-full}, we build on this to introduce the \emph{indistinguishability game} for proving equal homomorphism counts. The advantage of this game is that we play on the level of the equation system, rather than directly on the graph encoding.  This game is then specialized and simplified to make it more suitable for our lower bound proof in Section~\ref{sec:size-game-full}, where we introduce the \emph{size game}. In Section~\ref{sec:compressed-crossed-grids-full}, we introduce the concrete equation system which we use for our lower bound: the \emph{compressed crossed grid}. In Section~\ref{sec:lower-bound}, we prove our main result (Theorem~\ref{thm:main-conf}) by playing the size game on the compressed crossed grid.

\section{Preliminaries} \label{sec:prelims-full}

For integers $\ell\leq k$,
we write $[k]$ to denote the set $\set{1,\dots, k}$
and $[\ell,k]$ to denote the set $\set{\ell,\dots, k}$.

\subparagraph{Tuples.}
For a set $U$ and $k\in \nat$, we denote by $U^k$ the set of tuples of length $k$ with entries in $U$.
We write $\tup{u}$ for tuples and denote the $i$-th entry of $\tup{u}$ by $u_i$.
When it does not cause confusion, we write $v \in \tup{u}$ if $v$ appears in $\tup{u}$, that is, there is an $i\in[k]$ such that $u_i = v$.
The length of $\tup{u}$ is $|\tup{u}| = k$.
For $i\in[k]$ and $v\in U$, we set $\tup{u}[i/v] := (u_1, \dots, u_{i-1},v,u_{i+1},\dots, u_k)$ for the tuple obtained from $\tup{u}$ by replacing the $i$-th entry with $v$. 
The tuple $\tup{u}\setminus i$ is the $(k-1)$-tuple $(u_1,\dots, u_{i-1}, u_{i+1},\dots u_k)$ obtained from $\tup{u}$ by deleting the $i$-th entry.
The set of all tuples over $U$ of length at most $k$ is $U^{\leq k}$.
If $N$ is a finite set, we denote by $U^N$ the maps $N\to U$.
For $a \in U^N$ and $M\subseteq N$, the restriction of $a$ to $M$ is $\restrict{a}{M} \in U^M$.

\subparagraph{Graphs.} 
A \defining{graph} is a pair $G=(V,E)$ of a finite set $V$ of vertices and an edge set $E\subseteq \tbinom{V}{2}$. We write $V(G)$ for $V$ and $E(G)$ for $E$.
This paper considers undirected simple graphs unless noted otherwise. 
We use the letters $G$ and $H$ for graphs.
For a set $W\subseteq V$, the subgraph of $G$ induced by $W$ is denoted $G[W]$.
The \defining{order} $|G|$ of $G$ is $|V(G)|$.
The \defining{isomorphism type} of a tuple $\tup{u} \in V^k$
is the isomorphism class of the ordered graph $(G[\setcond{u_i}{i\in[k]}], \tup{u})$.
This means that for two graphs $G$ and $H$ two tuples
$\tup{u} \in V(G)^k$ and $\tup{v} \in V(H)^k$
have the same isomorphism type
if and only if the map $u_i \mapsto v_i$ is an isomorphism
of the induced subgraphs $G[\setcond{u_i}{i\in[k]}]$
and $H[\setcond{v_i}{i\in[k]}]$.

\subparagraph{Tree-decompositions.}
A \defining{tree-decomposition} $D=(T,\bag)$ of \defining{width} at most $k$ of a graph $H$ 
is a tuple of a tree $T$ and a \defining{bag}-function $\bag\colon V(T) \to V(H)^{k+1}$ such that
\begin{enumerate}
	\item for every edge $uv \in E(H)$, there is a bag $\bag(t)$ such that $u,v\in \bag(t)$ (so $u$ and $v$ appear as entries in $\bag(t)$)
	\item for every vertex $u \in V(H)$, the set of nodes $t$ such that $u \in \bag(t)$ induces a connected subgraph of $T$.
\end{enumerate}
Usually, one considers bags as sets of size at most $k+1$, but using tuples will be convenient later on.
Note that we can require all tuples to have the same length by repeating entries.
We also write $T_D$ and $\bag_D$ for $T$ and $\bag$, respectively.
The \defining{treewidth} of $H$ is the minimal width of a tree-decomposition of $H$.

A \defining{rooted tree-decomposition} is a pair $(D,r)$ of a tree-decomposition $D$
and a vertex $r \in V(T_D)$ that serves as the root of $T_D$.
In this case and for nodes $t,s \in V(T_D)$, we denote by $t \in s$ that $t$ is a child of $s$.

A tree-decomposition $D$ of width $k$ of a graph $H$ is \defining{smooth} if
\begin{enumerate}
	\item for every $s \in V(T_D)$, the bag $\bag_D(s)$ does not contain duplicate entries and
	\item for every $s \in V(T_D)$ and $t \in N_T(s)$ there is an $i\in[k]$ such that $\bag(s) \setminus i = \bag(t) \setminus i$.
\end{enumerate}
Every graph with a tree-decomposition of width $k$ also has a smooth tree-decomposition of width $k$~\cite{Bodlaender1996}.
Point~2 extends the usual notion of a smooth tree-decomposition by the additional replacement requirement, that only a single entry between adjacent bags is replaced. This is not restrictive and can always be achieved by  ordering the tuples appropriately.

\subparagraph{Graph Homomorphisms.}
Let $H$ and $G$ be graphs.
A map $h\colon V(H) \to V(G)$ is a \defining{homomorphism} $H\to G$ if
$uv \in E(H)$ implies $h(u)h(v) \in E(G)$ for all $u,v\in V(H)$.
We write $\Hom{H}{G}$ for the number of homomorphisms $H\to G$.
Let $D$ be a tree-decomposition of $H$.
A homomorphism $h \colon H \to G$ is \defining{bag-wise isomorphic} with respect to $D$ if, for every $s \in V(T_D)$, $h$ induces a local isomorphism $H[\bag_D(s)] \to G[h(\bag_D(s))]$.
The number of bag-wise isomorphic homomorphisms with respect to $D$ is denoted by $\bIso{(H,D)}{G}$.
Considering a rooted tree-decomposition $(D,r)$
and a set $U\subseteq V(G)^{|\bag_D(r)|}$,
we denote by $\bIso{(H,D,r)}{(G,U)}$ the number of bag-wise isomorphic homomorphisms $h$ such that $h(\bag_D(r)) \in U$.

We say that two graphs $G_1$ and $G_2$ are \defining{(homomorphism) distinguished} by a third graph $H$ if $\Hom{H}{G_1} \neq \Hom{H}{G_2}$. If $G_1$ and $G_2$ are distinguished by a graph of treewidth at most $k$ we say that $G_1$ and $G_2$ are \defining{treewidth-$k$-distinguishable}.

\subparagraph{The Weisfeiler-Leman Algorithm.}
The \defining{$k$-dimensional Weisfeiler-Leman ($k$-WL)} algorithm colors $k$-tuples of a graph $G$.
Initially, the tuples are colored by its isomorphism type.
The coloring is refined in each iteration until the partition into equally colored tuples does not get finer anymore.
This paper only required details on the one-dimensional algorithm.
Initially, the vertices of a graph $G$ are colored by its degree.
If the current coloring is $\chi\colon V(G) \to C$ for some finite set $C$ of colors,
the \defining{one-dimensional WL-refinement} of $\chi$ is the coloring of $V(G)$ defined 
for each $u\in V(G)$ by:
\[u \mapsto \left(\chi(u), \msetcondition{\chi(v)}{uv \in E(G)} \right),\]
where $\mset{a_1,\dots, a_m}$ denotes the multiset
containing the elements $a_1, \dots, a_m$.
Denote by $WL_k[G]$ the stable coloring of the $k$-dimensional algorithm on $G$.
The algorithm \defining{distinguishes} two graphs $G$ and $H$
if there is a color $c$ such that
$|\inv{WL}_k[G](c)| \neq |\inv{WL}_k[H](c)|$.
It distinguishes the graphs after $r$ iterations,
if there is such a color after $r$ rounds of the $k$-dimensional WL-refinement.
The power of the algorithm to distinguish graphs can be characterized in different ways.

\subparagraph{Finite Variable Counting Logic.}
Counting logic is the extension of first-order logic by the counting quantifier
$\exists^{\geq i}x. \phi$,
that expresses that there are at least $i$ different assignments to $x$
satisfying $\phi$.
The $k$-variable fragment of counting logic only uses $k$
distinct variables, which may be rebound.
The quantifier-depth-$d$ fragment only considers formulas
of quantifier depth at most $d$.
Two graphs $G$ and $H$ are equivalent in such a fragment,
if $G$ and $H$ satisfy the same formulas of it.

\subparagraph{The Bijective $k$-Pebble Game.}
This game characterizes treewidth-$k$-distinguishability.
It is played by Spoiler and Duplicator on two graphs $G$ and $H$.
If $|G|\neq |H|$, then Spoiler wins.
Otherwise, a position in the game is a pair $\tup{u},\tup{v}$ such that $\tup{u} \in V(G)^{\leq k}$, $\tup{v} \in V(H)^{\leq k}$ and $|\tup{u}| = |\tup{v}|$.
The initial position is the pair $\emptyset,\emptyset$ of empty tuples.
The game is played in rounds as follows:
\begin{enumerate}
	\item If $|\tup{u}| = k$, Spoiler picks $i \in [k]$ and the game proceeds to position $\tup{u}',\tup{v}' = \tup{u} \setminus i, \tup{v} \setminus i$.
	\item Duplicator provides a bijection $f \colon V(G) \to V(H)$.
	\item Spoiler picks a vertex $w \in V(G)$ and the game proceeds to position $\tup{u}'w,\tup{v}'f(w)$.
	\item If $\tup{u}'w$ and $\tup{v}'f(w)$ have different isomorphism types in $G$ and $H$, respectively, then Spoiler wins.
\end{enumerate}
Duplicator wins if Spoiler never wins.
This game slightly differs from the usual presentation using pebbles~\cite{Hella1996}, but it describes the same game.

\begin{theorem}[\cite{CaiFuererImmerman1992, Hella1996, DBLP:journals/jgt/Dvorak10, DellGroheRattan2018}]
	\label{thm:hom-eq-iff-biso-hom-eq}
	For all $k\geq 1$ and all finite undirected graphs $G$ and $G'$ the following are equivalent:
	\begin{enumerate}
		\item There is an undirected graph $H$ of treewidth at most $k$ such that $\Hom{H}{G} \neq \Hom{H}{G'}$.
		\item There is an undirected graph $H$ with a tree-decomposition $D$ of width at most $k$ such that $\bIso{(H,D)}{G} \neq \bIso{(H,D)}{G'}$.
		\item $k$-WL distinguishes $G$ and $G'$.
		\item The graphs $G$ and $G'$ are not equivalent in $(k+1)$-variable counting logic.
		\item Spoiler wins the bijective $(k+1)$-pebble game on $G$ and $G'$.
	\end{enumerate}
	Moreover for Cases~3 to~5,
	the number of iterations, the quantifier depth,
	and the number of rounds required
	to distinguish $G$ and $G'$
	differ by at most $k$.
\end{theorem}

\subparagraph{Linear Spaces.}
If $V \subseteq \FF_2^X$ is a linear space over $\FF_2$,
we write $U \leq V$ if $U$ is a subspace of $V$.
The zero element of $V$ is denoted $0_V$ or just $0$ if $V$ is clear from the context.
A set $W \subseteq \FF_2^X$ is an affine space if there is an $\alpha \in \FF_2^X$ and a linear space $V \leq \FF_2^X$ such that $W = \alpha + V$. We say that $V$ is the \defining{linear space of $W$}. It is easy to see that $W = \alpha' + V$ for all $\alpha' \in W$, i.e. that $V$ is unique.

Let $U \subseteq \FF_2^X$ and $W \subseteq \FF_2^Y$ be affine spaces.   
We write \[\exts{U}{W} :=  \setcond*{\alpha \in U}{ \text{there is a $\beta \in W$ such that } \restrict{\alpha}{U\cap W} = \restrict{\beta}{U\cap W} }\]
for the affine subspace of $U$ of all vectors extending to vectors in $W$.
When $W$ is a singleton set $\set{\beta}$, we may also write $\exts{U}{\beta}$. Observe that for all $E \subseteq S$ we have that $\sols{E^0} \leq \FF_2^{\vars{E}}$ and $\sols{E}$ is an affine space. We write $0_E$ to denote $0_{\sols{E^0}}$.

\subparagraph{Systems of Linear Equations over $\FF_2$.}
A \defining{$\FF_2$-system} $S$ on a set of variables $V$ is a set of linear equations over $V$.
We denote by $\homsys{S}$ the homogeneous system derived from $S$.
An equation $e \in S$ \defining{uses a variable} $v \in V$ if $v$ occurs with nonzero coefficient in $e$.
The set of variables used in $e$ is denoted $\vars{e}$. The arity of $e$ is $|\vars{e}|$ and the arity of $S$ is the maximal arity of its equations.
For a tuple $\tup{e} \in S^k$ or subset $F\subseteq S$, we denote the set of variables used in equations in $\tup{e}$ or $F$ by $\vars{\tup{e}}$ and $\vars{F}$, respectively.
The \defining{solution space} of an equation $e$ is
$\sols{e} := \setcond{\alpha\colon \vars{e} \to \FF_2}{\text{$\alpha$ satisfies $e$}}$.
For a tuple $\tup{e}$ or set $F$ we similarly write $\sols{\tup{e}} \subseteq \FF_2^{\vars{\tup{e}}}$ and $\sols{F}$, respectively.
For a subset $E \subseteq F$ and $\alpha \in \sols{F}$,
we denote by $\restrict{\alpha}{E} := \restrict{\alpha}{\vars{E}}$
the restriction of $\alpha$ to the variables of $E$.
We use the same notation for sets of solutions and for tuples and subtuples.

\section{CFI-Like Graphs for Equation Systems} \label{sec:CFI-graphs}

This section introduces CFI-like graphs for arbitrary equation systems over $\FF_2$.
The classical CFI graphs are obtained for so-called Tseitin-systems, which are defined from a so-called base graph~\cite{CaiFuererImmerman1992}. We use a more general construction which starts from an $\FF_2$ system with the following property.
\begin{definition}[Thin $\FF_2$-Systems]
An $\FF_2$-system $S$ is \defining{thin} if every pair $e\neq f \in S$ of distinct equations share at most one variable: $|\vars{e}\cap\vars{f}| \leq 1$.
\end{definition}

We can encode a thin $\FF_2$-system $S$ in a \defining{CFI graph $\CFI{S}$} as follows:
for each equation $e \in S$ and each $\alpha \in \sols{e}$, add a vertex $(e,\alpha)$.
Connect $(e,\alpha)$ and $(f,\beta)$ by an edge if $\vars{e}\cap\vars{f}=\set{v}$ for some $v\in V$ and $\alpha(v) = \beta (v)$.
The \defining{origin} of $(e,\alpha)$ is $e$
and the origin of a $k$-tuple of vertices of $\CFI{S}$ is the entry-wise $k$-tuple of origins.
When considering the CFI graphs $\CFI{S}$ and $\CFI{\homsys{S}}$ 
we will also say that $(\homsys{e},\alpha) \in V(\CFI{\homsys{S}})$ has origin $e$, so that the set of origins is equal for $\CFI{S}$ and $\CFI{\homsys{S}}$.
This is extended to tuples in the obvious way.

For convenience, we also write $|\CFI{S}|:= \sum_{e\in S} |\sols{e}|$ for non-thin systems. We do this since
 thinness is only needed to define the edges of the CFI graph,
not its vertex set.

\begin{lemma}
	\label{lem:cfi-size}
	Let $S$ be an $\FF_2$-system.
	If the 
	 arity of $S$ is $a$,
	then $|\CFI{S}| \leq |S| \cdot 2^{a-1}$.
\end{lemma}
\begin{proof}
	For every equation $e\in S$ of arity $r\leq a$, we add $|\sols{e}| = 2^{r-1}$ vertices. Thus $|\CFI{S}| \leq |S| \cdot 2^{a-1}$.
\end{proof}

Solvability of the system $S$ is equivalent to isomorphism of CFI graphs in the following sense.
We call an isomorphism $\Phi \colon \CFI{S} \to \CFI{\homsys{S}}$
\defining{origin-preserving} if, for all $u\in V(\CFI{S})$, the vertices $u$ and $\Phi(u)$ have the same origin.

\begin{lemma} \label{lem:thin-then-preserving-iso-is-solution}
	Let $S$ be a thin $\FF_2$-system.
	Then $S$ is solvable if and only if there is an origin-preserving isomorphism $\CFI{S} \to \CFI{\homsys{S}}$.
\end{lemma}
\begin{proof}
	Let $\phi \in \sols{S}$. Define a map $\Phi \colon V(\CFI{S}) \to V(\CFI{\homsys{S}})$ for each $e \in S$ and $\alpha \in \sols{e}$ via
	\[\Phi(e,\alpha) := (\homsys{e}, \alpha - \restrict{\phi}{\vars{e}}).\]
	Note that $\alpha - \restrict{\phi}{\vars{e}} \in \sols{\homsys{e}}$ because both $\alpha$ and $\restrict{\phi}{\vars{e}}$ are solutions to $e$ and thus $(\homsys{e}, \alpha - \restrict{\phi}{\vars{e}}) \in V(\CFI{\homsys{S}})$.
	
	We show that $\Phi$ is an isomorphism.
	If $(e,\alpha)(f,\beta) \in E(\CFI{S})$, then $\vars{e}\cap\vars{f} = \set{v}$ for some $v\in V$ and $\alpha(v) = \beta(v)$.
	This implies $(\alpha-\restrict{\phi}{\vars{e}})(v) = \alpha(v) - \phi(v) = \beta(v) - \phi(v) = (\beta - \restrict{\phi}{\vars{f}})(v)$.
	So $\Phi(e,\alpha)\Phi(f,\beta) \in E(\CFI{\homsys{S}})$.
	The case for non-edges is similar.
	It is clear that $\Phi$ is origin-preserving.
	
	Let $\Phi \colon V(\CFI{S}) \to V(\CFI{\homsys{S}})$ be an origin-preserving isomorphism
	and we aim to define a solution $\phi$.
	Let $v \in V$ be a variable and $e \in S$ such that $v \in \vars{e}$ and let $(e,\alpha) = \inv{\Phi}(\homsys{e}, 0)$ (the all zero map $0$ is clearly in $\sols{\homsys{e}}$).
	Define $\phi(v) := \alpha(v)$.
	We first show that this definition is independent of $e$.
	Let $e,f\in S$ such that $\vars{e}\cap \vars{f} = \set{v}$
	(note that because $S$ is thin, if $v \in \vars{e}\cap \vars{f}$, then $\vars{e}\cap \vars{f} = \set{v}$).
	Then $(\homsys{e},0)(\homsys{f},0)\in E(\CFI{\homsys{S}})$
	and so $\inv{\Phi}(\homsys{e}, 0)\inv{\Phi}(\homsys{f}, 0) \in E(\CFI{S})$.
	This implies that $\inv{\Phi}(\homsys{f}, 0) = (f,\beta)$ satisfies $\beta(v) = \alpha(v)$.
	
	To see that $\phi$ is a solution of $S$, let $e \in S$
	and $(e,\alpha) = \inv{\Phi}(\homsys{e}, 0)$.
	By the construction of $\phi$, we have that $\restrict{\phi}{\vars{e}} = \alpha$ and thus $\phi$ satisfies $e$.
\end{proof}

The restriction to thin $\FF_2$-systems and origin-preserving isomorphisms can be avoided.
Thinness can either be avoided by using edge colors or introducing gadgets not only for equations but also for variables (sometimes called inner and outer vertices of CFI gadgets).
Origin-preservation can be avoided by coloring the vertices of the CFI graph with its origin.
Since we aim for uncolored graphs, we modify the $\FF_2$-systems to be thin and only admit origin-preserving isomorphisms.
This has the benefit that we neither need two distinct types of gadgets nor an encoding of vertex colors on the graph level. Hence, all vertices of the CFI graph can be treated uniformly, simplifying later proofs.
Before we see these reductions, we will need to consider the relationship between a $\FF_2$-system and its CFI graph in more detail.

\subsection{Representing Vertex-Tuple Sets and Maps via Solutions}
Maps between vertices of CFI graphs can be represented by local solutions of the underlying equation system.
These local solutions naturally act on the vertices.
Using this action, certain sets of vertices (or of vertex-tuples) can be represented by linear or affine spaces. To formalize this we need some notation.

So let $S$ be a homogeneous $\FF_2$-system and $b,c \colon S\to \FF_2$.
For a vertex $u = (f,\alpha) \in V(\CFI{S+b})$ with origin $f \in S$,
a set of equations $E \subseteq S+(c-b)$ such that $f + (c-b) \subseteq E$,
and a solution $\beta \in \sols{E}$, we write $\beta + u$
for the vertex $(f, \alpha + \restrict{\beta}{\vars{f}})$
in $\CFI{S+c}$.
For a tuple $\tup{u} \in V(\CFI{S+b})$ with origin $\tup{f} \in S$,
a set $E \subseteq S+(c-b)$ such that all entries of $\tup{f} + (c-b)$ are in $E$,
and a solution $\beta \in \sols{E}$, we write $\beta + \tup{u}$ for the $k$-tuple
of vertices of $\CFI{S+c}$
whose $i$-th entry is $\beta + u_i$ (for all $i \in [k]$).
For an affine subspace $W = \alpha +V$ of $\sols{E}$, where $V \leq  \sols{\homsys{E}}$ is a subspace,
we write $W + \tup{u}$ for the set $\setcond{\gamma + \tup{u}}{\gamma \in W}$
of vertices of $\CFI{S+c}$.
Similarly, for a set of $k$-tuples $U \subseteq V(\CFI{S+b})$ such that all tuples in $U$ have origin $\tup{f}$,
we write $\alpha + U$ for $\setcond{\alpha + \tup{v}}{\tup{v} \in U}$.

Now let $S$ be an arbitrary $\FF_2$-system. Then $S = \homsys{S} + b$ for some $b$.
We mostly use the above notation in cases where either $c = b$ (to describe sets of vertices of $\CFI{S}$ using solutions of the homogeneous system $\homsys{S}$ as $V+ \tup{u}$ for $\tup{u} \in V(\CFI{S})^k$)
or where $c = 0$ (to describe such sets as $W+\tup{u}$
for $\tup{u} \in V(\CFI{\homsys{S}})^k$ and $W$ is an affine subspace of local solutions of $S$).
We now show that these operations indeed induce a meaningful action on the CFI graph.

\begin{lemma}
	\label{lem:solutions-preserve-isomorphism-type}
	Let $S$ be a thin $\FF_2$-system and $\eqTA \in S^k$.
	Two tuples $\tup{u} \in V(\CFI{\homsys{S}})^k$
	and $\tup{v} \in V(\CFI{S})^k$ of origin $\eqTA$ have the same isomorphism type in $\CFI{\homsys{S}}$ and $\CFI{S}$, respectively,
	if and only if there is an
	$\alpha \in \sols{\eqTA}$ such that
	$\tup{u} + \alpha = \tup{v}$.
\end{lemma}
\begin{proof}
	Let $\tup{u} \in V(\CFI{\homsys{S}})^k$ be of origin $\eqTA$.
	For the backward direction, let
	$\alpha \in \sols{\eqTA}$ and $\tup{v} = \tup{u} + \alpha$.
	We have to show that for every distinct $i,j$ the vertices $u_i$ and $u_j$ are adjacent in $\CFI{\homsys{S}}$ if and only if $v_i = u_i + \alpha$ and $v_j = u_j+ \alpha$ are adjacent in $\CFI{S}$.
	Let $u_i = (e,\beta)$ and $u_j = (f,\gamma)$.
	Then $u_i + \alpha = (e,\beta + \restrict{\alpha}{\vars{e}})$
	and similarly for $u_j$. Note that $\beta + \restrict{\alpha}{\vars{e}} \in \sols{e}$ because $\beta \in \sols{\homsys{e}}$ and $\restrict{\alpha}{\vars{e}} \in \sols{e}$.
	
	If $u_i$ and $u_j$ are adjacent, then there is a variable $v$ such that
	$\vars{e} \cap \vars{f} = \set{v}$ and $\beta(v) = \gamma(v)$.
	Thus, $(\beta + \restrict{\alpha}{\vars{e}})(v) = (\gamma + \restrict{\alpha}{\vars{e}})(v)$, meaning that $u_i + \alpha$ and $u_j + \alpha$ are adjacent in $\CFI{S}$.
	Similarly, if $u_i$ and $u_j$ are not adjacent, then neither are $u_i + \alpha$ and $u_j + \alpha$.
	
	For the forward direction,
	let $\tup{v} \in V(\CFI{S})^k$ be of origin $\eqTA$ have the same isomorphism type in $\CFI{S}$ as $\tup{u}$ has in $\CFI{\homsys{S}}$.
	There are solutions $\alpha_i \in \sols{\eqTAm_i}$
	such that $u_i + \alpha_i = v_i$ by the definition of CFI graphs.
	We need to show that these solutions agree on common variables.
	Because $S$ is thin, $|\vars{\eqTAm_i}\cap\vars{\eqTAm_j}| \leq 1$.
	If this is zero, there is nothing to show.
	Otherwise, $\vars{\eqTAm_i}\cap\vars{\eqTAm_j} =\set{v}$.
	Then $u_i = (\homsys{\eqTAm}_i, \beta_i)$, $v_i = (\eqTAm_i, \gamma_i)$
	and similar for $j$.
	Because $\tup{u}$ and $\tup{v}$ have the same isomorphism type,
	there is an edge $u_iu_j$ if and only if $v_iv_j$.
	Hence, $\beta_i(v) = \beta_j(v)$ if and only if $\gamma_i(v) = \gamma_j(v)$.
	This implies that $\alpha_i(v) = \alpha_j(v)$ because
	$\beta_i(v) + \alpha_i(v) = \gamma_i(v)$ and similarly for $j$.
\end{proof}

\subsection{The $k$-Equation Game}

The power of the $k$-dimensional Weisfeiler-Leman algorithm to distinguish CFI graphs is captured by the following game on the underlying equation systems played by Verifier and Falsifier.

Let $S$ be an $\FF_2$-system. A position in the \defining{$k$-equation game} is a pair
$(F,\alpha)$ of a set of equations $F\subseteq S$ of size at most $k$ and an $\alpha \in \sols{F}$. The initial position is empty.
Then the game is played as follows:
\begin{enumerate}
	\item If $|F| =k$, Falsifier chooses an equation $f \in F$ to forget
	and  updates $\alpha \leftarrow \restrict{\alpha}{\vars{F\setminus\set{f}}}$
	and $F \leftarrow F \setminus\set{f}$.
	\item Falsifier chooses an equation $e \in S$ and sets $E := F \cup \set{e}$.
	\item Verifier chooses a solution $\beta \in \exts{\sols{E}}{\set{\alpha}}$ and the game continues in position $(E,\beta)$.
	If no such solution exists, then Falsifier wins.
\end{enumerate}
Verifier wins if Falsifier never wins.

Since we need to consider origin-preserving isomorphisms of CFI graphs,
we for now assume that the Weisfeiler-Leman algorithm can distinguish different origins (and thus all isomorphisms have to be origin-preserving).

\begin{definition}
	We say that $k$-WL \defining{detects origins} of a thin $\FF_2$-system $S$ in $d$ iterations,
	if for all $G,H \in \set{\CFI{\homsys{S}}, \CFI{S}}$
	and all $u \in V(G)$ and $v \in V(H)$ of distinct origin
	Spoiler wins the $d$-round bijective $(k+1)$-pebble game on $G,H$ from position $(u), (v)$.
\end{definition}

The next two lemmas formalize the relationship between the $k$-equation game and the bijective $k$-pebble game (and hence via Theorem~\ref{thm:hom-eq-iff-biso-hom-eq} with the $k$-dimensional Weisfeiler-Leman algorithm and treewidth-$k$-distinguishability).

\begin{lemma}
	\label{lem:playerI-equation-game}
	Let $k \ge 2$ and $S$ be a thin $\FF_2$-system
	of which $(k-1)$-WL  detects origins in $d$ iterations.
	If Falsifier wins the $r$-round $k$-equation game on $S$,
	then Spoiler wins the $(r+d)$-round bijective $k$-pebble game on  $\CFI{\homsys{S}}$ and $\CFI{S}$. In particular,  $\CFI{\homsys{S}}$ and $\CFI{S}$ are treewidth-$(k-1)$-distinguishable.
\end{lemma}
\begin{proof}
	We show that Spoiler has a strategy to maintain the following invariant during the game:
	If $\tup{u},\tup{v}$ is the current position in the bijective pebble game
	and $r$ rounds are left,
	then
	\begin{enumerate}
		\item $\tup{u}$ and $\tup{v}$ have the same origin $\tup{f}$, in which no equation repeats.
		\item There is an $\alpha\in \sols{\tup{f}}$ such that $\tup{u} + \alpha = \tup{v}$.
		\item Falsifier wins in $r$ rounds from position $(F,\alpha)$ in the $k$-equation game on $S$, where $F$ is the set of entries of $\tup{f}$.
	\end{enumerate}
	Clearly, the starting position satisfies the invariant.
	The proof proceeds by induction on the number $r$ of rounds Falsifier needs to win in position $(F,\alpha)$. 
	
	So suppose that we are in position  $\tup{u},\tup{v}$ in the bijective $k$-pebble game satisfying the invariant. If $|F| = k$, then in the equation game Falsifier chooses an equation $f$ to forget according to their winning strategy and afterwards chooses 
	a new equation $e$. If $|F| <k$, then Falsifier just chooses a new equation $e$.
	In any case, we obtain a new equation set $E$.
	We can assume that $E$ is not a subset of $F$ since otherwise the game does not make progress.
	
	If $|F|=k$, there is a unique $i \in [k]$ such that $u_i$ and $v_i$ are of origin $f$.
	Spoiler chooses this $i$ to forget and we update $\tup{u},\tup{v} \leftarrow \tup{u}\setminus i, \tup{v}\setminus i$.
	Then the invariant is satisfied again, in this intermediate position.
	
	Duplicator then provides a bijection $h \colon V(\CFI{\homsys{S}}) \to V(\CFI{S})$.
	If $h$ is not origin-respecting, then there is a $w \in V(\CFI{S})$
	such that $w$ and $h(w)$ have distinct origins.
	Because $(k-1)$-WL detects origins, Spoiler can win by placing a pebble on $w$ and $h(w)$ and then playing another $d$ rounds.
	So we can assume that $h$ is origin-respecting.
	Falsifier chooses any vertex $w \in V(\CFI{\homsys{S}})$ with origin $e$.
	Spoiler places a pebble on $w$ and on $h(w)$.

	Assume that  there is an $\alpha' \in \exts{\sols{E}}{\set{\restrict{\alpha}{\vars{E\cap F}}}}$
	such that $\restrict{\alpha'}{\vars{e}} + w = h(w)$, so that $\alpha'$ satisfies (2) from the invariant.
	By the choice of $w$, invariant (1) is satisfied.
	Because we followed Falsifier's $r$-round winning strategy from position $(F, \alpha)$ in the $k$-equation game, Falsifier wins in $r-1$ rounds from position $(E, \alpha')$.
	Thus $\tup{u}w, \tup{v}h(w)$ also satisfies the invariant (3)
	and Spoiler wins by the inductive hypothesis.

	Otherwise, there is no such $\alpha'$ such that 
	$\restrict{\alpha'}{\vars{e}} + w = h(w)$.
	This in particular happens if $0$ rounds are left and Falsifier wins.
	Because $w$ and $h(w)$ have the same origin,
	every $\alpha' \in \sols{E}$ such that $\alpha' + \tup{u}w = \tup{v}h(w)$
	has to be in $\exts{\sols{E}}{\set{\restrict{\alpha}{\vars{E\cap F}}}}$
	since the image on $\vars{E\cap F}$ is fixed.
	Thus, by Lemma~\ref{lem:solutions-preserve-isomorphism-type},
	$\tup{u}w$ and $\tup{v}h(w)$ have different isomorphism types in
	$\CFI{\homsys{S}}$ and $\CFI{S}$, respectively and Spoiler wins the game.
	
	Treewidth-$(k-1)$-distinguishability follows from Theorem~\ref{thm:hom-eq-iff-biso-hom-eq}.
\end{proof}

\begin{lemma}
	\label{lem:playerII-equation-game}
	Let $k \ge 2$ and $S$ be a thin $\FF_2$-system.
	If Verifier wins the $r$-round $k$-equation game on $S$, then
	Duplicator wins the $r$-round bijective $k$-pebble game on $\CFI{\homsys{S}}$ and $\CFI{S}$.
\end{lemma}
\begin{proof}
	Assume that Verifier has a winning strategy in the $k$-equation game on $S$.
	We show that Duplicator has a winning strategy in the bijective $k$-pebble game played on $\CFI{\homsys{S}}$ and $\CFI{S}$,
	that maintains the following invariant in position $\tup{u},\tup{v}$ when $r$ rounds are left:
	\begin{enumerate}
		\item $\tup{u}$ and $\tup{v}$ have the same origin $\tup{f}$ in which no equation repeats.
		\item There is an $\alpha \in \sols{\tup{f}}$ such that $\tup{u} + \alpha = \tup{v} $.
		\item Verifier wins in $r$ rounds from position $(F,\alpha)$ in the $k$-equation game, where $F$ is the set of entries of $\tup{f}$. 
	\end{enumerate}
	The proof proceeds by induction on $r$. The initial position clearly satisfies the invariant.
	
	By Lemma~\ref{lem:solutions-preserve-isomorphism-type}, the tuples induce a local isomorphism.
	If $|\tup{u}|=k$, Spoiler chooses an index $i\in[k]$ and the game position is updated to position $\tup{u},\tup{v} \leftarrow \tup{u}\setminus i, \tup{v} \setminus i$.
	Duplicator updates $\tup{f}$ and $F$ accordingly.
	
	To provide a bijection, for each $e \in S$, Duplicator assumes that Falsifier chooses the set $F \cup \set{e}$ in the $k$-equation game, for which Verifier answers with $\alpha_e \in \sols{F\cup \set{e}}$.
	Duplicator chooses the bijection $h$ defined as $h((\homsys{e},\beta)) := (e,\beta + \restrict{\alpha_e}{\vars{e}})$.
	
	Spoiler chooses a vertex $\tup{w} = (\homsys{e}, \beta)$
	and the game proceeds to position $\tup{u}w, \tup{v}h(w)$.
	This position satisfies $\tup{u}w + \alpha_e = \tup{v}h(w)$ by the definition of $h$ and Falsifier wins from position $(F\cup \set{e}, \alpha_e)$ because Duplicator followed a winning strategy of Verifier in the $k$-equation game.
\end{proof}

\subsection{Making Systems Thin -- Recover Variable Gadgets} \label{ss:thin-cfi}
We now show how to make $\FF_2$-systems thin while
preserving the winner of the $k$-equation game.

Let $S$ be an $\FF_2$-system. Define the $\FF_2$-system $\thin{S}$ as follows:
For each equation $e \colon x_1+\dots + x_a = b$ in $S$,
add a copy on  fresh variables
$e' \colon x_1^e+\dots + x_a^e = b$ to $\thin{S}$.
For each $i \in [a]$, add the equation
$x_i^e = x_i$.
We call these equations the variable equations.

This construction essentially recovers the CFI-encoding
of $\FF_2$-systems, in which for every variable two vertices are added
and the vertices of the equations are connected to the variable vertices
(sometimes also known as inner and outer vertices).
This is done by the variable equations: 
The equation $x^e = x$ adds for $e$ a pair of vertices for the variable $x$,
which get connected to vertices of 
variable equations $x^f = x$.
The construction of the variable vertices as additional equations
allows us to treat all vertices in a uniform way.

\begin{lemma}
	\label{lem:thinnify-system}
	Let $S$ be an $\FF_2$-system of arity $a$. Then for all $k\geq 3$ and $r \in \nat$
	\begin{enumerate}
		\item the thin $\FF_2$-system $\thin{S}$ is of arity $a$ and satisfies $|\CFI{\thin{S}}| = O(|\CFI{S}|)$;
		\item if Falsifier wins the $r$-round $k$-equation game on $S$,
		then Falsifier wins the $(r+O(1))$-round $k$-equation game on $\thin{S}$; and
		\item if Falsifier wins the $r$-round $k$-equation game on $\thin{S}$,
		then Falsifier wins the $r$-round $k$-equation game on $S$.
	\end{enumerate}
\end{lemma}
\begin{proof}
	Let $S' = \thin{S}$.
	One easily sees that $S'$ is thin.
	For every $a$-ary equation in $S$, we add one $a$-ary equation to $S'$ along with 
	 $a$ binary equations.
	 For the $a$-ary equation, we add $2^{a-1}$ vertices to the CFI graph.
	For the additional equations, we add $2a \in O(2^{a-1})$ many vertices.
	Thus $|\CFI{\thin{S}}| = O(|\CFI{S}|)$.

	Given a winning strategy of Falsifier in the $r$-round $k$-equation game on $S$, 
	construct a winning strategy on $S'$ as follows:
	whenever Falsifier chooses an equation $e \in S$,
	let them choose $e' \in S'$.
	If on $S'$ Verifier picks an assignment $\alpha$
	that assigns different values to $x^e$ and $x^f$,
	then Falsifier can choose the equation $x^e = x$,
	forget the equation $e'$, and then choose $x^f = x$
	and win.
	Since $k\geq 3$ this is possible.
	If Verifier assigns the copy of the same variable consistently,
	the assignment on $S'$ translates to an assignment on $S$,
	where Falsifier can follow their winning strategy.
	Note that when Falsifier wins on $S$, then Verifier has to assign copied variables inconsistently, so Falsifier wins in 
	a constant number of additional rounds.
	
	For the converse direction,
	assume that Falsifier has a winning strategy on $S'$.
	Then Falsifier wins the game on $S$ as follows:
	If Falsifier chooses an equation $e'$ on $S'$, choose $e$ on $S$.
	If Falsifier chooses an equation $x^e = x$ on $S'$, choose the equation $e$ on $S$.
	The assignment on $S$ chosen by Verifier always induces an assignment on $S'$,
	so Falsifier can follow their winning strategy on $S'$ to win on $S$.
	Since these induced assignments always assign copied variables consistently,
	Falsifier wins  on $S$ when Verifier wins on $S'$ (or possibly earlier).
\end{proof}

\subsection{Coloring the CFI Graphs}
\label{sec:coloring-cfi}

Let $S$ be an $\FF_2$-system.
Let $e_1,\dots, e_m$ be an arbitrary enumeration of the equations of $S$.

Define the system $\cld{S}$ as follows:
Let $x_{-4},\dots, x_{m+1}, y_{1},\dots, y_{m}, z_1,\dots, z_m$ be fresh variables which we add to $\cld{S}$ as follows:
\begin{align*}
	x_{-4} &= 0 \\
	x_{i}  & = x_{i-1} & \text{for all }-3\leq i \leq m+1\\
	y_{i} & = x_{i}& \text{for all } i \in[m]\\
	z_{i} & = y_{i}& \text{for all } i \in[m]\\
	\cld{e}_i \colon  w_1+\dots+w_\ell + z_i &= a  &\text{for all }e_i \colon w_1+\dots+w_\ell = a, i \in [m]
\end{align*}

We call the additional $x_i$, $y_i$, and $z_i$ variables the \defining{new variables} and the ones from $S$ the \defining{old variables}.
Similarly, the $\cld{e}_i$ are the \defining{old equations} and all other equations in $\cld{S}$ are \defining{new equations}.
Clearly, the system $\cld{S}$ is satisfiable if and only if $S$ is satisfiable
because all new variables have to be set to zero by every solution
and then $\cld{e}_i$ is equivalent to $e_i$.
It is easy to see that $\homsys{(\cld{S})} = \cld{(\homsys{S})}$
and we just write $\homsys{\cld{S}}$.

\begin{lemma}
	\label{lem:cld-thinness}
	If $S$ is thin, then $\cld{S}$ is thin.
\end{lemma}
\begin{proof}
	All new equations are binary, sharing at most one variable with another equation (old or new).
	The old equations are just copies of the equations in $S$
	with one variable added, that is only shared with one new equation and no other. 
	Thus, thinness is preserved.
\end{proof}

\begin{lemma}
	\label{lem:coloring-size}
	$|\CFI{\cld{S}}| = O(|\CFI{S}|)$.
\end{lemma}
\begin{proof}
	Since the arity of each equation increases by one,
	the number of vertices for them doubles.
	There are $O(|S|)$ many new equations each of arity at most two.
	This adds $O(|S|)$ additional vertices.
\end{proof}

\begin{lemma}
	\label{lem:wl-detects-origin}
	$1$-WL detects the origins of $\cld{S}$ in $O(|S|)$ iterations.
\end{lemma}
\begin{proof}
	Consider $\CFI{\cld{S}}$ or $\CFI{\homsys{\cld{S}}}$ (the following argument does not depend on this).

	The vertices, whose origin is a new equation,
	induce two disjoint trees:
	one for assigning the new variables to $0$, and one for assigning them to one.
	The root of one tree is the unique vertex with origin $x_{-4}=0$,
	the other one the vertex with origin $x_{-3}=x_{-4}$, that assigns the variables to one.
	The leaves are the vertices with origin $y_i=z_i$ and the ones with origin $x_{m+1}=x_m$.
	Those vertices  with origin  $y_i=z_i$ are connected to the vertices with origin $\cld{e}_i$,
	since both equations use $z_i$.
	In particular, every vertex of origin $\cld{e}_i$ has a neighbor of origin
	$y_i=z_i$.
	
	We argue that there is no path of length $5$ in which all vertices
	are of degree two and one  vertex has origin $\cld{e}_i$.
	This is the case because every such vertex has a neighbor of origin  $y_i=z_i$,
	so can have at most one other neighbor with origin $\cld{e}_j$.
	Since all vertices with origin $x_i=x_{i-1}$ for $i \in [m]$ have degree $3$,
	such a path indeed does not exist.
	
	However a path of length 5 in which all vertices are of degree two exists at the root of the two induced trees,
	so the roots get a unique color. Therefore, the tree vertices get a unique color by 1-WL,
	and so the vertices of origin $\cld{e}_i$ get different colors than the ones of origin $\cld{e}_j$.
	This means that 1-WL detects the origins of $\cld{S}$.
	Because this tree has $O(|S|)$ vertices, it does so in $O(|S|)$ iterations.
\end{proof}

\begin{lemma}
	\label{lem:color-reduction-equation-game}
	Let $k\geq 2$.
	If Falsifier wins the $r$-round $k$-equation game on $S$,
	then Falsifier wins the $r+O(1)$-round $k$-equation game on $\cld{S}$.
\end{lemma}
\begin{proof}
	By Lemma~\ref{lem:playerI-equation-game},
	it suffices to show that $(k-1)$-WL detects the origins of $\cld{S}$ and that 	Falsifier wins the $k$-equation game on $\cld{S}$. By Lemma~\ref{lem:wl-detects-origin} 1-WL can detect the origins of $\cld{S}$, therefore so can $(k-1)$-WL.
	
	\let\cldold\cld
	\renewcommand{\cld}[1]{\dot{#1}}
	We now write $\cld{S}$ for $\cldold{S}$.
	To show that Falsifier wins the $k$-equation game on $\cld{S}$, we adapt Falsifier's winning strategy on $S$:
	Falsifier follows the strategy of $S$ as long as Verifier sets all new variables to zero. Then Falsifier will eventually win.
	If Verifier sets a new variable to one, Falsifier follows the binary equations to $x_{-4}$ to derive the contradiction. In detail:
	
	Let $\cld{F} \subset \cld{S}$ be a set of equations and let $\cld{\alpha} \in \sols{\cld{F}}$.
	We call $(\cld{F},\cld{\alpha})$ good if $\cld{\alpha}$ always sets new variables to zero.
	We say that a good pair $(\cld{F},\cld{\alpha})$ corresponds to the pair $(F,\alpha)$ defined via $F = \setcond{e_i}{\cld{e}_i \in \cld{F}}$ and $\alpha = \restrict{\cld{\alpha}}{\vars{F}}$.
	
	We claim that Falsifier can maintain the following invariant:
	If the current position $(\cld{F},\cld{\alpha})$ on $\cld{S}$ is good,
	then Falsifier has a winning strategy from the corresponding position $(F,\alpha)$ on $S$.
	Clearly, the initial position is good and satisfies the invariant,
	so the claim implies the lemma.
	
	We prove the claim by induction on the number of rounds Falsifier needs to win in the corresponding position.
	Assume that Falsifier wins on the corresponding position $(F,\alpha)$ in $r+1$ rounds.
	Suppose that on $S$ Falsifier chooses a set $E\subseteq S$ such that $|E\setminus F|\leq1$ according to Falsifier's winning strategy.
	On $\cld{S}$ we let Falsifier choose $\cld{E}  :=\setcond{\cld{e}}{e\in E}$, which 
	satisfies $|\cld{E} \setminus \cld{F}| \leq 1$.
	Now Verifier chooses $\cld{\beta} \in \exts{\sols{E}}{\set{\cld{\alpha}}}$.
	If  $\cld{\beta}$ sets all new variables (in its domain) to zero,
	we have $\beta := \restrict{\cld{\beta}}{\vars{E}} \in \exts{\sols{E}}{
		\set{\alpha}}$.
	Then $(E,\beta)$ corresponds to $(\cld{E},\cld{\beta})$.
	If Verifier chooses $\beta$ on $S$, then Falsifier wins in $r-1$ rounds from $(E,\beta)$.
	Hence, by the inductive hypothesis, Falsifier wins on $S$ from $(\cld{E},\cld{\beta})$.
	
	If $\cld{\beta}$ sets a new variable to one,
	then there is a chain of new equations $f_1,\dots, f_\ell$
	such that $f_1$ contains this new variable,
	$f_\ell \colon x_{-4} = 0$, and $f_i$ and $f_{i+1}$ share a new variable.
	Falsifier chooses $f_1$ in the next round (forgetting one arbitrary other equation) and the assignment chose by Verifier has to set both new variables of $f_1$ to one.
	Then Falsifier follows the chain:
	If the current position contains $f_1$ and the assignment sets both variables to one, then Falsifier chooses $f_2$ and the assignment chosen by Verifier
	has to set both variables of $f_2$ to one.
	Finally, Falsifier will arrive at $f_\ell$,
	where Verifier cannot choose a valid extension anymore
	and Falsifier wins.
\end{proof}

 It is not hard to see that strategies of Verifier also translate from $S$ to $\cld{S}$
by always setting new variables to zero.
\begin{lemma}
	\label{lem:color-reduction-equation-game-fi}
	Let $k\geq 2$.
	If Verifier wins the $r$-round $k$-equation game on $S$,
	then Verifier wins the $r$-round $k$-equation game on $\cld{S}$.
\end{lemma}

Combining the thinness with the coloring reduction yields:
\begin{lemma}
	\label{lem:main-wl-iff-equation-game}
	For every $\FF_2$-system $S$ of arity $a$, there is an $\FF_2$-system $S'$
	such that
	\begin{enumerate}
		\item $S'$ is of arity $a+1$ such that $|\CFI{S'}| = O(|\CFI{S}|)$, and for all positive integers $k\geq 3$ and $r$
		\item if Verifier wins the $r$-round $k$-equation game on $S'$,
		then $\CFI{S}$ and $\CFI{\homsys{S}}$ 
		are equivalent in quantifier-depth-$r$ $k$-variable counting logic,
		\item  if Falsifier wins the $r$-round $k$-equation game on $S'$,
		then $\CFI{S}$ and $\CFI{\homsys{S}}$ 
		are not equivalent in quantifier-depth-$(r+O(|S|))$ $k$-variable counting logic,
		\item if Falsifier wins the $r$-round $k$-equation game on $S$,
		then Falsifier wins the $r+O(1)$-round $k$-equation game on $S'$, and
		\item if Falsifier wins the $r$-round $k$-equation game on $S'$,
		then Falsifier wins the $r$-round $k$-equation game on $S$.
	\end{enumerate}
\end{lemma}
\begin{proof}
	By Lemma~\ref{lem:thinnify-system},
	the thin $\FF_2$-system $S'' := \thin{S}$ is of arity $a$
	such that $|\CFI{S''}| = O(|\CFI{S}|)$.
	The winner of the $k$-equation game is preserved,
	from $S$ to $S''$, Falsifier may need $O(1)$ additional rounds.
	
	We then apply the color reduction to $S''$.
	Let $S' := \cld{S''}$.
	This increases the arity by one
	and we have $|\CFI{S'}| = O(|\CFI{S''}|) = O(|\CFI{S}|)$ by Lemma~\ref{lem:coloring-size}.
	The winner of the $k$-equation game is again preserved by Lemmas~\ref{lem:color-reduction-equation-game} and \ref{lem:color-reduction-equation-game-fi}. 
	
	The color reduction preserves thinness (Lemma~\ref{lem:cld-thinness}).
	So if Falsifier wins the $r$-round $k$-equation game on $S'$,
	then Spoiler wins the bijective $k$-pebble game on $\CFI{S}$ and $\CFI{\homsys{S}}$  in $r+O(|S|)$ rounds by Lemma~\ref{lem:playerI-equation-game}
	because $1$-WL detects origins in $O(|S|)$ iterations (Lemma~\ref{lem:wl-detects-origin}).
	If Verifier wins the $k$-equation game,
	then Duplicator wins the bijective $k$-pebble game in the same number of rounds by Lemma~\ref{lem:playerII-equation-game}.
\end{proof}

\subsection{An Easy Exponential Lower Bound} \label{ss:for-free-lower-bound}

We now show how an exponential lower bound on the order of a graph $H$ of treewidth $k$ required to distinguish two graphs $G$ and $G'$ can be obtained from the literature.
It is based on a robust round number lower bound for $k$-WL on $k$-ary relational structures~\cite{GroheLichterNeuen2025}.
This bound is based on a relational CFI encoding of $k$-ary $\FF_2$-systems.
The hardness of the relational CFI encoding is captured by the following $k$-variable game: 
It is played on an $\FF_2$-system $S$ by two players Verifier and Falsifier.
A position is a set $X \subseteq \vars{S}$ of size at most $k$ and an assignment $\alpha \colon X \to \FF_2$. The initial position is empty.
\begin{itemize}
	\item Falsifier chooses a variable $x \in \vars{S}$ and a set $X' \subseteq X \cup \set{x}$ of size at most $k$.
	\item Verifier chooses an assignment $\alpha' \colon X' \to \FF_2$ that agrees with $\alpha$ on their common domain.
	If this assignment violates an equation in $S$, then Falsifier wins.
\end{itemize}
Verifier wins, if Falsifier never wins.
We also consider the $r$-round game: the game is limited to $r$ rounds and Verifier wins if Falsifier does not win in $r$ rounds.

What remains is to relate the $k$-variable game to the graph-based CFI encoding to avoid $k$-ary relations. Here again, the robustness will be useful.
\begin{lemma}
	\label{lem:variable-equation-game}
	Let $S$ be an $a$-ary $\FF_2$-system.
	\begin{enumerate}
		\item 	If Falsifier wins the $r$-round $k$-equation game on $S$,
		then Falsifier wins the $ar$-round $ak$-variable game on $S$.
		\item If Falsifier wins the $r$-round $k$-variable game on $S$,
		then Falsifier wins the $r$-round $k$-equation game on $S$.
	\end{enumerate}
	
\end{lemma}
\begin{proof}
	For (2), observe that in the equation game Falsifier
	can choose an equation containing a variable $x$ instead of choosing $x$ itself. By doing this, for every move in Falsifier's winning strategy in the variable game, Falsifier will win the $k$-equation game. 
	
	For (1),
	we show the contraposition:
	If Verifier wins the $ar$-round $ak$-variable game on $S$,
	then Verifier wins the $r$-round $k$-equation game on $S$.
	We show by induction on $r$ that Verifier maintains the following invariant:
	If the $k$-equation game on $S$ is in position $(E, \alpha)$
	and $r$ rounds remain,
	then Verifier has an $ar$-round winning strategy  in the $ak$-variable game on $S$ in position $(\vars{E}, \alpha)$.
	Because the invariant is clearly satisfied by the initial empty positions,
	proving this claim proves the lemma.
	
	If $r=0$, then Verifier wins the $k$-equation game.
	So assume $r\geq 0$,
	that the $k$-equation game is in position $(E, \alpha)$
	and that Verifier has an $ar$-round strategy in the $ak$-variable game from position $(\vars{E},\alpha)$.
	In the $k$-equation game,
	Falsifier chooses an equation $f\in S$ and a subset $F\subseteq E \cup \set{f}$ of size at most $k$.
	To determine the solutions that Verifier chooses,
	let $\vars{E} \setminus \vars{F}$
	and let $\vars{F} \setminus \vars{E}=\set{x_1,\dots,x_\ell} \subseteq \vars{f}$.
	
	Let Falsifier in the $ak$-variable game choose $X'_0 = \vars{F} \setminus \set{f} \subseteq E$.
	Then Verifier has no choice and the game proceeds to $(X'_0, \restrict{\alpha}{X'_0})$.
	Now let Falsifier iteratively choose $X'_{i+1} = X'_{i} \cup \set{x_i}$ until $X'_\ell = \vars{F\cup \set{f}}$.
	Note that $|X'_\ell| \leq ak$ because $|F\cup \set{f}|\leq k$
	and every equation is of arity at most $a$.
	Also note that $\ell \leq a$.
	Because Verifier has an $ar$-round strategy from position $(\vars{E},\alpha)$, the
	$ak$-variable game builds up a solution
	$\beta \in \sols{F}$ 
	and reaches position $(\vars{F}, \beta)$,
	from which Verifier has an $a(r-1)$-round strategy.
	
	In the $k$-equation game, Verifier chooses this assignment $\beta$
	and the game reaches position $(F,\beta)$.
	Verifier wins the next $r-1$ rounds by the inductive hypothesis.
\end{proof}

\begin{theorem}[{\cite[Theorem 1.2]{GroheLichterNeuen2025}}]
	\label{thm:robust-wl-round-numbers}
	There are absolute constants $k_0 \in \nat$, $\alpha>0$, and $1> \epsilon > 0$ such that, for every $\ell \geq k \geq k_0$ and every $n \geq \alpha \ell^8k^6$, there is an $\FF_2$-system $S$ of arity at most $k$ on 
	 $n$ variables such that 
	\begin{itemize}
		\item Falsifier wins the $k$-variable game on $S$, 
		but
		\item Verifier  wins the $n^{\epsilon k}$-round $\ell$-variable game on $S$.
	\end{itemize}
\end{theorem}

\begin{corollary}
	\label{cor:robust-wl-round-numbers-equation}
	There are absolute constants $k_0 \in \nat$, $\alpha>0$, and $1> \epsilon > 0$ such that, for every $\ell \geq k \geq k_0$ and every $n \geq \alpha \ell^8k^{14}$, there is an $\FF_2$-system $S$ of arity at most $k+1$ on at least $n$ variables
	such that $|\CFI{S}|$ is of order $O(2^kn^k)$ and
	\begin{itemize}
		\item  $\CFI{S}$ and $\CFI{S^0}$ are not equivalent in $k$-variable counting logic, and
		\item $\CFI{S}$ and $\CFI{S^0}$ are equivalent in quantifier-depth-$n^{\epsilon k}$ $\ell$-variable counting logic.
	\end{itemize}
\end{corollary}

\begin{proof}
	Let $k_0$, $\alpha$, and $\epsilon$ be the constants of Theorem~\ref{thm:robust-wl-round-numbers}.
	Let $\ell \geq k \geq k_0$ and $n \geq\alpha (k\ell)^8k^6 = \alpha \ell^8k^{14}$.
	Set $\epsilon'$ to be sufficiently smaller than $\epsilon$
	such that $n^{\epsilon' k} \leq \inv{k}n^{\epsilon k}$.
	Then by Theorem~\ref{thm:robust-wl-round-numbers}
	there is an $\FF_2$-system $S$ of arity at most $k$ on 
	$n$ variables such that 
	\begin{itemize}
		\item Falsifier wins the $k$-variable game on $S$, 
		\item Verifier  wins the $n^{\epsilon k}$-round $k\ell$-variable game on $S$.
	\end{itemize}
	Using Lemma~\ref{lem:variable-equation-game} we move to the equation game:
	\begin{itemize}
		\item Falsifier wins the $k$-equation game on $S$, 
		\item Verifier  wins the $n^{\epsilon' k}$-round $\ell$-equation game on $S$.
	\end{itemize}
	By Lemma~\ref{lem:main-wl-iff-equation-game},
	there is an $\FF_2$-system $S'$ such that
	\begin{enumerate}
		\item $S'$ is of arity at most $k+1$, size $|S'| = O(k|S|)$ and $S'$ contains at most $|S|$ equations of arity larger than $3$, and for all positive integers $k\geq 3$ and $r$
		\item  $\CFI{S}$ and $\CFI{\homsys{S}}$  are not equivalent in $k$-variable counting logic
		\item $\CFI{S}$ and $\CFI{\homsys{S}}$  are equivalent  in quantifier-depth-$n^{\epsilon'k}$ $\ell$-variable counting logic.
	\end{enumerate}
	It remains to consider the order on $\CFI{S}$.
	$S'$ contains at most $|S|$ equations of arity larger than three
	and at most $O(k|S|)$ other equations.
	Since $S$ is on $n$ variables, there can be at most $n^k$ equations.
	This means that $\CFI{S}$ has order $O(2^{k}n^k + 2^2kn^k) = O(2^kn^k)$.
\end{proof}

We use the robustness of the system to increase the number of equations in the game,
to apply the following result of Bodlaender to obtain tree-decompositions of logarithmic depth:

\begin{lemma}[\cite{Bodlaender1988}]
	\label{lem:tree-decomp-log-depth}
	Every graph of treewidth at most $k$ has a tree-decomposition $D$ of width at most $3k+2$ such that $T_D$ is a binary tree of depth at most $2\lceil \log_{5/4} (2n) \rceil$.
\end{lemma}

\begin{corollary} 
	There are constants $k_0 \in \nat$ and $0< \epsilon<1$, such that for every $k\geq k_0$  and $n \in \Omega(k^{22k})$ there are two graphs $G$ and $G'$ of order $O(n)$ that satisfy the following:
	\begin{enumerate}
		\item $G$ and $G'$ are treewidth-$k$-distinguishable.
		\item Every graph $H$ of treewidth $k$ such that
		 $\Hom{H}{G} \neq \Hom{H}{G'}$ is of order $|H| = 2^{\Omega(n^\epsilon)}$.
	\end{enumerate}
\end{corollary}
\begin{proof}
	Pick $k_0,\alpha,\epsilon $ to be the constants of Corollary
	\ref{cor:robust-wl-round-numbers-equation}.
	Let $k+1\geq k_0$ and set $\ell := 3k+3$.
	Let $n \geq (\alpha\ell^8{(k+1)}^{14})^k = \Theta(k^{22k})$ such that $m = \lceil n^{-k} \rceil $.
	Then $m \geq \alpha\ell^8{(k+1)}^{14}$.
	Let $S$ be the $k$-ary $\FF_2$-system given by the Corollary~\ref{cor:robust-wl-round-numbers-equation} for $m$ variables.
	Because $\CFI{\homsys{S}}$ and $\CFI{S}$ are not equivalent in $(k+1)$-variable counting logic,
	there is a graph of treewidth $k$ such that $\Hom{H}{\CFI{\homsys{S}}} \neq \Hom{H}{\CFI{S}}$ by Theorem~\ref{thm:hom-eq-iff-biso-hom-eq}.
	
	Let the order of $H$ be $N$.
	Then by Lemma~\ref{lem:tree-decomp-log-depth}, $H$ has a tree-decomposition $D$ of width at most $3k + 2$  such that $T_D$ is a binary tree of depth at most $d:= 2\lceil \log_{5/4} (2N) \rceil = 2\lceil \log_{2} (2N)/\log_{2}(5/4) \rceil$.
	Then there is a $(3k+3)$-variable counting logic formula of quantifier depth at most $d$ that is satisfied by $\CFI{S}$ but not by $\CFI{\homsys{S}}$~\cite{FluckSeppeltSpitzer2024}.
	Since $3k+3=\ell$, the properties of $S$ imply $d \geq m^\epsilon$.
	Because $d = 2\lceil \log_{2} (2N)/\log_{2}(5/4) \rceil$, this yields
	$\log(2N) \geq  \log_{2}(5/4)/2 \cdot(m^\epsilon -1)$.
	Because $m^{\epsilon k} \geq n^{\epsilon}$,
	we obtain
	Hence, $N = 2^{\Omega(n^{\epsilon})}$.
	
	Finally, we need to bound the order of $\CFI{S}$ and $\CFI{\homsys{S}}$.
	Because $|S| \leq m^k$,
	these graphs have order $O(2^km^k)$.
	Since $m^k \geq (n^{-k}+1)^k \leq (2n^{-k})^k \leq 2^kn$,
	the CFI graphs have order $O(2^k2^kn)=O(n)$.
\end{proof}

Using the graph encoding instead of the relational encoding of $k$-ary systems
costs us the $k$ in the exponent of the $n^{\epsilon k}$ round lower bound,
thus only giving us an $2^{O(n^\epsilon)}$ size lower bound (note that the structures from~\cite{GroheLichterNeuen2025} indeed contain $\Theta(n^k)$ tuples in the relations).
That is why we consider games based on $k$-equations instead of on $k$-variables
and can directly consider the graph based CFI encodings.
However, for this setting, no super exponential lower bound robust in the number of equations is known.

\section{Homomorphisms of Bounded Treewidth Graphs} \label{sec:homo-bounded-treewidth-full}

This section establishes techniques to show that two graphs $G_1$ and $G_2$ admit the same number of homomorphisms from another graph $H$ of bounded treewidth.
While having equal homomorphism counts from some $H$ is a global property, this section establishes methods to show equal counts from $H$ by following the tree-decomposition of $H$. 

We introduce notation to select subdecompositions of a tree-decomposition.
Let $H$ be a graph with a rooted tree-decomposition $(D,r)$, 
let $s\in r$ be a child of $r$, and let  $S\subseteq r$ be a set of children of $r$.
\begin{itemize}
	\item The subdecomposition rooted at the child $s$ is denoted $D_s$.
	\item The subdecomposition rooted at $r$ only including the children in $S$ and their subtrees is denoted $D^r_S$.
	\item The subdecomposition rooted at $r$, in which all children in $S$ and their subtrees are deleted, is denoted $D^r_{- S}$. 
\end{itemize}
When $S = \set{s}$, we may only write $D^r_s$ and $D^r_{-s}$.
Note that $D^r_{- S} = D^r_{r\setminus S}$ following our convention that $r$ represents the set of its children.
For a subdecomposition $D'$ of $D$, the corresponding induced subgraph of $H$ is 
$H[D'] := H[\bigcup_{t \in V(T_{D'})} \bag_{D'}(t)]$.
In the former cases, we write $H_s := H[D_s]$, $H^r_S := H[D^r_S]$, and $H^r_{-S} := H[D^r_{-S}]$.
We also apply these operations to trees and define for a tree $T$ with root $r$ the trees $T_s$, $T^r_S$ and $T^r_{-S}$ for $s\in r$ and $S\subseteq r$ similarly.

The next lemma refines a result of~\cite{DellGroheRattan2018} and shows that we can focus on counting bag-wise isomorphic homomorphisms.
\begin{lemma}
	\label{lem:biso-indistinguishability-implies-indistinguishability}
	Let $G$ and $G'$ be graphs.
	If there is a graph $H$  of treewidth $k$ such that $\Hom{H}{G} \neq \Hom{H}{G'}$,
	then
	there is a graph $H'$ of order  $|H'|\leq |H|$ with a tree-decomposition $D'$ of width $k$ such that
	$\bIso{(H',D')}{G} \neq \bIso{(H',D')}{G'}$.
\end{lemma}
\begin{proof}
	For the forward direction, define the following (infinite) matrices $\text{Surj}$ and $\text{bExt}$. Their rows and columns are indexed by pairs $(F,D)$ of a graph $F$ and a tree-decomposition $D$ of $F$ of width at most $k$.
	\begin{itemize}
		\item $\text{Surj}$:
		The matrix entry $\Surj{(F,D)}{(F',D')}$ is the number of homomorphisms $F \to F'$ that are vertex- and edge-surjective and map every bag of $D$ to a bag of $D'$ yielding a surjective homomorphism from $T_D$ to $T_{D'}$.
		\item $\text{bExt}$: The matrix entry $\bExt{(F,D)}{(F',D')}$ is the number of bag-wise extensions of $(F,D)$ isomorphic to $(F',D')$: A bag-wise extension of $(F,D)$ is a tree-decomposed graph $(F'',D'')$ with $V(T_{D''}) = V(T_D)$ such that $F''[\bag_{D''}(t)]$ is an extension of $F[\bag_D(t)]$ for every $t \in V(T_D)$ (this means that $F[\bag_D(t)]$ is a subgraph of $F''[\bag_{D''}(t)]$ while $\bag_D(t) = \bag_{D''}(t)$, so $F''[\bag_{D''}(t)]$ can have more edges).
	\end{itemize}
	We denote by $\Aut{F}{D}$ the number of automorphisms of $(F,D)$,
	which are automorphisms of $F$ that also stabilize $D$ (bags seen as sets, so tuples may be permuted).

	Let $D$ be a width $k$ tree-decomposition of $H$.
	We have (see~\cite{DellGroheRattan2018} and~\cite[Proof of Theorem~24]{DellGroheRattan2018arxiv})
	\begin{align*}
		\Hom{H}{G} &= \sum_{(H',D')} (\text{Surj} \cdot \text{bExt})((H,D),(H',D')) \cdot \bIso{(H',D')}{G}/\Aut{H'}{D'} \\
		&=\sum_{(H',D')} \left(\sum_{(H'',D'')}  \Surj{(H,D)}{(H'',D'')} \cdot \bExt{(H'',D'')}{(H',D')}\right)\\
		&\hspace{2cm} \cdot \bIso{(H',D')}{G}/\Aut{H'}{D'}
	\end{align*}
	We see that $\Surj{(H,D)}{(H'',D'')}$ is nonzero only if $|H''|\leq |H|$ and that $\bExt{(H'',D'')}{(H',D')}$ is nonzero only if $|H''| = |H'|$.
	Now, if $\Hom{H}{G}\neq \Hom{H}{G'}$, then for one $(H',D')$ the summand
	has to be different for $G$ and $G'$. But since the left factors do not depend on $G$ or $G'$, the right factors have to be different.
	Hence, $\bIso{(H',D')}{G}/\Aut{H'}{D'} \neq \bIso{(H',D')}{G'}/\Aut{H'}{D'}$ which implies
	$\bIso{(H',D')}{G} \neq \bIso{(H',D')}{G'}$.
\end{proof}

We now establish techniques to prove that $\bIso{(H,D)}{G_1} = \bIso{(H,D)}{G_2}$  by decomposing these homomorphism counts into counts from subgraphs of $H$ and the subdecompositions of $D$.
Let $G$ and $H$ be graphs and
$(D,r)$ be a smooth rooted tree-decomposition  of $H$ of width $k-1$.
A tuple $\tup{u} \in V(G)^k$ is \defining{possible for} $(D,r)$ if $\tup{u}$ has the same isomorphism type in $G$ as $\bag(r)$ in $H$.
A set of tuples $\tup{U} \subseteq V(G)^k$ is possible for $(D,r)$ if each tuple in $U$ is possible for $(D,r)$.

Let $\tup{u} \in V(G)^k$ be possible for $(D,r)$.
For a child $s\in r$,
an \defining{$s$-neighbor} of $\tup{u}$ with respect to $(D,r)$ is a 
tuple $\tup{v} \in V(G)^k$ such that:
\begin{enumerate}
	\item $\tup{v}$ has the same isomorphism type in $G$ as $\bag(s)$ in $H$.
	\item If $\bag(r)_i = \bag(s)_j$, then $u_i = v_j$ for all $i,j \in [k]$.
\end{enumerate}
When $(D,r)$ is clear from the context,
$N_G^s(\tup{u})$ denotes the set of $s$-neighbors of $\tup{u}$ with respect to $(D,r)$.
For a set $U \subseteq V(G)^k$ which is possible for $(D,r)$, we set $N_G^s(U) := \bigcup_{\tup{u} \in U} N_G^s(\tup{u})$.
Note that every bag-wise isomorphic homomorphism $h\colon (H,(D,r)) \to G$ has to map
$\bag_D(s)$ to an $s$-neighbor of $h(\bag_D(r))$ in $G$.
In particular, $N_G^s(U)$ is possible for $(D_s, s)$.

\begin{lemma}
	\label{lem:biso-children}
	Let $G, H$ be graphs, let $(D,r)$ be a smooth rooted tree-decomposition of width $k-1$  of $H$ such that $r$ is not a leaf, and let $U \subseteq V(G)^k$ be possible for $(D,r)$.
	Then 
	\begin{align*}
	\bIso{(H,D,r)}{(G,U)} &= \sum_{\tup{u} \in U} \prod_{s \in r} \bIso{(H_s,D_s,s)}{(G,N_{G}^s (\tup{u}))}\\
	&=   \sum_{\tup{u} \in U} \prod_{s \in r} \bIso{(H^r_s, D^r_s, r)}{(G,\tup{u})}.
	\end{align*}
\end{lemma}
\begin{proof}
	Since $\bIso{(H,D,r)}{(G,U)} = \sum_{\tup{u} \in U} \bIso{(H,D,r)}{(G,\tup{u})}$,
	it suffices to consider the case that $U=\set{\tup{u}}$.
	Moreover, for $s \in r$, the
	tuples $N_{G}^s (\tup{u})$ are exactly the possible images of $\bag(s)$ under a bag-wise isomorphic homomorphism $h$ with $h(\bag_D(r)) = \tup{u}$,
	because by definition these tuples are the only ones with the matching isomorphism type where the correct vertices get replaced with respect to $\bag(r)$.
	
	Let $s_1,\dots, s_\ell$ be the children of $r$.
	Let $\phi_1, \dots, \phi_\ell$ be bag-wise isomorphic homomorphisms $(H_{s_i},D_{s_i},s_i) \to (G,N_{G}^{s_i} (\tup{u}))$.
	Because the image of the root $r$ is fixed, they all agree on mapping vertices in $\bag(r)$ (in case they are in $V(H_{s_i})$).
	By the properties of a tree-decomposition, $V(H_{s_i}) \cap V(H_{s_j}) \subseteq \bag(r)$ for all distinct $i,j$.
	Thus, we can compose the $\phi_1, \dots, \phi_\ell$ to a well-defined map $\phi \colon V(H) \to G$.
	Because all edges of $H$ are covered by the bag $r$ or the ones in the $D_{s_i}$, this map is actually a bag-wise isomorphic homomorphism.
	
	Conversely, every bag-wise isomorphic homomorphism $\phi \colon(H,D,r) \to (G,\tup{u})$ induces such a sequence $\phi_1, \dots, \phi_\ell$ of bag-wise isomorphic homomorphisms of the $H_{s_i}$.
	This means that there is a one-to-one correspondence between
	bag-wise isomorphic homomorphisms $\phi \colon V(H) \to G$
	and these sequences.
	Clearly, there are  $\prod_{s \in r} \bIso{(H_s,D_s,s)}{(G,N_{G}^s (\tup{u}))}$ many such sequences.
	
	Finally,
	observe that $\bIso{(H_s,D_s,s)}{(G,N_{G}^s (\tup{u}))} =   \bIso{(H^r_s, D^r_s, r)}{(G,\tup{u})}$
	because $N_{G}^s (\tup{u})$ contains exactly the possible images of $\bag(s)$ under a bag-wise isomorphic homomorphism.
\end{proof}

\begin{lemma}[Factorization Lemma]
	\label{lem:hom-one-step-graphs-2}
	Let $(D,r)$ be a smooth tree-decomposition of a graph $H$ of width $k-1$ such that $r$ is not a leaf.
	Let $G_1$ and $G_2$ be graphs and $U_i \subseteq V(G_i)^k$ be possible for $(D,r)$.
	Assume that there is a set $R \subseteq r$ such that all $s\in R$ satisfy for all $i\in[2]$ and $\tup{u},\tup{u}' \in U_i$ that
	\begin{enumerate}
		\item 	$\bIso{(H^r_s,D^r_ s, r)}{(G_i,\tup{u})} = \bIso{(H^r_s, D^r_s, r)}{(G_i, \tup{u}')} =: c_s^i$,
		\item $\prod_{s\in R}c_s^1 = \prod_{s\in R} c_s^2$, and
		\item $\bIso{(H^r_{-R},D^r_{-R},r)}{(G_1, U_1)}= \bIso{(H^r_{-R},D^r_{-R},r)}{(G_2, U_2)}$.
	\end{enumerate}
	Then $\bIso{(H,D,r)}{(G_1, U_1)}= \bIso{(H,D,r)}{(G_2, U_2)}$.
\end{lemma}
\begin{proof}
	We first apply Lemma~\ref{lem:biso-children}, factor the terms
	$\bIso{(H^r_s,D^r_s,r)}{(G_i, \tup{u})} =c^1_s$ for each $s \in R$ out, and
	apply the equation  $\prod_{s\in R}c_s^1 = \prod_{s\in R} c_s^2$.
	\begin{align*}
		\bIso{(H,D,r)}{(G_1,U_1)} 
		&=\sum_{\tup{u} \in U_1}\prod_{s \in r} \bIso{(H^r_s, D^r_s, r)}{ (G_1, \tup{u})}\\
		&=\sum_{\tup{u} \in U_1} \left( \prod_{s\in R} c_s^1 \cdot \prod_{s \in r \setminus R}\bIso{(H^r_s, D^r_s, r)}{ (G_1, \tup{u})}  \right)\\
		&=  \prod_{s\in R} c_s^1\cdot \left(\sum_{\tup{u} \in U_1}\prod_{s \in r \setminus R} \bIso{(H^r_s, D^r_s, r)}{ (G_1, \tup{u})} \right) \\
		&= \prod_{s\in R} c_s^2 \cdot \left(\sum_{\tup{u} \in U_2}\prod_{s \in r \setminus R} \bIso{(H^r_s, D^r_s, r)}{ (G_2, \tup{u})} \right) \\
		&=\bIso{(H,D,r)}{(G_2,U_2)}
	\end{align*}
	In the penultimate step we used that \[\left(\sum_{\tup{u} \in U_i}\prod_{s \in r \setminus R} \bIso{(H^r_s, D^r_s, r)}{ (G_i, \tup{u})} \right) = \bIso{(H^r_{-R},D^r_{-R},r)}{(G_i,U_i)}\] for both $i\in[2]$ by Lemma~\ref{lem:biso-children}.  
\end{proof}

\begin{lemma}[Unique-Child Lemma]
		\label{lem:hom-one-step-graphs-3}
	Let $(D,r)$ be a smooth tree-decomposition of a graph $H$ of width $k-1$ such that $r$ has a unique child $s$.
	Let $G_1$ and $G_2$ be graphs and $U_i \subseteq V(G_i)^k$ be possible for $(D,r)$.
	Assume, for both $i\in[2]$, there are partitions $P_i$ of $N_{G_i}^s(U_i)$
	and a bijection $f \colon P_1 \to P_2$
	such that for all $p \in P_1$,
	\begin{enumerate}
		\item all $\tup{v} \in p$ and $\tup{w} \in f(p)$ satisfy $|\setcond{\tup{u} \in U_1}{\tup{v} \in N_{G_1}^s(\tup{u})}| = |\setcond{\tup{u} \in U_2}{\tup{w} \in N_{G_2}^s(\tup{u})}| := c_p$ and
		\item
		$\bIso{(H_s,D_s,s)}{(G_1, p)} = \bIso{(H_s,D_s,s)}{(G_2, f(p))}$.
	\end{enumerate}
	Then $\bIso{(H,D,r)}{(G_1, U_1)}= \bIso{(H,D,r)}{(G_2, U_2)}$.
\end{lemma}
\begin{proof}
	For both $i\in[2]$ we have by Lemma~\ref{lem:biso-children}
	\begin{align*}
		\bIso{(H,D,r)}{(G_i,U_i)} 
		&= \sum_{\tup{u} \in U_i} \sum_{\tup{w} \in N^s_{G_i}(\tup{u})}   \bIso{(H_s, D_s, s)}{ (G_i, \tup{w})} \\
		&= \sum_{\tup{w} \in N^s_{G_i}(U_i)} |\setcond{\tup{u} \in U_i}{\tup{w} \in N_{G_i}^s (\tup{u})}|  \cdot \bIso{(H_s, D_s, s)}{ (G_i, \tup{w})}\\
		&= \sum_{p \in P_i} \sum_{\tup{w} \in p} |\setcond{\tup{u} \in U_i}{\tup{w} \in N_{G_i}^s (\tup{u})}|  \cdot \bIso{(H_s, D_s, s)}{ (G_i, \tup{w})}\\
		&=  \sum_{p \in P_i} c_p \cdot \sum_{\tup{w} \in p} \bIso{(H_s, D_s, s)}{ (G_i, \tup{w})} \\
		&= \sum_{p \in P_i} c_p \cdot \bIso{(H_s, D_s, s)}{ (G_i, p)} .
	\end{align*}
	We see that this expression is equal for both $i \in [2]$
	by identifying the blocks of the partitions by $f$.
\end{proof}

\begin{lemma}[Singleton-Set Lemma]
	\label{lem:singleton-set-lemma}
	Let $(D,r)$ be a smooth tree-decomposition of a graph $H$ of width $k-1$ such that $r$ is not a leaf.
	Let $G_1$ and $G_2$ be graphs and $\tup{u}_i \in V(G_i)^k$ be possible for $(D,r)$ for both $i\in[2]$.
	If every $s \in r$ satisfies
	\[
	\bIso{(H_s,D_s,s)}{(G_1, N_{G_1}^s(\tup{u}_1))} = \bIso{(H_s,D_s,s)}{(G_2, N_{G_2}^s(\tup{u}_2))},
	\]
	then $\bIso{(H,D,r)}{(G_1, \tup{u}_1)}= \bIso{(H,D,r)}{(G_2, \tup{u}_2)}$.
\end{lemma}

\begin{proof}
	The claim follows from Lemma~\ref{lem:biso-children}:
	\begin{align*}
		\bIso{(H,D,r)}{(G_1,\tup{u}_1)}
		&=\prod_{s \in r} \bIso{(H_s, D_s, s)}{ (G_1, N^s_{G_1}(\tup{u}_1))}\\
		&=\prod_{s \in r} \bIso{(H_s, D_s, s)}{ (G_2, N^s_{G_2}(\tup{u}_2))}\\
		&=\bIso{(H,D,r)}{(G_2,\tup{u}_2)}.\qedhere
	\end{align*}
\end{proof}

\begin{lemma} [Partition Lemma]
	\label{lem:bijection-lemma}
	Let $(D,r)$ be a smooth tree-decomposition of a graph $H$ of width $k-1$ such that $r$ is not a leaf.
	Let $G_1$ and $G_2$ be graphs and $U_i \subseteq V(G_i)^k$ be possible for $(D,r)$.
	If $P_i$ is a partition of $U_i$ and $f \colon P_1 \to P_2$ is a bijection such that for all $V \in P_1$
	\[
	\bIso{(H,D,r)}{(G_1, V)} = \bIso{(H,D,r)}{(G_2, f(V))},
	\]
	then $\bIso{(H,D,r)}{(G_1, U_1)}= \bIso{(H,D,r)}{(G_2, U_2)}$.
\end{lemma}
\begin{proof}
	\begin{align*}
		\bIso{(H,D,r)}{(G_1, U_1)} &= \sum_{V\in P_1} \bIso{(H,D,r)}{(G_1, V)}\\
		& = \sum_{V\in P_1} \bIso{(H,D,r)}{(G_2, f(V))} \\
		& = \sum_{V\in P_2} \bIso{(H,D,r)}{(G_2, V)} \\
		& = \bIso{(H,D,r)}{(G_2, f(U_2))}.
	\end{align*}
	The first equation holds because $P_1$ is a partition of $U_1$, the second by the assumption of the lemma, the third because $f$ is a bijection, and the last because $P_2$ is a partition of $U_2$.
\end{proof}

 The key idea is to think about the four prior lemmas as a proof system for showing equal homomorphisms counts: each lemma reduces the task of showing equal homomorphism counts to subgoals. The Factorization Lemma can reduce the number of children of the root, so that the Unique-Child Lemma may become applicable. Similarly, the Partition Lemma can reduce the size of $U$, so that the Singleton-Set Lemma may become applicable.
This system is probably not complete but, by the lemmas, it is sound.
With this perspective in mind we will now apply the standard approach of turning this proof system into a Prover-Verifier like game.

\newcommand{\extTo}[2]{#1^{#2}}
\newcommand{\subO}[2]{O_{#1}^{#2}}
\newcommand{\elmA}[2]{A_{#1}^{#2}}
\newcommand{\elmAhat}[2]{\hat{A}_{#1}^{#2}}	
\newcommand{\varc}[3]{c_{#1,#3}^{#2}}
\newcommand{\simne}[3]{\sim_{#1}^{#2,#3}}
\newcommand{\partPe}[3]{P_{#1}^{#2,#3}}
\newcommand{\partP}[2]{P_{#1}^{#2}}
\newcommand{\const}[2]{\kappa_{#1}^{#2}}
\newcommand{\bijLambe}[2]{\lambda_{#1}^{#2}}
\newcommand{\bijLamb}[1]{\lambda_{#1}}
\newcommand{\repK}[4]{K_{#1,#4}^{#2,#3}}
\newcommand{\blockp}[4]{p_{#1,#4}^{#2,#3}}
\newcommand{\classC}[4]{C_{#1,#4}^{#2,#3}}

\section{A Game for Proving Equal Homomorphism Counts} \label{sec:ind-game-full}
This section develops a Prover-Verifier-style game based on the lemmas in the former section.
The game will be specific for CFI graphs:
arguing on the level of homomorphism counts is translated to the level of local isomorphisms of CFI graphs and thereby to local solutions of the underlying linear system.
The game also abstracts away from the graph $H$ we count homomorphisms from and only considers the tree of a tree-decomposition of $H$.

For an integer $k\geq 2$,
the \defining{locally-indistinguishable $k$-equation game} is played between Prover and Verifier
on an $\FF_2$-system $S$ and a tree $T$.
A position in the game is a tuple
$(\eqTA, W, T, r)$
of a $k$-tuple $\eqTA \in S^k$ of equations, 
an affine linear space $W \subseteq \sols{\eqTA}$
and a tree $T$ with root $r$.
The game proceeds in rounds as follows. If $|T| = 1$ at the start of the current round, then the game ends and Prover wins.
Otherwise, Prover chooses whether to play a 
\defining{singleton-set}, \defining{partition}, \defining{factorization}, or \defining{unique child} round.
Let $ V \leq \sols{\homsys{\eqTA}}$ be the linear space of $W$,
that is $W = \alpha +V$ for every $\alpha \in W$.
These rounds proceed as follows:

\begin{description}

	\item [Partition Round] ~
	
		Prover chooses an affine subspace $W' \subseteq W$.
		The game continues in position $(\eqTA, W', T, r)$.
	\item [Factorization Round] ~
	\begin{enumerate}
		\item 
		Prover picks a subset $R \subsetneq r$ of children of $r$.
		\item Verifier decides between the following:
		\begin{enumerate}
			\item Verifier picks $s \in R$ and $\alpha' \in V$.
			 The game continues on $\homsys{S}$ in position $(\homsys{\eqTA}, \set{\alpha'},  T^r_s,r)$.
			 \item Verifier picks $s \in R$.
			 The game continues from position $(\eqTA, W,T^r_s, r)$.
			\item The game continues in position $(\eqTA, W, T^r_{-R}, r)$.
		\end{enumerate}
	\end{enumerate}
		
	\item [Singleton-Set Round] Prover may choose this round only if $W = \set{\alpha}$.
	
	Verifier chooses $\ell \in [k]$, $\eqTB \in N_S^\ell(\eqTA)$, and $s\in r$.
	If $W' := \exts{\sols{\eqTB}}{\forgt{\alpha}{\ell}} = \emptyset$, then Verifier wins.
	Otherwise, the game continues in position $(\eqTB, W', T_s, s)$.
	
	\item [Unique-Child Round] Prover may only choose this round if 
	$r$ has a unique child $s$. 
	\begin{enumerate}
		\item Verifier chooses $\ell \in [k]$ and $\eqTB \in N_S^\ell(\eqTA)$.
		\item If $\exts{\forgt{W}{\ell}}{\sols{\eqTB}} = \emptyset$, Verifier wins.
		Otherwise, Prover chooses an affine subspace $W' \subseteq  \exts{\forgt{W}{\ell}}{\sols{\eqTB}}$.
		With $W'' := \exts{\sols{\eqTB}}{W'}$,
		the game continues in position $(\eqTB, W'',  T_s, s)$.
	\end{enumerate} 
	\end{description}

\begin{lemma}
	\label{lem:indistinguishability-to-hom-indistinguishability}
	Let $S$ be a thin $\FF_2$-system, 
	$b \colon S \to \FF_2$,
	$\eqTA \in S^k$,
	$V \leq \sols{\homsys{\eqTA}}$ a linear space, $\alpha \in \sols{\eqTA}$
	and $T$ be a tree with root~$r$.
	Set  $G_1 := \CFI{\homsys{S}+b}$ and $G_2 :=\CFI{S+b}$.
	Let $U_1 \subseteq V(G_1)^k$ be of origin $\eqTA$
	such that, for all $\tup{u} \in U_1$, we have that $U_1 = \tup{u} +V$.
	Set $U_2 := \alpha + U_1 \subseteq V(G_2)^k$.
	
	If Prover has a winning strategy in the $k$-indistinguishability game in position $(\eqTA, \alpha + V, T, r)$ on $S$,
	then, every graph $H$ with a smooth rooted
	 tree-decomposition
	 $(D,r)$ of width $k-1$ satisfying $T_D = T$ satisfies $\bIso{(H, D,r)}{(G_1, U_1)} = \bIso{(H, D,r)}{(G_2, U_2)}$.
\end{lemma}

\subparagraph{On Neighbourhoods in CFI Graphs}

In order to prove Lemma~\ref{lem:indistinguishability-to-hom-indistinguishability} we first need technical lemmas about CFI graphs. Essentially they describe how neighborhoods behave in CFI graphs and in particular that they form affine spaces. These are only used to show Claim~\ref{clm:neighborhood} in the proof of Lemma~\ref{lem:indistinguishability-to-hom-indistinguishability} and a reader may wish to skip over them on a first parse. We now give some definitions before proving these auxiliary lemmas.

Let $\eqTA \in S^k$.
A \defining{neighbor-type} of $\eqTA$ is a tuple $(\ell, \eqTB, \tau)$
of a number $\ell\in[k]$, a tuple $\eqTB \in N_S^\ell(\eqTA)$ and an isomorphism type $\tau$ of $k$-tuples in a graph. Recall that $\eqTB \in N_S^\ell(\eqTA)$ means that there is an $\eqTBm \in S$ such that $\eqTB = \eqTA [\ell / \eqTBm]$.
Let $\tup{u} \in \CFI{S}^k$ be of origin $\eqTA$.
A tuple $\tup{v} \in V(\CFI{S})^k$ is an \defining{$(\ell,\eqTB,\tau)$-neighbor} of $\tup{u}$
if $\tup{u} \setminus \ell= \tup{v} \setminus \ell$, $\tup{v}$ has origin $\eqTB$, and the isomorphism type of $\tup{v}$ in $\CFI{S}$ is $\tau$.
The set of all $(\ell, \eqTB,\tau)$-neighbors of $\tup{u}$ is denoted
$N_{\CFI{S}}^{\ell, \eqTB,\tau}(\tup{u})$.
For a set $U \subseteq V(\CFI{S})^k$, 
set $N_{\CFI{S}}^{\ell, \eqTB,\tau}(U) := \bigcup_{\tup{u} \in U} N_{\CFI{S}}^{\ell, \eqTB,\tau}(\tup{u})$.
When $\ell$ is clear from the context, we also just call $\tau$ an $(\eqTB, \tau)$ neighbor and write
$N_{\CFI{S}}^{\eqTB,\tau}(\tup{u})$ and $N_{\CFI{S}}^{\eqTB,\tau}(U)$
for $N_{\CFI{S}}^{\ell, \eqTB,\tau}(\tup{u})$ and $N_{\CFI{S}}^{\ell, \eqTB,\tau}(U)$, respectively.

\begin{lemma}
	\label{lem:neighbors-space}
	Let $S$ be an $\FF_2$-system,
	$\eqTA \in S^k$,
	$\tup{u} \in V(\CFI{\homsys{S}}) ^k$ be of origin $\eqTA$,
	$\eqTB \in N_S^\ell(\eqTA)$,
	and	$\tau$ be an isomorphism type of $k$-tuples.
	Let $\tup{v}$ be an $(\eqTB,\tau)$-neighbor of $\tup{u}$
	and $V' = \exts{\sols{\homsys{\eqTB}}}{0_{\vars{\tup{f}\setminus \ell}}}$.
	Then $N_{\CFI{S}}^{\eqTB, \tau} (\tup{u}) = V' + \tup{v}$.
\end{lemma}
\begin{proof}
	We show the first inclusion.
	Let $\tup{w}$ be another $(\eqTB,\tau)$-neighbor of $\tup{u}$.
	By Lemma~\ref{lem:solutions-preserve-isomorphism-type},
	there is a $\gamma \in \sols{\homsys{\eqTB}}$
	such that $\tup{v} + \gamma = \tup{w}$.
	Since $(\tup{v}+\gamma)\setminus \ell = \tup{w}\setminus\ell$, we have that
	$\gamma$ is zero on variables in $\vars{\eqTB \setminus\ell} = \vars{\eqTA \setminus\ell}$ which implies $\gamma\in V'$.
	
	The converse direction follows also from Lemma~\ref{lem:solutions-preserve-isomorphism-type}:
	$\tup{v} + \gamma$ has isomorphism type $\tau$ for every $\gamma \in V' \subseteq \sols{\homsys{\eqTB}}$.
	Since $\gamma$ is zero on $\vars{\eqTB \setminus\ell} = \vars{\eqTA \setminus\ell}$, we have $(\tup{v}+\gamma)\setminus \ell = \tup{u}\setminus\ell$ and so $\tup{v} + \gamma$ is an $(\eqTB, \tau)$-neighbor of $\tup{u}$.
\end{proof}

\begin{lemma}
	\label{lem:neigbhor-iff-extends}
	Let $S$ be a thin $\FF_2$-system,
	$\eqTA \in S^k$,
	$\tup{u} \in V(\CFI{\homsys{S}})^k$ be of origin $\eqTA$,
	$\alpha \in \sols{\eqTA}$,
	$\eqTB \in N_S^\ell(\eqTA)$,
	$\alpha' \in \sols{\eqTB}$,
	$\tau$ an isomorphism type of $k$-tuples,
	and $\tup{v} \in V(\CFI{S}) ^k$
	be an $(\eqTB,\tau)$-neighbor of $\tup{u}$.
	
	Then $\alpha' +\tup{v}$ is an $(\eqTB,\tau)$-neighbor of $\alpha + \tup{u}$
	if and only if
	 $\alpha' \in \exts{\sols{\eqTB}}{\forgt{\alpha}{\ell}}$.
\end{lemma}
\begin{proof}
	Let $u_i = (\eqTAm_i, \beta_i)$ for all $i\in[k]$ and $v_\ell = (\eqTBm_\ell, \gamma)$.
	Clearly, $\alpha + u_i = (\eqTAm_i, \beta_i + \restrict{\alpha}{\vars{\eqTAm}})$ and similar for $\alpha'$.
	For readability, we will write $\beta_i + \alpha$ for $\beta_i + \restrict{\alpha}{\vars{\eqTAm}}$.
	
	First let $\alpha' \in \exts{\sols{\eqTB}}{\forgt{\alpha}{\ell}}$.
	Then all $i\in[k]\setminus \set{\ell}$ satisfy
	 $\alpha + u_i = \alpha'+ v_i$
	since  $u_i = v_i$ because $\tup{v}$ is an  $(\eqTB,\tau)$-neighbor of $\tup{u}$,
	 and $\restrict{\alpha}{\vars{\eqTAm_i}} = \restrict{\alpha}{\vars{\eqTBm_i}}$ because $\eqTAm_i = \eqTBm_i$
	and $\alpha'$ extends $\restrict{\alpha}{\eqTA \setminus \ell}$.
	Hence, $(\alpha + \tup{u})\setminus \ell = (\alpha' + \tup{v})\setminus\ell$.
	By Lemma~\ref{lem:solutions-preserve-isomorphism-type},
	$\tup{v}$ and $\alpha' +\tup{v}$ have the same isomorphism type $\tau$.
	Hence,  $\alpha' +\tup{v}$ is a $(\eqTB,\tau)$-neighbor of $\alpha + \tup{u}$.
	
	Consider the reverse direction and let $\alpha' +\tup{v}$ be an $(\eqTB,\tau)$-neighbor of $\alpha + \tup{u}$.
	Thus, $(\alpha' + \tup{v}) \setminus \ell = (\alpha + \tup{u}) \setminus \ell$.
	Since $ \tup{v} \setminus \ell = \tup{u} \setminus \ell$
	because $\tup{v}$ is a $(\eqTB,\tau)$-neighbor of $\tup{u}$,
	the action of $\alpha'$ and $\alpha$ coincide on all variables in $\vars{\eqTA \setminus \ell} = \vars{\eqTB \setminus \ell}$.
	Thus, $\alpha'$ extends $\restrict{\alpha}{\eqTA \setminus \ell}$.
\end{proof}

\begin{lemma}
	\label{lem:neighbors-subspace-local-solution}
	Let $S$ be an $\FF_2$-system,
	$\eqTA \in S^k$,
	$\tup{u} \in V(\CFI{\homsys{S}})^k$ be of origin $\eqTA$,
	$V \leq \sols{\homsys{\eqTA}}$,
	$\alpha \in \sols{\eqTA}$,
	$\eqTB \in N_S^\ell(\eqTA)$,
	$\tau$ an isomorphism type of $k$-tuples,
	and
	$\alpha' \in \exts{\sols{\eqTB}}{\forgt{(\alpha + V)}{\ell}}$.
	Then $\alpha' + N_{\CFI{\homsys{S}}}^{\eqTB, \tau}(V +\tup{u}) = N_{\CFI{S}}^{\eqTB, \tau}(\alpha + V + \tup{u})$.
\end{lemma}
\begin{proof}
	
	There is some $\alpha'' \in \alpha + V$ such that $\alpha'$ extends to $\restrict{\alpha''}{\eqTA\setminus \ell}$ because 	$\alpha' \in \exts{\sols{\eqTB}}{\forgt{(\alpha + V)}{\ell}}$.
	
	We show the first inclusion.
	Let $\tup{w} \in \alpha' + N_{\CFI{\homsys{S}}}^{\eqTB, \tau}(V +\tup{u})$.
	Then there is $\beta \in V$ such that 
	$\tup{w} \in \alpha' + N_{\CFI{\homsys{S}}}^{\eqTB,\tau}(\beta + \tup{u})$
	and some $\tup{w}' \in  N_{\CFI{\homsys{S}}}^{\eqTB,\tau}(\beta + \tup{u})$
	such that 
	$\tup{w} = \alpha' + \tup{w}'$.
	Because $\tup{w}' \in  N_{\CFI{\homsys{S}}}^{\eqTB,\tau}(\beta + \tup{u})$,
	we also have $\tup{w} = \alpha' + \tup{w}' \in  N_{\CFI{\homsys{S}}}^{\eqTB,\tau}(\alpha'' + \beta + \tup{u})$ by Lemma~\ref{lem:neigbhor-iff-extends}.
	Since $\alpha''+\beta \in \alpha + V$, it follows that 
	 $\tup{w} \in N_{\CFI{S}}^{\eqTB,\tau}(\alpha + V + \tup{u})$.
	
	We now show the converse direction.
	Let $\tup{w} \in N_{\CFI{S}}^{\eqTB,\tau}(\alpha + V + \tup{u})$.
	Then there is a $\beta \in V$ such that $\tup{w} \in N_{\CFI{S}}^{\eqTB,\tau}(\alpha + \beta + \tup{u})$.
	By Lemma~\ref{lem:neigbhor-iff-extends},
	$\alpha' +\tup{w} \in N_{\CFI{\homsys{S}}}^{\eqTB,\tau}(\alpha +\alpha'' + \beta + \tup{u})$ and so 
	$\alpha' +\tup{w} \in N_{\CFI{\homsys{S}}}^{\eqTB,\tau}(V + \tup{u})$.
	Thus, $\tup{w} \in \alpha' + N_{\CFI{\homsys{S}}}^{\eqTB,\tau}(V + \tup{u})$.
\end{proof}

\begin{lemma}
	\label{lem:neighbors-subspace-mem}
	Let $S$ be an $\FF_2$-system,
	$\eqTA \in S^k$,
	$\tup{u} \in V(\CFI{S})^k$ be of origin $\eqTA$,
	$\ell \in [k]$,
	$\eqTB \in N_S^{\ell}(\eqTA)$,
	$V \leq \sols{\homsys{\eqTA}}$ such that 
	$V' := \exts{\sols{\homsys{\eqTB}}}{\forgt{V}{\ell}}$
	satisfies $\exts{\forgt{V}{\ell}}{V'} = \forgt{V}{\ell}$
	, and
	let $\tau$ be an isomorphism type of $k$-tuples.
	Every $\tup{v} \in N_{\CFI{S}}^{\eqTB,\tau}(V+ \tup{u} )$ satisfies
	$\tup{v} + V' = N_{\CFI{S}}^{\eqTB,\tau}(V+\tup{u})$. 
\end{lemma}
\begin{proof}
	Because $\tup{v} \in N_{\CFI{S}}^{\tau}(V+ \tup{u} )$,
	there is a $\beta \in V$ such that $\tup{v} \in N_{\CFI{S}}^{\eqTB,\tau}(\beta+ \tup{u} )$.
	Since the claim of the lemma only depends on $V + \tup{u}$ and not $\tup{u}$ itself,
	we may assume that $\beta = 0$ and  $\tup{v} \in N_{\CFI{S}}^{\eqTB,\tau}(\tup{u} )$.
	
	To show the first inclusion,
	let $\tup{w} \in \tup{v} + V'$.
	Then there is a $\beta' \in V'$
	such that $\tup{w} = \tup{v} + \beta'$.
	Because $V' = \exts{\sols{\homsys{\eqTB}}}{\forgt{V}{\ell}}$,
	there is a $\beta \in V$ such that $\beta'$ extends to $\forgt{\beta}{\ell}$.
	Because $\tup{v} \in N_{\CFI{S}}^{\eqTB,\tau}(\tup{u} )$,
	Lemma~\ref{lem:neigbhor-iff-extends} implies that 
	$\tup{w} = \tup{v} + \beta' \in N_{\CFI{S}}^{\eqTB,\tau}(\beta + \tup{u} ) \subseteq  N_{\CFI{S}}^{\eqTB,\tau}(V + \tup{u} )$.
	
	For the reverse inclusion,
	let $\tup{w} \in N_{\CFI{S}}^{\eqTB,\tau}(V + \tup{u} )$.
	So there is a $\beta \in V$  such that $\tup{w} \in N_{\CFI{S}}^{\eqTB,\tau}(\beta + \tup{u} )$.
	Because $\exts{\forgt{V}{\ell}}{V'} = \forgt{V}{\ell}$,
	there is a $\beta' \in V'$ that extends to $\forgt{\beta}{\ell}$.
	Lemma~\ref{lem:neigbhor-iff-extends} implies that 
	$\beta' + \tup{w} \in N_{\CFI{S}}^{\eqTB,\tau}(\tup{u} )$.
	By Lemma~\ref{lem:neighbors-space}
	there is a $\beta''\in \sols{\homsys{\eqTB}}$ such that $\beta'' + \tup{w} = \beta' +\tup{v}$ and $\beta''$ extends $0_{\vars{\eqTA \setminus \ell}}$.
	This implies that $\beta'' \in V'$
	and so $\tup{w} \in V' +\tup{v}$.
\end{proof}

\subparagraph{Correctness of the Indistinguishability Game}

\begin{proof}[Proof of Lemma~\ref{lem:indistinguishability-to-hom-indistinguishability}]
Let $H$ be a graph with a tree-decomposition $(D,r)$ as required.
If $U_1$ is not possible for $(D,r)$,
then also $U_2$ is not possible for $(D,r)$
because $\alpha$ defines a local isomorphism by Lemma~\ref{lem:solutions-preserve-isomorphism-type}.
Thus, in this case, $0 = \bIso{(H, D, r)}{(G_1, U_1)} = \bIso{(H, D,r)}{(G_2, U_2)}$.
So we now  consider the case that $U_1$ and $U_2$ are possible for $(D,r)$.
		
The proof proceeds by induction on $(|T|, |W|)$ (using the lexicographic order).
Consider the case that $|T| = 1$.
In this case the game ends and Prover wins.
By Lemma~\ref{lem:solutions-preserve-isomorphism-type}, all tuples in $U_1$ and $U_2$ have the same isomorphism type in $G_1$ and $G_2$, respectively.  Because $U_2 = \alpha +U_1$, it follows that $|U_1| = |U_2|$.
Hence, $\bIso{(H, D,r)}{(G_1, U_1)} = \bIso{(H, D,r)}{(G_2, U_2)}$.

So consider the case $|T| > 1$ and set $W = \alpha + V$.
For $s \in r$, we partition the $s$-neighborhood $N_{G_i}^s(U_i)$ of $U_i$ according to the origin $\eqTB$;
let $N_{G_i}^{s,\eqTB}(U_i) \subseteq N_{G_i}^s(U_i)$
be the $s$-neighbors of $U_i$ of origin $\eqTB$.
This indeed partitions the $s$-neighborhood by its neighbor-type because
the $s$-neighbors are all tuples with the same isomorphism type
in which a single entry is replaced.
We use similar notation for a single tuple in $U_i$.
We first need some observations about the neighborhoods of the $U_i$.
They are all consequences of Lemmas~\ref{lem:neighbors-subspace-local-solution} and~\ref{lem:neighbors-subspace-mem} and the definition of the $s$-neighborhood.
\begin{claim}
	\label{clm:neighborhood}
	Let $i\in[2]$, $\ell \in [k]$, $\eqTB \in N_S^\ell(\eqTA)$, and $s \in r$. 
	\begin{enumerate}[(a)]
		\item \label{itm:neighborhood_single_vertex_auto}  Let $\tup{u} \in U_i$, $\tup{v} \in N_{G_i}^{s, \eqTB}(\tup{u})$, and $V' := \exts{\sols{\homsys{\eqTB}}}{\set{0_{\forgt{V}{\ell}}}}$.
		Then $\tup{v} + V' = N_{G_i}^{s, \eqTB}(\tup{u})$. 
			\item \label{itm:neigbors-subspace-step}	Let $\tup{u} \in U_i$, $\tup{v} \in N_{G_i}^{s,\eqTB}(\tup{u})$,
		$X \leq V$ and $X'= \exts{\sols{\homsys{\eqTB}}}{\forgt{X}{\ell}}$
		such that $\exts{\forgt{X}{\ell}}{X'} = \forgt{X}{\ell}$.
		Then $  \tup{v} + X' = N_{G_i}^{s, \eqTB}(X + \tup{u})$. 
				\item \label{itm:neighbrood_single_vertex_iso} Let $\tup{u} \in U_1$,
		$\tilde{\alpha} \in \alpha + V$, and  $\alpha' \in \exts{\sols{\eqTB}}{\set{\forgt{\tilde{\alpha}}{\ell}}}$. Then  $\alpha' + N_{G_1}^{s,\eqTB}(\tup{u})= N_{G_2}^{s, \eqTB}(\tilde{\alpha} + \tup{u})$.

	\end{enumerate}
\end{claim}
\begin{claimproof}
	\begin{enumerate}[(a)]
		\item Follows from Lemma~\ref{lem:neighbors-subspace-mem} for $V:= 0$.
		\item 	Follows from Lemma~\ref{lem:neighbors-subspace-mem}, using $V = X$.
		
		\item 	Follows from Lemma~\ref{lem:neighbors-subspace-local-solution}
		using $S:=S$, $V:= \set{0}$, $\alpha := \tilde{\alpha}$, and $\alpha' := \alpha'$. \claimqedhere
	\end{enumerate} 
\end{claimproof}

 Prover chooses, according to their winning strategy, the round type to play.

\subparagraph*{Singleton-Set.}
Because $W=\set{\alpha}$, we have $V=\set{0}$ and $U_i = \set{\tup{u}_i}$.
We apply the Singleton-Set Lemma (Lemma~\ref{lem:singleton-set-lemma}).
We need to show, for every $s \in r$, that
\[\bIso{(H_s,D_s,s)}{(G_1,N_{G_1}^s(\tup{u}_1))} = \bIso{(H_s,D_s,s)}{(G_2,N_{G_2}^s(\tup{u}_2))}.\]
So let $s\in r$.
We first partition the $s$-neighborhood by the neighbor-type
and  then apply the Partition Lemma. 
Let $\ell \in [k]$ be such that $\bag(s)\setminus \ell = \bag(r) \setminus \ell$. Such an $\ell$ exists since $D$ is a smooth tree-decomposition.
Thus, we need to show for every $\eqTB \in N_S^\ell(\eqTA)$
that 
\[\bIso{(H_s,D_s,s)}{(G_1,N_{G_1}^{s,\eqTB}(\tup{u}_1))} = \bIso{(H_s,D_s,s)}{(G_2,N_{G_2}^{s,\eqTB}(\tup{u}_2))}.\]
For this, let $\eqTB \in N_S^\ell(\eqTA)$.
Assume that Verifier chooses this $s$, $\ell$, and $\eqTB$ in the game.
Then there is an $\alpha' \in \exts{\sols{\eqTB}}{\forgt{\alpha}{\ell}} =:W'$ since Prover has a winning strategy
and the game reaches position $(\eqTB, W', T_s, s)$.
Let $V' :=  \exts{\sols{\homsys{\eqTB}}}{\set{0_{\forgt{V}{\ell}}}}$ be the linear space of $W'$.
By Claim~\ref{clm:neighborhood}\ref{itm:neighbrood_single_vertex_iso},
$\alpha' + N_{G_1}^{s,\eqTB}(\tup{u}_1) = N_{G_2}^{s,\eqTB}(\tup{u}_2)$
because $\tup{u}_2 = \alpha + \tup{u}_1$.
Claim~\ref{clm:neighborhood}\ref{itm:neighborhood_single_vertex_auto} shows
$N_{G_i}^{s,\eqTB}(\tup{u}_i) = \tup{v} + V'$ for every $\tup{v} \in N_{G_i}^{s,\eqTB}(\tup{u}_i)$.
Because we followed the strategy of Prover, they win from position
$(\eqTB, \alpha' +V', T_s, s)$.
Hence, $\bIso{(H_s,D_s,s)}{(G_1,N_{G_1}^{s,\eqTB}(\tup{u}_1))} = \bIso{(H_s,D_s,s)}{(G_2,N_{G_2}^{s,\eqTB}(\tup{u}_2))}$ by the inductive hypothesis.
By the Singleton-Set Lemma we conclude $\bIso{(H, D, r)}{(G_1, U_1)} = \bIso{(H, D, r)}{(G_2, U_2)}$.

\subparagraph*{Partition.}
Prover picks, according to their winning strategy, an affine subspace $W' \subseteq W$.
We can assume that $|W'| < |W|$ since otherwise the game makes no progress.
Let $V' \leq V$ be the linear space of $W'$. Recall, that this means $W' = \alpha' + V'$ for every $\alpha' \in W'$.
The space $V'$ defines a partition $P_i$ of the $U_i$ whose blocks are the elements of $U_i/V'$ (formally two tuples $\tup{u},\tup{v} \in U_i$ are in the same block of $P_i$ if there is some $\beta \in V'$ such that $\tup{u} + \beta = \tup{v}$).
This is indeed a partition of $U_i$ since $U_i = \tup{u} + V$ for all $\tup{u} \in U_i$ and so in particular $\tup{u} + V' \subseteq U_i$ for all $\tup{u} \in U_i$. We define a bijection $P_1\to P_2$:
A set $\tup{u} + V' \in P_1$ is mapped to $\tup{u} + W' \in P_2$.
Because $\tup{u} + V' = \tup {v} +V'$ for all $\tup{v} \in \tup{u}+V'$, this is well-defined. Since, $W' + V' = W'$, the set $\tup{u} + W'$ is indeed in $P_2$ because $\alpha + \tup{u} \in U_2$ and $W' \subseteq W$.
So the map is indeed a bijection.

To show $\bIso{(H, D, r)}{(G_1, U_1)} = \bIso{(H, D, r)}{(G_2, U_2)}$
we apply the Partition Lemma (Lemma~\ref{lem:bijection-lemma}).
For every $U\in P_1$, we need to show 
$\bIso{(H, D, r)}{(G_1, U)} = \bIso{(H, D, r)}{(G_2, f(U))}$.
By the definition of the $P_i$, we have that $V' + U = U$ for every $U \in P_i$.
Because $f(U) = W' + U$, each $\alpha' \in W'$ satisfies $\alpha' + U = f(U)$.
Then the claim follows by the inductive hypothesis because Prover wins in the reached position $(\eqTA, W', T, r)$.

\subparagraph*{Factorization.} 
Prover picks, according to their winning strategy, some $R \subsetneq r$.
Note that we can assume that Prover never picks $R = \emptyset$, since then case (c) of the round is the same position. Since $R \subsetneq r$ there are at least two children of $r$.

We want to use the Factorization Lemma and prove the assumption of the lemma.
\begin{enumerate}
	\item We first show that all $s\in R$ satisfy for all $i\in[2]$ and $\tup{u},\tup{u}' \in U_i$ that
	\[\bIso{(H^r_s, D^r_s,r)}{(G_i,\tup{u})} = \bIso{(H^r_s,D^r_s, r)}{(G_i, \tup{u}')} =: c_s^i.\]
	We let in subround (2a) Verifier choose the $\alpha' \in V$ such that $\tup{u} + \alpha' = \tup{u}'$.
	The game proceeds to the homogeneous system $\homsys{S}$ to position $(\homsys{\eqTA}, \set{\alpha'}, T^r_s, r)$.
	Because the size of the tree decreased, the claim follows from the inductive hypothesis.
	
	\item 
	We next show that $\prod_{s \in R} c_s^1 = \prod_{s \in R} c_s^2$,
	for which we will use case (2b) of the round.
	We will show the stronger claim that $c_s^1 = c_s^2$ for all $s \in S$.
	By the previous case, all $s\in S$ satisfy
	$\bIso{(H^r_s,D^r_s,r)}{(G_i, U_i)} = |U_i| \cdot c_s^i$.
	Since $|U_1| = |U_2|$, it suffices to show for
	all $s \in S$ that 
	\[\bIso{(H^r_s,D^r_s,r)}{(G_1, U_1)} = \bIso{(H^r_s,D^r_s,r)}{(G_2, U_2)}.\]
	To show this, we let Verifier pick this $s$ and the game proceeds to position
	$(\eqTA, W,T^r_s, r)$.
	Because Prover has a winning strategy from this position,
	the inductive hypothesis implies the claim.
	\item 
	We finally show the following:
	\[\bIso{(H^r_{-R},D^r_{-R},r)}{(G_1, U_1)}= \bIso{(H^r_{-R},D^r_{-R},r)}{(G_2, U_2)}.\] 
	Since Prover has a winning strategy from position $(\eqTA, W, T^r_{-R}, r)$, this follows immediately from the induction hypothesis. 
\end{enumerate}
We have shown all assumptions of the Factorization-Lemma (Lemma~\ref{lem:hom-one-step-graphs-2}) and conclude that  $\bIso{(H,D,r)}{(G_1, U_1)}= \bIso{(H,D,r)}{(G_2, U_2)}$, as required.

\subparagraph*{Unique-Child.}
In this case $r$ has a unique child $s$ and we use the Unique-Child Lemma.

Let $\ell\in[k]$ such that $\bag(r)\setminus \ell = \bag(s) \setminus \ell$
and let $\eqTB \in N_S^\ell(\eqTA)$.
We define partitions $P_i^{\eqTB}$ of $N_{G_i}^{s,\eqTB}$ for each $i\in [2]$.
Let $W'_{\eqTB} \leq \exts{\forgt{W}{\ell}}{\sols{\eqTB}}$ be the choice of Prover when Verifier chooses $\ell$ and $\eqTB$.
Let $V'_{\eqTB}$ be the linear space of $W'_{\eqTB}$.

Set $X_{\eqTB}' := \exts{\sols{\homsys{\eqTB}}}{V'_{\eqTB}}$.
Define $P_i^{\eqTB} := \setcond{\tup{v} + X_{\eqTB}' }{\tup{v} \in N_{G_i}^{s,\eqTB}(U_i)}$.
To show that this defines a partition of $N_{G_i}^{s,\eqTB}(U_i)$,
we show that $\tup{v} + X_{\eqTB}' \subseteq  N_{G_i}^{s,\eqTB}(U_i)$ for every $\tup{v} \in N_{G_i}^{s,\eqTB}(U_i)$.
Let $\tup{v} \in N_{G_i}^{s,\eqTB}(\tup{u})$ for some $\tup{u} \in U_i$.
Let $X := \exts{V}{V'_{\eqTB}}$.
Then $\forgt{X}{\ell} = V'_{\eqTB}$ because  $V'_{\eqTB} \leq \exts{\forgt{V}{\ell}}{\sols{\homsys{\eqTB}}}$.
So $\exts{\forgt{X}{\ell}}{X_{\eqTB}'} = \forgt{X}{\ell}$.
Claim~\ref{clm:neighborhood}\ref{itm:neigbors-subspace-step} implies $X_{\eqTB}' + \tup{v} = N_{G_i}^{s,\eqTB}(X+\tup{u}) \subseteq N_{G_i}^{s,\eqTB}(V+\tup{u})$.
Finally, $W^{\eqTB}$ defines a bijection $P_1^{\eqTB} \to P_2^{\eqTB}$ via
$\tup{v} + V_{\eqTB}' \mapsto   \tup{v} + W''_{\eqTB}$, where $W''_{\eqTB} := \exts{\sols{\eqTB}}{W'_{\eqTB}}$.

We combine the partitions $P_i^{\eqTB}$ 
into a partition $P_i$ of $N_{G_i}^{s}$ (by the disjoint union of the $P_i^{\eqTB}$).
The bijections defined by $W^{\eqTB}$ induce a bijection $P_1 \to P_2$.
We show that this partition satisfies the requirement of the Unique-Child Lemma:
\begin{enumerate}
	\item 
	We show that all $\eqTB \in N_S^\ell(\eqTA)$, $p \in P_1^{\eqTB}$, $\tup{v}_1 \in p$, and $\tup{v}_2 \in 
	 p+ W''_{\eqTB}$
	satisfy
	\[\left|\setcond*{\tup{w} \in U_1}{\tup{v}_1 \in N_{G_1}^{s}(\tup{w})}\right| =  \left|\setcond*{\tup{w} \in U_2}{\tup{v}_2 \in N_{G_2}^{s}(\tup{w})}\right|.\]
	Because $\tup{v_1} \in p \in P_1^{\eqTB}$,
	if $\tup{v}_i \in N_{G_i}^{s}(\tup{w})$,
	then $\tup{v}_i \in N_{G_i}^{s,\eqTB}(\tup{w})$.
	
	For $\eqTB \in N_S^\ell(\eqTA)$,	define $K^{\eqTB}:= \exts{V}{0_{\forgt{\sols{\homsys{\eqTB}}}{\ell}}}$.
	Now let $\eqTB \in N_S^\ell(\eqTA)$, $p \in P_i^{\eqTB}$, $\tup{v}_i \in p$.
	It suffices to show that $|\setcond{\tup{w} \in U_i}{\tup{v} \in  N_{G_i}^{s, \eqTB}(\tup{w})}| = |K^{\eqTB}|$.

	Because $\tup{v} \in p \subseteq N_{G_i}^{s,\eqTB}(U_i)$,
	there is a $\tup{u} \in U_i$ such that $\tup{v} \in N_{G_i}^{s,\eqTB}(\tup{u})$.
	We show $\setcond{\tup{w} \in U_i}{\tup{v} \in  N_{G_i}^{s,\eqTB}(\tup{w})} = K^{\eqTB} +\tup{u}$, which implies the claim.
	
		We have $\setcond{\tup{w} \in U_i}{\tup{v} \in  N_{G_i}^{s,\eqTB}(\tup{w})} = N_{G_i}^{r,\eqTA}(\tup{v}) \cap U_i$.
		Then we have that $N_{G_i}^{r,\eqTA}(\tup{v}) = \tup{u} + \exts{\sols{\homsys{\eqTA}}}{\set{0_{\eqTA \setminus \ell}}}$ by Claim~\ref{clm:neighborhood}\ref{itm:neighborhood_single_vertex_auto} (where the roles of $\eqTA$ and $\eqTB$ are switched).
		Thus, $N_{G_i}^{r,\eqTA}(\tup{v}) \cap U_i = \tup{u} + \exts{V}{\set{0_{\forgt{\sols{\homsys{\eqTA}}}{\ell}}}} = \tup{u}+K^{\eqTB}$.
	
	\item 	We show that all $\eqTB \in N_S^\ell(\tup{f})$ and all $p \in P_i^{\eqTB}$ satisfy
	\[\bIso{(H_s, D_s, s)}{(G_1, p)} = \bIso{(H_s, D_s, s)}{(G_2, W''_{\eqTB} + p)}.\]
	Let $\eqTB \in N_S^\ell(\tup{f})$ and $p \in P_i^{\eqTB}$.
	By definition of the partition, we have that $p = \tup{v} + V_{\eqTB}'$ for all $\tup{v} \in p$.
	Because Prover has a winning strategy from position $(\eqTB, W''_{\eqTB}, T_s, s)$ (the position reaches after Prover chose $W'_{\eqTB}$)
	the claim follows.
\end{enumerate}

With the Unique-Child Lemma (Lemma~\ref{lem:hom-one-step-graphs-3})
we conclude that $\bIso{(H,D,r)}{(G_1, U_1)}= \bIso{(H,D,r)}{(G_2, U_2)}$.
\end{proof}

\subparagraph{Adding an Initial Position.}
Currently, the indistinguishability game does not have an initial empty position.
In the empty position $(T,r)$, the game proceed as follows:
Verifier picks a tuple $\eqTA \in S^{k}$ and Prover picks a solution $\alpha \in \sols{\eqTA}$. If such a solution does not exist, then Verifier wins immediately.
Otherwise, the game continues in the position $(\eqTA, \set{\alpha}, T, r)$.

\begin{lemma}
	\label{lem:indistinguisability-game-to-biso-initial}
	Let $S$ be an $\FF_2$-system and $T$ be a tree with root $r$.
	If Prover has a winning strategy in the $k$-indistinguishability game in position $(T, r)$,
	then every graph $H$ with a smooth tree-decomposition $(D,r)$ of width $k-1$ such that $T_D = T$ satisfies 
	\[\bIso{(H,D)}{\CFI{S}} = \bIso{(H,D)}{\CFI{\homsys{S}}}.\]
\end{lemma}
\begin{proof}
	For every $\eqTA\in S^{k}$, let $\alpha_{\eqTA} \in \sols{\eqTA}$ be a solution picked by Prover according to the winning strategy when Verifier chooses $\eqTA$.
	Then $\alpha_{\eqTA}$ defines a bijection between the vertices of $\CFI{S}$ and $\CFI{\homsys{S}}$ of origin $\eqTA$ and $\homsys{\eqTA}$, respectively.
	Combine these bijections into a bijection $V(\CFI{S})^k \to V(\CFI{\homsys{S}})^k$.
	
	We trivially have
	$\bIso{(H,D)}{\CFI{S}} =  \bIso{(H,D,r)}{(\CFI{S},V(\CFI{S}))}$
	and similarly for $\CFI{\homsys{S}}$.
	By the Partition Lemma (Lemma~\ref{lem:bijection-lemma}),
	it suffices to show for every $\eqTA\in S^k$ and $\tup{u} \in V(\CFI{S})^k$ of origin $\eqTA$ that
	$ \bIso{(H,D,r)}{(\CFI{S},\set{\tup{u}})} = \bIso{(H,D,r)}{(\CFI{\homsys{S}},\set{\alpha_{\eqTA} + \tup{u}})}$.
	Lemma~\ref{lem:indistinguishability-to-hom-indistinguishability} implies this using  $V=\set{0}$
	because Prover wins  the $k$-indistinguishability game on $S$ in position $(\eqTA, \set{\alpha_{\eqTA}}, T, r)$.
\end{proof}

\subsection{Color Reduction and the Indistinguishability Game}

We now need to revisit the construction which ``colors'' the CFI graphs from Section~\ref{sec:coloring-cfi}.
We need to show that Prover strategies for an $\FF_2$-system $S$ also translate to
the ``colored'' system $\cld{S}$.
The idea is simple: always set new variables to zero.
Due to the complexity of the indistinguishability game, the proof is quite involved even though the underlying idea is very simple.

\begin{lemma}
	\label{lem:color-reduction-indistinguishability-game}
	Let $T$ be a tree with root $r$.
	If Prover wins the $k$-indistinguishability game from position $(T,r)$ on $S$,
	then Prover wins the $k$-indistinguishability game from position $(T,r)$ on~$\cld{S}$.
\end{lemma}
\begin{proof}
	\let\cldold\cld
	\renewcommand{\cld}[1]{\dot{#1}}
	We again write $\cld{S}$ for $\cldold{S}$.
	We first consider non-initial positions.
	Let $(\cld{\eqTA},\cld{W}, T, r)$ be a position in the game in $\cld{S}$ (or $\homsys{\cld{S}}$).
	\begin{enumerate}
		\item 	We call the position $(\cld{\eqTA}, \cld{W}, T, r)$ good if all $\cld{\alpha} \in \cld{W}$ assign all new variables to zero.
		\item  A tuple $\eqTA \in S^k$ witnesses $\cld{\eqTA}$,
		if for every old equation in $\cld{\eqTA}$, the tuple $\eqTA$ contains the corresponding old equation (without the new variables) at the same position.
		\item A good position $(\cld{\eqTA},\cld{W}, T, r)$  on $\cld{S}$ (or $\homsys{\cld{S}})$ is witnessed by a position $(\eqTA,W, T , r)$ on $S$ (or $\homsys{S}$, respectively),
		if $\eqTA$ witnesses $\cld{\eqTA}$, and $\restrict{W}{\vars{\cld{\eqTA}}} := \restrict{\cld{W}}{\vars{\eqTA}}$,
		that is, the positions agree on the common old variables.
	\end{enumerate}
	
	We show that Prover has a strategy that maintains the following invariant:
	The current position on $\cld{S}$ (or $\homsys{\cld{S}}$) is good and witnessed by a position $(\eqTA,W, T , r)$ on $S$ (or $\homsys{S}$, respectively), from which Prover has a winning strategy.
	The proof proceeds by induction first on $|T|$ and second on $|V|$.
	
	So assume that the current position is $(\cld{\eqTA},\cld{W}, T, r)$ on $\cld{S}$ (the following is also true for $\homsys{\cld{S}}$).
	Assume that the position is good and witnessed by the position $(\eqTA, W, T, r)$.
	On $S$, Prover decides which round type to play.
	We let Prover play the same round type on $\cld{S}$:
	\begin{enumerate}
		\item Assume that Prover plays a singleton-set round on $S$.
		Prover also plays a singleton-set round on $\cld{S}$
		and Verifier picks $\ell\in[k], \cld{\eqTB} \in N_{\cld{S}}^\ell(\cld{\eqTA})$ and $s\in r$.
			
		Then there is a $\eqTB \in N_S^\ell(\eqTA)$ that witnesses $\cld{\eqTB}$:
		\begin{itemize}
			\item If the $\ell$-th position in $\cld{\eqTB}$ is an old equation $\cld{g}$, choose the $\ell$-th position of $\eqTB$ to be $g$.
			\item Otherwise, choose $\eqTB = \eqTA$.
		\end{itemize}
		We let on $S$ Verifier choose $\ell$,  $\eqTB$, and $s$.
		Then game proceeds to position $(\eqTB, W', T_s, s)$,
		where $W' = \exts{\sols{\eqTB}}{\forgt{\alpha}{\ell}}$.
		
		The game on $\cld{S}$ proceeds to position $(\cld{\eqTB}, \cld{W}' , T_s, s)$, where $\cld{W}':=\exts{\sols{\cld{\eqTB}}}{\set{\cld{\alpha}}}$.
		We now let Prover on $\cld{S}$ play a partition round choosing the subspace $\cld{W}'' \subseteq  \cld{W}'$ of all assignments that 
		\begin{itemize}
			\item set all new variables to zero and
			\item extend to assignments in $W'$.
		\end{itemize}
		The game on $\cld{S}$ proceeds to position $(\cld{\eqTB}, \cld{W}'', T_s, s)$.
		This position is good and witnessed by the position $(\eqTB, W', T_s, s)$ on $S$.
		Because Prover wins from $(\eqTB, W', T_s, s)$ on $S$,
		the invariant is satisfied.
		Thus, Prover wins from $(\cld{\eqTB}, \cld{W}'', T_s, s)$ on $\cld{S}$ by the inductive hypothesis.
		
		We will see this additional partition round to restrict $\cld{W}'$
		as above also in the following cases.
		We will refer to this as the \emph{cleanup round} without repeating the details again.

		\item Assume that Prover plays a partition round on $S$ choosing the affine subspace $W' \subseteq W$.
		We can assume that $W' \neq W$  since otherwise this round is unnecessary.
		The game continues in position $(\eqTA, W', T, r)$.
		
		On $\cld{S}$ Prover plays a partition round and chooses the subspace $\cld{W'} \subseteq \cld{W}$ of assignments that extend to assignments in $W'$ and set new variables to zero.
		The game on $\cld{S}$ continues in position $(\cld{\eqTA}, \cld{W}', T, r)$.
		This position is good and witnessed by $(\eqTA, W', T, r)$.
		
		Because Prover wins in this position on $S$, the invariant is satisfied and Prover wins on $\cld{S}$ by the inductive hypothesis (note that $|W'| < |W|$).
		
		\item Assume that Prover plays a factorization round on $S$
		and chooses $R \subsetneq r$.
		On $\cld{S}$, Prover chooses the same set $R$.
		We again can assume that $s$ has at least two children,
		otherwise Prover does not benefit from this round.

		Now Verifier decides for one of the following options:
		\begin{enumerate}
				\item 
			In the game on $\cld{S}$, Verifier picks an $\cld{\alpha}' \in \cld{V}$ and an $s \in R$.
			The game proceed on $\homsys{\cld{S}}$ to position
			$(\cld{\eqTA},\set{\cld{\alpha}'}, T^r_s, r)$.
			In the game on $S$, we let Verifier pick the same $s$ and the $\alpha' \in V$ that agrees with $\cld{\alpha}'$ on the common variables.
			The game proceed on $\homsys{S}$ in position
			$(\cld{\eqTA},\set{\alpha'}, T_s^r, r)$.
			\item Verifier picks  $s \in R$.
			On $\cld{S}$, the game proceeds to position $(\cld{\eqTA}, \cld{W}, T_s^r, r)$.
			On $S$, we let Verifier pick  the same and $s$.
			The game on $S$ continues in position $(\eqTA, W, T_s^r, r)$,
			which clearly witnesses the position on $\cld{S}$.
			\item 
			The game on $\cld{S}$ continues in position $(\cld{\eqTA}, \cld{W}, T^r_{- R}, r)$,
			which is good and witnessed by
			the position $(\eqTA, W, T^r_{ - R}, r)$ reached on $S$.
		\end{enumerate}
		In all cases, the invariant is satisfied and Prover wins by the inductive hypothesis. 
		\item 
		Assume that Prover plays a Unique-Child round on $S$. Then $r = \set{s}$.
		Prover also plays a Unique-Child round on $\cld{S}$.
		Verifier chooses a $\ell \in [k]$ and $\cld{\eqTB} \in N_{\cld{S}}^\ell({\cld{\eqTA}})$.
		Let $\eqTB \in N^{\ell}_S(\eqTA)$ witness $\cld{\eqTB}$.
		Suppose that on $S$  Verifier picks $\eqTB$. Then Prover replies according to their winning strategy by picking
		a non-empty affine subspace $W' \subseteq \exts{\forgt{W}{\ell}}{\sols{\eqTB}}$.
		Let $W'' = \exts{\sols{\eqTB}}{W'}$ 
		The game on $S$ continues in position $(\eqTB, W'', T_s, s)$.
		
		Define the affine subspace $\cld{W}' \subseteq \exts{\forgt{\cld{W}}{\ell}}{\sols{\cld{\eqTB}}}$ as above to be the set of assignments extending to $W'$ and that set all new variables to zero.
		In the game on $\cld{S}$, we let Prover choose $\cld{W}'$.
		Let $\cld{W}'' = \exts{\sols{\cld{\eqTB}}}{\cld{W}'}$.
		This affine subspace is non-empty by the same reasons as before:
		extend elements in $W''$ by assigning new variables to zero and restrict appropriately.
		The game on $\cld{S}$ reaches position $(\cld{\eqTB}, \cld{W}'', T_s, s)$.
		This position is good and witnessed by $(\eqTB, W'', T_s, s)$.
		So Prover wins on $\cld{S}$ by the inductive hypothesis.
	\end{enumerate}	
	Finally, we have to deal with the initial position.
	When Verifier chooses a tuple $\cld{\eqTA} \in \cld{S}^k$, on $S$
	we let Verifier choose a tuple $\eqTA \in S^k$ witnessing $\cld{\eqTA}$.
	Then Prover chooses an $\alpha \in \sols{\eqTA}$ on $S$
	and we pick $\cld{\alpha} \in \sols{\cld{\eqTA}}$,
	that is obtained from extending $\alpha$ by the zero assignment on new variables (and possibly restricting).
	The resulting position $(\cld{\eqTA}, \set{\cld{\alpha}}, T, r)$ on $\cld{S}$ is good and witnessed by the position $(\eqTA, \set{\alpha}, T, r)$ on $S$.
	Thus Prover wins from $(\cld{\eqTA}, \set{\cld{\alpha}}, T, r)$ on $\cld{S}$.
\end{proof}

In the same way we have shown that the color reduction preserves strategies of Prover, we also have the same for the thinness reduction.
The strategy is the same as for the $k$-equation system in Lemma~\ref{lem:thinnify-system} and can be carried
out in a similar way as in the former proof.
We omit a full formal proof.

\begin{lemma}
	\label{lem:thinness-indistinguishability-game} 
	Let $T$ be a tree with root $r$ and $S$ be an $\FF_2$-system. 
	If Prover wins the $k$-indistinguishability game from position $(T,r)$ on $S$,
	then Prover wins the $k$-indistinguishability game from position $(T,r)$ on $\thin{S}$.
\end{lemma} 

\begin{theorem}
	\label{thm:indistingusihability-game-equation-game-summary}
	For every $\FF_2$-system $S$, there is an $\FF_2$-system $S'$ such that for all $k\ge 3$:
	\begin{enumerate}
		\item $|S'| = O(|S|)$,
		\item the arity of $S'$ is the arity of $S$ plus one,
		\item for all trees $T$ it holds that if Prover wins the $k$-indistinguishability game on $S$ in position $(T,r)$,
		then all graphs $H$ of treewidth $k-1$ with a smooth  tree-decomposition $D$ such that $T_D = T$ satisfy $\bIso{(H,D)}{\CFI{S'}} = \bIso{(H,D)}{\CFI{\homsys{(S')}}}$, and
		\item if Falsifier wins the $k$-equation game on $S$, then 
		there is a graph $H$ of treewidth $k-1$ such that $\Hom{H}{\CFI{S'}} \neq \Hom{H}{\CFI{\homsys{(S')}}}$. 
	\end{enumerate}
\end{theorem}
\begin{proof}
	Let $S$ be an $\FF_2$-system and consider the $S' := \cld{\thin{S}}$.
	Clearly, $|S'| = O(|S|)$  by Lemmas~\ref{lem:thinnify-system} and~\ref{lem:coloring-size} and the maximal arity increased by one.
	
	If Prover wins the $k$-indistinguishability game from $(T,r)$ on $S$,
	then Prover wins from $(T,r)$ on $S'$ by Lemmas~\ref{lem:color-reduction-indistinguishability-game} and~\ref{lem:thinness-indistinguishability-game}.
	Thus, by Lemma~\ref{lem:indistinguisability-game-to-biso-initial},
	$\bIso{(H,D)}{\CFI{S'}} = \bIso{(H,D)}{\CFI{\homsys{(S')}}}$
	for all graphs $H$ of treewidth $k$ as required.
	
	If Falsifier wins the $k$-equation game on $S$,
	then Falsifier wins the $k$-equation game on $S'$ by Lemma~\ref{lem:color-reduction-equation-game} and~\ref{lem:thinnify-system}.
	Thus, Spoiler wins the bijective $k$-pebble game on $\CFI{\homsys{(S')}}$ and $\CFI{S'}$
	by Lemma~\ref{lem:playerI-equation-game}.
	Hence, there is a graph of treewidth $k-1$ distinguishing the two CFI graphs via homomorphism counts by Theorem~\ref{thm:hom-eq-iff-biso-hom-eq}.
\end{proof}

\section{The Size Game} \label{sec:size-game-full}

We now simplify the game of the previous section.
First, instead of considering a tree $T$, we only consider its size.
Second, we only allow for more restrictive rules, making the game harder for Prover. This makes the game simpler and still suffices to show our lower bound later. 
Finally, we change our perspective. We no longer remember a tuple of equations but instead the underlying set. Moreover, while our tuples have length $k+1$ the sets will only have size at most $k$ (in the position at the end of the round). This is more convenient for our lower bound proof and is achieved via a simple trick.
In the indistinguishability game Verifier often chooses a $(k+1)$-tuple $\eqTB \in N_S^\ell(\eqTA)$.
We can view this as Verifier first choosing an entry of $\eqTA$ to forget and then choosing an equation to fill the gap.
In the size game we effectively end the round after the first choice is made and start the next round with the second choice. 

These changes make the game more similar to the bijection/equation games. Reflecting this in this game we will use the term Spoiler instead of Verifier and Duplicator instead of Prover. We now give the formal details.

For an $\FF_2$-system $S$, a set $F\subseteq S$, an equation $f \in S$,
and a set $W \subseteq \sols{F}$, define $W^{+f}:= \exts{\sols{F \cup \set{f}}}{W}$.
If $V \subseteq \sols{\homsys{F}}$, we write $V^{+f}$ for $V^{+\homsys{f}}$. For a positive integer $k$ we say that $E \subseteq F \cup \set{f}$ of size at most $k$ 
is an $(f,k)$-neighbor of $F$.
 If $k$ is clear/fixed from the context so we just call $E$ an $f$-neighbor of $F$. In particular, whenever we are playing the width-$k$ size game we consider $k$ to be fixed.

For an integer $k \geq 1$,
the \defining{width-$k$ size game} is played between Spoiler and Duplicator on an $\mathbb{F}_2$-system $S$.
A position in the game is a tuple $(F, W, n)$ consisting of:
\begin{enumerate}
	\item a set $F \subseteq S$ of size at most $k$,
	\item an affine space $W \subseteq \sols{F}$, and
	\item a positive number $n \in \nat$.
\end{enumerate}
The initial position is $(\emptyset, \set{0}, n)$. 
The game is played between Spoiler and Duplicator as follows.

If $F=\emptyset$, Spoiler picks a non-empty set $E\subseteq S$ of size 
at most $k$ and Duplicator picks a solution $\alpha \in \sols{E}$. If such a solution does not exist, Spoiler wins immediately.
The game continues in position $(E, \set{\alpha}, n)$.

Otherwise $|F| >0 $ and a round proceeds as follows. If any $\alpha \in W$ violates an equation of $S$, then Spoiler wins.
Otherwise, if $n=1$, then Duplicator wins. If $n>1$, 
then Duplicator chooses whether to play a \defining{regular round} or a \defining{subtree round}.

\begin{description}
	\item[Regular Round] ~
	\begin{enumerate}
		\item Spoiler chooses an equation $f \in S$.
		\item Duplicator chooses $\alpha \in W^{+f}$.
		\item Spoiler chooses an $f$-neighbor $E$ of $F$.
		The game continues in position $(E, \set{\restrict{\alpha}{E}}, n-1)$.
	\end{enumerate}
	\item [Subtree Round]~
	\begin{enumerate}
		\item Spoiler chooses an equation $f \in S$.
		\item Duplicator chooses an affine subspace $W^+ \subseteq W^{+f}$.
		\item Spoiler chooses an $f$-neighbor $E$ of $F$. Define $W^- := \restrict{W^+}{E}$. Spoiler also decides how to continue:
		\begin{description}
			\item[Big Subtree Round] Duplicator chooses an affine space $W' \subseteq W^{-}$.
			The game continues on $S$ from position $(E, W', n-1)$.
			\item [Small Subtree Round] Spoiler picks $\alpha \in  V^{-}$, where $V^{-}$ is the linear space of $W^{-}$.
			The game continues on the homogeneous system $\homsys{S}$ in position $(\homsys{E}, \set{\alpha}, \lfloor n/2 \rfloor)$.
		\end{description}

	\end{enumerate}
	
\end{description}

 Note that Spoiler can only win at the beginning of a round.
Moreover, if Spoiler does not win at the start of a round,
then $W^{+f}$ is non-empty.
This means that the above description is well-defined.
In detail: if Duplicator does not lose immediately we have $W \subseteq \sols{S[\vars{F}]}$. Let $\alpha \in W$.
Clearly if $\vars{f} \subseteq \vars{F}$, then $\alpha \in W^{+f} = W$.
Otherwise, $\vars{f}$ has at least one variable $z \notin \vars{F}$, so we can 
extend $\alpha$ on $z$ to satisfy $f$.
This extension is in $W^{+f}$.

\begin{lemma}
	\label{lem:size-game-to-indistinguishability-game}
	Let $S$ be an $\FF_2$-system, $F \subseteq S$ of size at most $k$, $W \subseteq \sols{F}$ be an affine space, and $n \in \nat$.
	If Duplicator has a winning strategy from position $(F, W, n)$  in the width-$k$ size game on $S$ such that $F\neq \emptyset$,
	then for 
	\begin{enumerate}
		\item  every tuple $\eqTA\in S^{k+1}$ 
		whose entry set is $F \cup \set{f}$ for some $f\in S$ and 
		\item  every tree  $T$ of size at most $n$ with root $r$
	\end{enumerate}
	Prover has a winning strategy in the $(k+1)$-indistinguishability game on $S$ from position $(\eqTA, W^{+f}, T, r)$. 
\end{lemma}

\begin{proof}
	 We proceed by induction on $|T|$. For the base case $|T|=1$, since Duplicator wins we saw above that $W^{+f}$ is non-empty for every $f \in S$, so $(\eqTA, W^{+f}, T, r)$ is indeed a well-defined position of the $(k+1)$-indistinguishability game. Since $|T| = 1$, Prover immediately wins from this position.
	
	So now suppose that $|T|>1$ and that Duplicator has a winning strategy in the width-$k$ size game from position $(F,W, n)$.
	Let $f \in S$ be an equation such that 
	the entry set of $\eqTA$ is $F \cup \set{f}$.
	Again since Duplicator does not lose immediately, $W^{+f}$ is non-empty. Duplicator decides whether to play a regular round or a subtree round according to their winning strategy.
	 We make a case distinction based on this decision.
	
	\subparagraph*{Case 1: Regular Round}
	 Suppose that Spoiler decides to pick the equation $f$.
	Then Duplicator picks $\alpha \in W^{+f}$ according to their winning strategy. 
	In the $(k+1)$-indistinguishability game starting from the position $(\eqTA,W^{+f}, T, r)$, Prover plays a partition round where they choose the affine subspace $W' := \set{\alpha}$.
	The indistinguishability game reaches the position $(\eqTA, \set{\alpha} , T,r)$.
	
	Next, Prover chooses a singleton set round. 
	Verifier chooses $\ell \in [k+1]$, $\eqTB \in N_S^\ell(\eqTA)$ and $s\in r$.
	We need to show that  $W'' := \exts{\sols{\eqTB}}{\forgt{\alpha}{ \ell}}$ is non-empty and 
	that Prover has a winning strategy in the indistinguishability game from position $(\eqTB, W'', T_s,s)$.
	
	To this end, suppose that in the size game Spoiler chooses $E$ to be the entry set of $\tup{e}\setminus \ell = \tup{f} \setminus \ell$. Then $|E| \leq k$.
	The size game reaches the position $(E, \set{\restrict{\alpha}{E}}, n -1)$, from which Duplicator has a winning strategy.
	Therefore, $\forgt{\alpha}{ \ell} = \restrict{\alpha}{E} \in \sols{S[\vars{E}]}$, since otherwise Spoiler would win at the start of the next round.
	It follows that the affine space
	$W''$ is non-empty (applying the same argument to show that the game is well-defined). 
	
	Next observe that $E$ is equal to the entry set of $\eqTB\setminus \ell$ and
	 $|T_s| < |T| \leq n$. Therefore, by the definition of $W''$ and the inductive hypothesis we conclude that Prover has a winning strategy in the indistinguishability game in position
	$(\eqTB, W'', T_s, s)$ and therefore also from position $(\eqTA, W^{+f}, T, r)$.

	\subparagraph*{Case 2: Subtree Round}
	
	Suppose that Spoiler chooses the equation $f$ and that Duplicator chooses, according to their winning strategy, $W^+\subseteq  W^{+f}$.	
	In the indistinguishability game, Prover chooses to play a partition round and picks the affine space $W^+$. Then the indistinguishability game reaches position $(\eqTA, W^+, T, r)$.

	Next Prover chooses to play a factorization round.  
	Let $s^* \in r$ be a child of $r$  such that $|T_{s^*}|$ is maximal.
	Note that if $|T_{s^*}| \geq |T|/2$, the child $s^*$ is unique.
	Prover chooses $R :=  r \setminus \set{s^*}$.
	Now Verifier chooses how to continue in the indistinguishability game:
	\begin{enumerate}[(a)]

		\item Verifier picks $s \in R$ and $\alpha \in V^{+}$,
		where $V^{+}$ is the linear space of $W^+$.
		The indistinguishability game reaches the  position $(\homsys{\eqTA}, \set{\alpha}, T^r_s, r)$ on the homogeneous system $\homsys{S}$.
		From here Prover plays a singleton-set round. In this round Verifier chooses $\ell \in [k+1]$, and $\eqTB	\in N_{\homsys{S}}^{\ell}(\homsys{\eqTA})$.
		Let $W':= \exts{\sols{\eqTB}}{\forgt{\alpha}{\ell}}$.
		 We claim that  $W' \neq \emptyset$ (so Verifier does not immediately win) and that Prover has a winning strategy from the resulting position  $(\eqTB, W', T_s, s)$.
		
		To show these  claims, we suppose that in the size game Spoiler picks $E$ to be the set of entries of $\eqTB \setminus \ell$, chooses to play a small subtree round, and then picks $\restrict{\alpha}{E} = \forgt{\alpha}{\ell}$.
		 The size game continues on the homogeneous system $\homsys{S}$ from position $(\homsys{E}, \set{\forgt{\alpha}{\ell}} , \lfloor n/2 \rfloor)$, from which Duplicator has a winning strategy.
		 
		Therefore, $\forgt{\alpha}{\ell} \in \sols{\homsys{S}[\vars{E}]}$, since otherwise Spoiler would win the size game at the start of the next round. This implies that $\forgt{\alpha}{\ell}$ extends to a solution to $\eqTB$, so $W' \neq \emptyset$.
		Furthermore, the entry set of $\eqTB$ is equal to $E \cup \set{e}$, for some $e \in \homsys{S}$. Also since $s \neq s^{*}$, we have $|T_s| \leq \lfloor |T|/2 \rfloor \leq \lfloor n/2 \rfloor$. Therefore, by the induction hypothesis Prover has a winning strategy in the indistinguishability game from position $(\eqTB, W', T_s, s)$ on $\homsys{S}$ and therefore also from position $(\eqTA, W^{+f}, T, r)$ on $S$.
		
		\item
		Verifier picks $s \in R$ and the indistinguishability game proceeds to position
		$(\eqTA, W^+, T^r_s, r)$.
		Now Prover plays a unique-child round (the unique child is $s$).
		Verifier picks $\ell \in [k+1]$ and $\eqTB \in N_S^\ell(\eqTA)$.
		Let $W^{-}:= \forgt{(W^{+})}{\ell}$. We will show that $W'' := \exts{\sols{\eqTB}}{W^{-}}$ is non-empty. This implies that
		the indistinguishability game moves to position $(\eqTB, W'', T_s, s)$, from which we will show Prover has a winning strategy.

		Towards this, suppose that in the subtree round of the size game Spoiler chooses $E$ to be the entry set of $\eqTB \setminus \ell$ and also to play a big subtree round.
		Then Duplicator chooses $W' \subseteq W^{-}$ according to their winning strategy. Then the size game reaches the position $(E, W', n-1)$ from which Duplicator has a winning strategy. Therefore, as we have seen many times before $\exts{\sols{\eqTB}}{W'} \subseteq W''$ is non-empty, so the indistinguishability game does indeed reach the position $(\eqTB, W'', T_s, s)$. 

		It remains to show that Prover can win from position $(\eqTB, W'', T_s, s)$. To do this Prover chooses another partition round and picks $\exts{\sols{\eqTB}}{W'} \subseteq W''$, so that the game reaches the position $(\eqTB, \exts{\sols{\eqTB}}{W'}, T_s, s)$. By the induction hypothesis, Prover has a winning strategy from this position and therefore also from $(\eqTB, W'', T_s, s)$ and $(\eqTA, W^{+f}, T, r)$.

		\item This is similar to $(b)$. The indistinguishability game continues in  position $(\eqTA, W^+, T^r_{-R}, r)$. Observe that in $T^r_{-R}$, the unique child of $r$ is $s^{*}$. From this position, Prover chooses a unique child round and 
		Verifier picks $\ell \in [k+1]$ and  $\eqTB \in N_S^{\ell}(\eqTA)$. Let $W^{-}:= \forgt{(W^{+})}{\ell}$. Then we need to show that $W'' := \exts{\sols{\eqTB}}{W^{-}}$ is non-empty and that Prover has a winning strategy in   
		the indistinguishability game from the result position $(\eqTB, W'', T_s, s)$. This follows by an identical argument to that given in (b).
		 \qedhere
	\end{enumerate}
	
\end{proof}

\begin{lemma}
	\label{lem:size-game-to-indistinguishability-game-initial}
	Let $S$ be an $\FF_2$-system.
	If Duplicator has a winning strategy in the width-$k$ size game on $S$ from the initial position $(\emptyset, \set{0}, n)$ where in the first round Duplicator chooses a regular round,
	then Prover has a winning strategy in the $(k+1)$-indistinguishability game on $S$ from position $(T, r)$ for every tree $T$ of order at most $n$.
\end{lemma}
\begin{proof}
	Let $T$ be a tree of order at most $n$ with root $r$.
	Let $\eqTA \in S^{k+1}$ be the tuple chosen by Verifier in the indistinguishability game. 	Let  $E$ be the set of entries of $\eqTA$ and let $F$ be any subset of $E$ of size $k = |E|-1$. Let $f$ be the unique element of $E \setminus F$, so that $F \cup \set{f} = E$.
	Suppose Spoiler chooses $F$ in the size game and that Duplicator responds according to their winning strategy with $\alpha \in \sols{F}$. Then we reach the position $(F, \set{\alpha}, n)$ in the size game. In Duplicator's winning strategy from $(F, \set{\alpha}, n)$ in the size game they next choose a regular round (by assumption). Suppose Spoiler chooses the equation $f$ in this round. In this round Duplicator then chooses according to their winning strategy some $\alpha^{+} \in W^{+f}$.
	
	So in the indistinguishability game Prover chooses $\alpha^{+}$ and the game moves to position $(\eqTA, \set{\alpha^{+}}, T, r)$. Then we proceed as in the regular round case in the proof of Lemma~\ref{lem:size-game-to-indistinguishability-game} to show that Prover has a winning strategy from this position; for completeness we repeat the details.

	Prover chooses a singleton set round. 
	Verifier chooses $\ell \in [k+1]$, $\eqTB \in N_S^\ell(\eqTA)$ and $s\in r$.
	We need to show that  $W'' := \exts{\sols{\eqTB}}{\forgt{\alpha}{ \ell}}$ is non-empty and 
	that Prover has a winning strategy in the indistinguishability game from position $(\eqTB, W'', T_s,s)$.
	
	To this end, suppose that in the size game Spoiler chooses $E'$ to be the entry set of $\tup{e}\setminus \ell = \tup{f} \setminus \ell$. Then %
	$|E'| \leq k$.
	The size game reaches the position $(E', \set{\restrict{\alpha}{E'}}, n -1)$, from which Duplicator has a winning strategy.
	Therefore, $\forgt{\alpha}{ \ell} = \restrict{\alpha}{E'} \in \sols{S[\vars{E'}]}$, since otherwise Spoiler would win at the start of the next round.
	It follows that the affine space
	$W''$ is non-empty (applying the same argument to show that the game is well-defined).

	Next observe that $E'$ is equal to the entry set of $\eqTB\setminus \ell$ and
	 $|T_s| < |T| \leq n$. Therefore, by the definition of $W''$ and the inductive hypothesis we conclude that Prover has a winning strategy in the indistinguishability game in position
	$(\eqTB, W'', T_s, s)$ and therefore also from position $(\eqTA, \set{\alpha^{+}}, T, r)$ and hence from position $(T,r)$.
\end{proof}

\section{Compressed Crossed Grids} \label{sec:compressed-crossed-grids-full}

Via the indistinguishability and the size game, to prove our main lower bound it suffices to define for every integer $w$ an $\FF_2$-system $S^w$ such that Falsifier wins the $(w+1)$-equation game on $S^w$ and Duplicator wins the width-$w$ size game on $S^w$. In this section, we define this system $S^w$ and show that Falsifier wins the $(w+1)$-equation game on it.
The system $S^w$ will be a compressed crossed grid. These are closely related to, but crucially distinct from, compressed Tseitin systems~\cite{RezendeFJN025}.
We first introduce Tseitin systems and the notion of compressing an equation system before formally defining these compressed crossed grids.

\subsection{Components and Tseitin Systems} \label{ss:tseitin}
\newcommand{\boundary}[2][]{\partial_{#1}(#2)}
\newcommand{\dangling}[1]{\delta(#1)}
\newcommand{\cmpnt}{\kappa}

Let $S$ be an $\FF_2$-system and $V \subseteq \vars{S}$ be a set of variables.
Think of $V$ as a set of variables assigned to a fixed value and which are thereby ``removed'' from $S$.
The system $S$ is \defining{$V$\nobreakdash-connected} if for every subsystem $T\subseteq S$ such that $\emptyset \neq T \neq S$, it holds that $\vars{T} \setminus V$ and $\vars{S\setminus T}$ are non-disjoint.
A \defining{$V$-component} of $S$ is a maximal $V$-connected subsystem $\cmpnt \subseteq S$
such that $\vars{\cmpnt} \setminus V$ and $\vars{S \setminus \cmpnt}$ are disjoint.
The $V$-component of a variable $v$ in $S$ is $\set{v}$ if $v \notin \vars{S}$ or $v\in V$ and these components are \defining{trivial}, where we set $\vars{\set{v}} = \set{v}$.
Otherwise, the component of $v$ is the $V$-component $\cmpnt$ such that $v \in \vars{\cmpnt}$.
The \defining{boundary} of a $V$-component $\cmpnt$ is the set of variables $\boundary[V]{\cmpnt} := \vars{\cmpnt} \cap V$.
We omit the subscript $V$ if $V$ is clear from the context.
We also use these notions for sets of equations $E \subseteq S$, where we implicitly consider the variable set $\vars{E}$.
We also allow $E$ to contain equations not in $S$, in which case we mean the $(\vars{E} \cap \vars{S})$-component of $S$.
In the case that $V=\emptyset$, we just say component instead of $V$-component.

An $\FF_2$-system $S$ is a \defining{Tseitin-system} if it only has one component and each variable $v\in \vars{S}$ occurs in at most two equations in $S$.
Variables, that are used in only one equation, are \defining{dangling}.
The set of dangling variables of $S$ is denoted $\dangling{S}$.
The \defining{parity} of $S$ is the sum of all ``right-hand sides'' of the equations in $S$ modulo two (so of all $a$ for each equation $x_1+\dots+x_\ell = a$ in $S$). If the parity is zero we also say that $S$ has \defining{even parity} and similarly if the parity is one we say that $S$ has \defining{odd parity}.
Usually, a Tseitin-system (without dangling variables) 
 is obtained from a connected base graph $G$, the so called base graph, and a map $f \colon V(G) \to \FF_2$~\cite{Tseitin68}.
The edges of $G$ become variables and each vertex $u \in V(G)$ is turned into an equation on its incident edges $E(G,u)$, where $f$ specifies the right hand side:
$\sum_{e\in E(G,u)} e = f(u)$.
However, we do not need this translation from graphs to Tseitin-systems formally. 
We allow dangling variables to consider Tseitin $V$-components of non-Tseitin systems.

The following lemma records the key properties of Tseitin-systems.

\begin{lemma}
	\label{lem:sol-Tseitin-extend}
	Let $S$ be a Tseitin-system of parity $p$.
	\begin{enumerate}[a)]
		\item An assignment $\alpha\colon \dangling{S}\to \FF_2$ extends to a solution of $S$
		\label{itm:tseitin-boundary-parity}
		if and only if $\sum_{v \in \dangling{S}} \alpha(v) =  p$.
		\item In particular, if $W\subsetneq\dangling{S}$, then every assignment $\alpha \colon W \to \FF_2$ extends to a solution of $S$.%
		\label{itm:tseitin-boundary-unassigned}
		\item For every $e \in S$, every assignment $\alpha\colon \dangling{S}\to \FF_2$ extends to an assignment $\beta \colon \vars{S} \to \FF_2$ that satisfies all equations of $S$ apart from possibly $e$.\label{itm:tseitin-move-invonsistency}
		\item If $S$ is homogeneous, $V\subseteq \vars{S}$, $\cmpnt$ is a $V$-component of $S$ such that no dangling variables of $S$ are in $\vars{\cmpnt}$, and $\alpha \in \sols{S}$, then $\sum_{v \in \boundary[V]{\cmpnt}}  \alpha(v)$ is even.
		\label{itm:tseitin-even-boundary}
	\end{enumerate}
\end{lemma}
\begin{proof}
	It is well-known that Tseitin-systems without dangling variables are solvable if and only if their parity is zero~\cite{Tseitin68}.
	Since assigning a dangling variable essentially changes the right-hand side of the equation in which it occurs, Assertion~1 follows.
	
	Assertion~2 follows from Assertion~1 via assigning the non-assigned dangling variables to satisfy the parity condition of Assertion~1.
	
	To show Assertion~3, let $e\in S$ and $\alpha \colon \dangling{S} \to \FF_2$. Extend $e$ by a new dangling variable to yield equation $e'$ and the system $F'$ (all other equations remain unchanged).
	Then by Assertion~2, $\alpha$ extends to a solution of $F'$. Since we only modified $e$, this solution only possibly violates $e$ in $S$.
	
	To prove Assertion~4, let $S$ be homogeneous,  $V\subseteq \vars{S}$, $\cmpnt$ be a $V$-component of $S$, and $\alpha \in \sols{S}$.
	Then $\cmpnt$ is a Tseitin system of even parity whose dangling variables are
	exactly the ones in $\boundary[V]{\cmpnt}$ because no dangling variables of $S$
	are in $\vars{\cmpnt}$.
	Thus, $\sum_{v \in \boundary[V]{\cmpnt}}  \alpha(v)$ is even by Assertion~1.
\end{proof}

\subsection{Compressing Equation Systems}

\begin{definition}[Compression]
	A \defining{compression} of an $\FF_2$-system $S$ is an equivalence relation on $\vars{S}$.
\end{definition}

For a variable $x$, its $\compEquiv$-equivalence class is denoted $x/_\compEquiv$.
The compression is lifted to equations $e \colon x_1 + \dots + x_\ell = a$ in $S$
via $e/_\equiv \colon x_1/_\equiv + \dots + x_\ell/_\equiv = a$.
The notation is extended to sets $V$ of variables or $E$ of equations:
$V/_\compEquiv := \setcond{x/_\compEquiv}{x\in V}$ and similarly for $E$.
We sometimes identify $V/_\compEquiv$ or $E/_\compEquiv$ with the set of variables or equations in the classes in $V/_\compEquiv$ or $E/_\compEquiv$, respectively. That is, we identify $V/_\compEquiv$ with $\bigcup_{x\in V} x/_\compEquiv$. We will only do this when this does not cause any ambiguity. 
Let $E \subseteq S$ and suppose $\alpha\in\sols{E}$ is such that $\alpha(y) = \alpha(z)$ for all variables $y,z \in \vars{E}$ with $y \compEquiv z$. Then 
we write $\alpha/_\compEquiv$ for the unique assignment $\vars{E/_\compEquiv} \to \FF_2$ such that $\alpha/_\compEquiv(\eclass{y}) = \alpha(y)$ for all $y \in \vars{E}$. Otherwise, $\alpha/_\compEquiv$ is undefined.

\begin{definition}[Compressed $\FF_2$-System]
	Let $S$ be an $\mathbb{F}_2$-system and $\equiv$ be a compression of $S$. Then $S/_\equiv := \setcond{e/_\equiv}{e \in S}$ is the \defining{compressed system}.
\end{definition}

\subsection{Crossed Grids and Their Compression}

We now introduce the concrete systems we will study. Afterwards we will compress them.

\begin{figure}
		\centering
	\begin{tikzpicture}[scale = 0.50]
		\def\w{3}
		\def\h{4}
		\def\c{6}
		\tikzstyle{eq} = [fill=black,circle, inner sep = 2pt, minimum width = 2.2mm]
		\tikzstyle{unary} = [draw=black,circle, font=\tiny, inner sep = 0pt, minimum width = 2.2mm]
		\tikzstyle{invisible} = [inner sep=0pt]
		\tikzstyle{var} = [black, thick]

		\definecolor{colA}{RGB}{30,120,255}
		\definecolor{colB}{RGB}{60,200,230}
		\definecolor{colC}{RGB}{60,180,75}
		\definecolor{colD}{RGB}{140,140,140}
		\definecolor{colE}{RGB}{245,130,48}
		\definecolor{colF}{RGB}{145,30,180}
		\definecolor{colG}{RGB}{240,50,230}%
		\definecolor{colH}{RGB}{170,110,40}
		
		\foreach \k in {1,...,\w} {
			\foreach \j in {0,1} {
				\node[unary] (0\c\k\j) at ($(\k+ \j*\w + 1*\j, \c -1)$) {0};
			}
		}

		\foreach \b in {1,...,\h}{
			\draw [decorate,decoration={brace,amplitude=5pt,raise=4ex}]
			(7,\b*\c+\c-0.3) -- node[midway,xshift=1.5cm]{block $\b$} (7,\b*\c-0.2) ;
			
			\ifnumequal{\b}{\h}{
				\pgfmathtruncatemacro{\cmax} {int(\c+1)}
			}{
				\xdef\cmax{\c}
			}
			
			\def\highlightedVEdges{3/1/0/colA, 2/1/1/colD, 2/2/1/colD, 2/3/1/colD, 1/1/0/colE, 1/1/1/colE, 1/2/0/colE, 1/2/1/colE, 1/3/0/colE, 1/3/1/colE}
			
			\foreach \n in {1,...,\cmax}{
				\foreach \k in {1,...,\w} {
					\foreach \j in {0,1} {
						
						\pgfmathtruncatemacro{\bminus} {int(\b-1)}
						\pgfmathtruncatemacro{\nminus} {int(\n-1)}
						\pgfmathtruncatemacro{\kminus}{int(\k-1)}
						\pgfmathtruncatemacro{\otherj}{int(1-\j)}

						\ifthenelse {  \equal{\b\n\j}{\h\cmax0} } {
							\def\v{0}
							\ifthenelse{ \equal{\b\n\k}{\h\cmax1}}{
								\def\v{1}
							}{}
							\node[unary] (\b\n\k\j) at ($(\k+ \j*\w + 1*\j, \b*\c + \n -1)$) {\v};
						} { \ifthenelse {  \equal{\b\n\j}{\h\cmax1} } {
								\node[invisible] (\b\n\k\j) at ($(\k+ \j*\w + 1*\j, \b*\c + \n -1)$) {};
							}{
								\node[eq]  (\b\n\k\j) at ($(\k+ \j*\w + 1*\j, \b*\c + \n -1)$) {};
						}}

						\ifnumequal{\n}{1}{
							\def\styleA{black}
							\def\styleB{black}
							\foreach \bh/\kh/\jh/\c in \highlightedVEdges{
								\ifthenelse{\bminus=\bh \AND \j = \jh \AND \k = \kh}{
									\xdef\styleA{\c}
								}{}
								\ifthenelse{\bminus=\bh \AND \otherj = \jh \AND \k = \kh}{
									\xdef\styleB{\c}
								}{}
							}
							\draw[var, \styleA] (\b\n\k\j) edge (\bminus\c\k\j);
							\draw[var, \styleB] (\b\n\k\j) edge ($(\bminus\c\k\otherj)+(0,0.3)$);
						}{
							\draw (\b\n\k\j) edge (\b\nminus\k\j);
						}	
						\ifnumequal{\k}{1}{
						}{
							\ifthenelse {  \equal{\b\n}{\h\cmax} } {
							} {
								\draw[var] (\b\n\k\j) edge (\b\n\kminus\j);
							}
						}

					}
				}
			}
		}
		
		\node[below=5mm of 0620] {side $1$};
		\node[below=5mm of 0621] {side $2$};
		
		\node[eq, colF] (eqhigh) at (4310) {};
		\node[right, colF, left=1mm of eqhigh] {$e^1_{4,3,1}$};
		\node[right, colA, left=1mm of 3610, yshift=3mm] {$x^1_{3,6,1}$};
		
		\def\internalSep{3410,3520,3430}
		\foreach \u in \internalSep{
			\node[eq, colC] (eqhigh) at (\u) {};
		}
		\node[right, colC, left=1mm of 3410, yshift=3mm, text width=2.5cm, align=right] {block-$3$ internal separator};
		
		\def\crossSep{3110,2621,3131}
		\foreach \u in \crossSep{
			\node[eq, colD] (eqhigh) at (\u) {};
		}
		\node[right, colD, left=1mm of 2610, yshift=3mm, text width=2.8cm, align=right] {block-$2$ internal separator};
		
		\def\dcrossSep{2110,2120,2131}
		\foreach \u in \dcrossSep{
			\node[eq, colE] (eqhigh) at (\u) {};
		}
		\node[right, colE, left=1mm of 1610, yshift=3mm, text width=2.8cm, align=right] {block-$1$ double-crossing separator};
		
		\node[right, left=1mm of 1110] {row $(1,1)$};
		\node[right, left=1mm of 1210] {row $(1,2)$};
		\node[right, left=1mm of 1310] {row $(1,3)$};
		
		\node[right, left=1mm of 4610] {row $(4,6)$};
		\node[right, left=1mm of 4510] {row $(4,5)$};
		\node[right, left=1mm of 4410] {row $(4,4)$};
	\end{tikzpicture}
	
	\caption{The crossed grid $\cwgrid{w}{h}$ for $w=3$ and $h=4$. For size reasons, we set $c = 2w$.
	Equations are drawn as nodes and variables as edges.
	An equation sums over the incident edges.
	Note that at the crossings edges can have 3 endpoints.
	For unary equations, the sum of the variables is given as a
	number, for all others the sum is always 0.
	Horizontal variables in  column $w$, which join equations in column $w$ and $1$, are for clarity not drawn.
	}
	\label{fig:crossed-grid}
\end{figure}
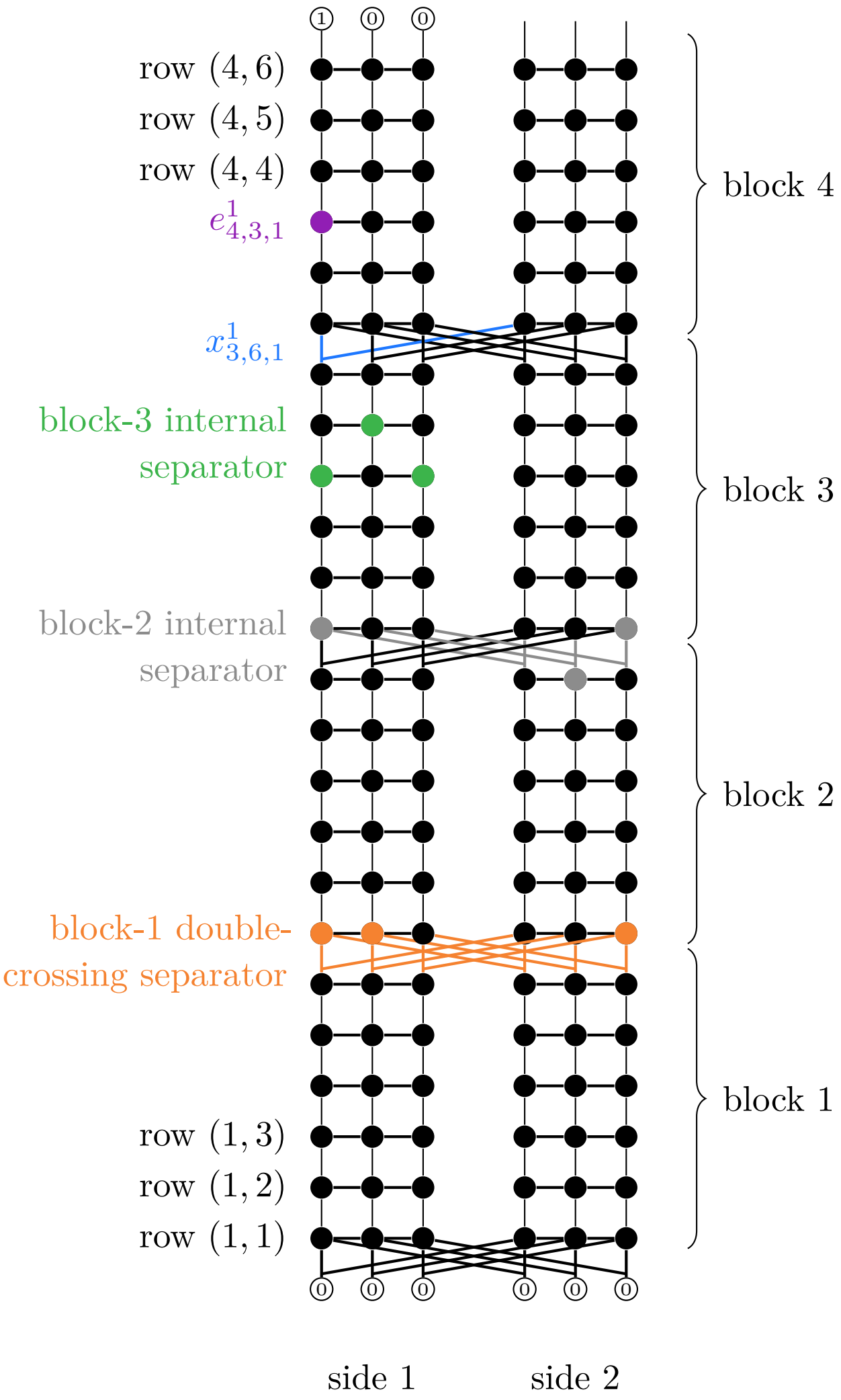

 Let $h, w \ge 3$ be integers and define $c:=3w$. A \defining{crossed grid of width $w$ and height $h$} is the following $\FF_2$-system $\cwgrid{w}{h}$.
For every $j\in[2], b \in [h], n \in [c]$, and $k \in [w]$, there are variables $x^j_{b,n,k}$ and $y^j_{b,n,k}$.
We also add the $x^j_{b,n,k}$ variables for $b=0$, $n=c$, and all $j\in[2]$ and  $k\in [w]$.
For every $j\in[2], b \in [h], n \in [c]$, and $k \in [w]$,
we have the equation
\begin{align*}
	e^j_{b,n,k}\colon
	\begin{cases}
	x^j_{b,n,k} = x^j_{b,n-1,k} + y^j_{b,n,k} + y^j_{b,n,k+1} & \text{if } n >1, \\
	x^j_{b,1,k} = x^1_{b-1,c,k} + x^2_{b-1,c,k} + y^j_{b,n,k} + y^j_{b,n,k+1} & \text{if } n = 1.
	\end{cases}  
\end{align*}
We use addition on the last index in a modulo-like manner, for example,
$y^j_{b,n,w+1}$ is $y^j_{b,n,1}$.  Finally, for 
$k \in [w]$ we have the equations
\begin{align*}
f_{k}  &\colon \begin{cases}
x^1_{h,c,k} = 1 & \text{if }  k =1, \\
x^1_{h,c,k} = 0 & \textrm{otherwise,}
\end{cases} 
\end{align*}
and for $j \in[2]$ and $k\in[w]$ we have the equations
\begin{align*}
	f^j_k & \colon x^j_{0,c,k} = 0.
\end{align*}
This completes the description of $\cwgrid{w}{h}$, see Figure~\ref{fig:crossed-grid} for a graphical representation.
Intuitively, this system consists of two $w \times ch$ grids (more precisely, Tseitin-systems obtained from $w \times ch$ grid-graphs)
connected by additional `crossings' every $c$ rows.  
Note that because of these crossings $\cwgrid{w}{h}$ is not a Tseitin-system since
the variables $x^j_{b,c,k}$ are used in three equations $e^j_{b,c,k}$, $e^1_{b+1,1,k}$, and $e^2_{b+1,1,k}$.

We say that variables $x^j_{b,n,k}$ and $y^j_{b,n,k}$, as well as equations $e^j_{b,n,k}$, are on side $j$, in row block $b$, in row $n$, and in column $k$.
The equations $f^k$ belong to block $(1,h)$ and the $f^j_k$ equations to block $(j,1)$.
\defining{Block $b$} is the subsystem $\cwgrid{w}{h}[b]$ induced by all equations in block $b$ and \defining{block $(j,b)$} consists of the set of equations in block $b$ on side $j$, the corresponding subsystem is $\cwgrid{w}{h}[(j,b)]$.
Similarly, \defining{row $(b,n)$}  is the subsystem induced by all equations in block $b$ and row $n$ and  \defining{row $(j,b,n)$} is the set of equations in row $(b,n)$ on side $j$.
For $b \leq b'$, the subsystem induced by all blocks $b$ to $b'$ is denoted $\cwgrid{w}{h}[b,b']$.
For $n,n' \in [1,c]$, the subsystem induced by all rows from row $(b,n)$ up to row $(b',n')$ is denoted  $\cwgrid{w}{h}[(b,n),(b',n')]$ (if $b = b'$ we require that $n \le n'$). Moreover, for $j \in [2]$ we define $\cwgrid{w}{h}[(j,b,n),(j,b',n')]$ to be those equations in $\cwgrid{w}{h}[(b,n),(b',n')]$ on side $j$.
Note that $\cwgrid{w}{h}[b,b'] = \cwgrid{w}{h}[(b,1),(b',1)]$.

We call row $c$ the top of a block $b$ and row $1$ its bottom.
The bottom row of each block consists of \defining{crossed equations}.
Crossed equations connect both sides of the crossed grids every $c$ rows.
Block $h$ is the \defining{top block} and block $1$ the \defining{bottom block}.
Reflecting the intuitive geometry of our grids we call variables of the form $x^j_{b,n,k}$ \defining{vertical variables} and variables of the form $y^j_{b,n,k}$ \defining{horizontal variables}.

We will now `compress' these systems similarly  to in~\cite{DBLP:journals/jacm/GroheLNS25, RezendeFJN025}. 
In order to define our equivalence relation on the variables of $\cwgrid{w}{h}$ (for suitable $h$) we define the following parameters. Let $w \geq 3$, recall that $c=3w$, and let $t$ be a sufficiently large integer. Let $p_1, \dots, p_{w}$
be pairwise coprime integers such that 
\[
\frac{t}{2} < p_i < t
\]
for all $i \in [w]$. Such numbers exist for every sufficiently large $t$~\cite{DBLP:journals/jacm/GroheLNS25}. We set $h:= p_1 \cdot \ldots \cdot p_{w}$.

We now use these parameters to define our compression $\compEquiv$ of $\cwgrid{w}{h}$. It implicitly depends on $w$ and $h$, which will always be clear from the context.
Firstly, each variable indexed by the top or bottom block is only $\compEquiv$-equivalent to itself.
Outside of that, each $x^j_{b,n,k}$ or $y^j_{b,n,k}$ variable is only identified with variables of the form $x^j_{b', n,k}$ or $y^j_{b',n,k}$, respectively.
That is, horizontal or vertical variables are only identified with horizontal or vertical variables, respectively, on the same side, in the same column, in the same row of a block. 

We start with an auxiliary relation that identifies variables in all blocks. It will be relevant later. Define for all $j\in [2]$, $b,b'\in[h]$, $n \in [c]$, and $k\in[w]$, as well as for  $b = 0$, $n=c$ and $k \in [w]$,
\begin{align*}
x^j_{b,n,k} \equiv^+ x^j_{b',n,k} & \iff (b' - b) \textit{ is divisible by } p_k \cdot p_{k+1}\\
y^j_{b,n,k} \equiv^+ y^j_{b',n,k} &\iff (b' - b) \textit{ is divisible by } p_{k}.
\intertext{
The actual compression is then obtained by making the top and bottom block singleton classes:}
x^j_{b,n,k} \equiv x^j_{b',n,k} & \iff b=b' \text{ or } \bigl(x^j_{b,n,k} \equiv^+ x^j_{b',n,k} \text{ and } b,b' \in [2,h-1]\bigr)\\
y^j_{b,n,k} \equiv y^j_{b',n,k} &\iff b=b' \text{ or } \bigl(y^j_{b,n,k} \equiv^+ y^j_{b',n,k} \text{ and } b,b' \in [2,h-1]\bigr).
\end{align*}
This is -- apart from the two sides with crossings and the explicit partition into blocks -- the same compression as in~\cite{DBLP:journals/jacm/GroheLNS25}, but formulated on an equation system level (and not on the level of the base graph of a CFI graph or Tseitin system as in~\cite{DBLP:journals/jacm/GroheLNS25, RezendeFJN025}).
This completes the description of compression and thus of $\comcwgrid{w}{h}$.
Note that a horizontal variable $y^j_{b,n,k+1}$ is used by equations in columns $k$ and $k+1$ and that $p_{k+1}$ is the largest common divisor of the periods used to compress vertical variables in columns $k$ and $k+1$ (more precisely, the period of their blocks).

\begin{lemma}
	\label{lem:compressed-crossed-size}
	$|\comcwgrid{w}{h}| = |\homsys{(\comcwgrid{w}{h})}| \in \Theta(w^2 t^2)$.
	Their maximal arity is $5$.
\end{lemma}
\begin{proof}
	The arity bound follows from the definition of $\cwgrid{w}{h}$ because compressions do not change the arity.
	We need to count $|\comcwgrid{w}{h}|$.
	The equations in the top and bottom block are in singleton equivalence classes. There are at most $2 (w (c+1)) =  \Theta(w^2)$ many.
	Now consider the equations $e^j_{b,n,k} \colon x^j_{b,n,k} = x^j_{b,n-1,k}  + y^j_{b,n,k} + y^j_{b,n,k+1}$ and 
	$e^{j}_{b',n,k} \colon  x^j_{b',n,k} = x^j_{b',n-1,k} +  y^j_{b',n,k} + y^j_{b',n,k+1}$ for $n>1$.
	They are in the same equivalence class if and only if $x^j_{b,n,k} \compEquiv x^j_{b',n,k}$, $x^j_{b,n-1,k} \compEquiv x^j_{b',n-1,k}$,
	 $y^j_{b,n,k} \compEquiv y^j_{b',n,k}$, and $y^j_{b,n,k+1}  \compEquiv y^j_{b',n,k+1}$.
	 For the vertical variables, this happens if and only if $b'-b$ is divisible by $p_kp_{k+1}$.
	 This implies that also the horizontal variables are equivalent, because for them $b'-b$ needs to be divisible by $p_k$ and $p_{k+1}$, respectively.
	 The same argument applies to equations in row $1$.
	 Since $cp_kp_{k+1} \in \Theta(w t^2)$,
	 there are $\Theta(w^2 t^2)$ non-singleton equivalence classes (since sides and columns are compressed independently).
	 In total, there are $\Theta(w^2) + \Theta(w^2 t^2) = \Theta(w^2 t^2)$ equations in $\comcwgrid{w}{h}$ (and $\homsys{(\comcwgrid{w}{h})}$).
\end{proof}

\subsection{Separators in Crossed Grids}

For Tseitin-systems, separators of the underlying graph are important to isolate and then eventually ``catch'' an inconsistency. Crossed grids require an additional notion of a separator, which intuitively are those sets which fix the parity of assignments to variables in rows $(j,n)$ for $n \in \set{1,c}$. The next two lemmas highlight the importance of such parities

\begin{lemma} \label{lem:block_sol_preserves_parity}
	Let $b \in [h]$, $j \in [2]$, and $\beta \in \sols{\cwgrid{w}{h}[(j,b)]}$.
	Then  $\sum_{k\in[w]} \beta(x^{j}_{b,c,k})= \sum_{k\in[w]} \beta(x^{j}_{b,1,k})$. 
\end{lemma}
\begin{proof}
	Consider der Tseitin-system $\cwgrid{w}{h}[(j,b,2),(j,b,c)]$ and remove all its unary equations.
	It is homogeneous and its dangling variables
	are $\setcond{x^{j}_{b,n,k}}{n \in \set{1,c}, k \in [w]}$.
	So by Lemma~\ref{lem:sol-Tseitin-extend}, their sum under each solution has to be zero.
	So we obtain $\sum_{k\in[w]} \beta(x^{j}_{b,c,k})= \sum_{k\in[w]} \beta(x^{j}_{b,1,k})$, as required. 
\end{proof}

\begin{lemma}
	\label{lem:parity-crossing}
	Let $b \in [1,h-1]$, $j \in [2]$, and $\beta \in \sols{\cwgrid{w}{h}[(b,c),(b+1,1)]}$.
	Then $\sum_{k\in[w]} \beta(x^j_{b+1,1,k}) = \sum_{k\in[w], i\in[2]} \beta(x^{i}_{b,c,k})$.
\end{lemma}
\begin{proof}
	The result follows by summing up the equations in row $(j,b+1,1)$.
\end{proof}

It may be that assigning some set of equations determines the parities of the top/bottom row of a block or of multiple blocks. We will call sets of equations for which this occurs separators. Because we have blocks, which are locally Tseitin systems, and then crossings between blocks we have two types of separators; we now give the formal definitions. See Figure~\ref{fig:crossed-grid} for some visual examples.

\subparagraph{Double-Crossing Separators.}

 For $b>1$ we say that a set of equations $E\subseteq \cgrid{}$ is \defining{block-$b$ double-crossing} if all vertical variables in row $(b-1,c)$ are in $\vars{E}$; in symbols:
 $\setcond{x^j_{b-1,c,k}}{j \in [2], k \in [w]} \subseteq \vars{E}$. Observe, that if $|E| = w$ this occurs if and only if,
for each $k\in [w]$, exactly one of $e^1_{b,1,k}$ and $e^2_{b,1,k}$ is in $E$. Intuitively, such a separator transfers parity from block $b$ to block $b-1$. Formally, let $E$ be a block-$b$ double-crossing separator and $\alpha \in \sols{E}$.
For $j\in[2]$,
the \defining{side-$j$ parity} of $(E,\alpha)$ is
$\sum_{k \in [w]} \alpha(x^{j}_{b-1,c,k})$.
The \defining{parity of $(E,\alpha)$ is odd} if, for some $j\in [2]$, the side-$j$ parity of $(E,\alpha)$ is odd.
Otherwise, the parity of $(E, \alpha)$ is \defining{even}.

\subparagraph{Block Internal Separators.}
All other separators will be called internal separators. Formally, let
$E \subseteq \cgrid{}$.
For $j\in[2]$, $b\in[h]$, and $n\in[c]$,
a variable $v \in \vars{\cgrid{}[(j,b)]}$ in a block $(j,b)$ equation (which is either a variable on block $(j,b)$ or a vertical variable in row $(b-1,c)$)
is \defining{$E$-separated} from row $(j,b,n)$ if the $E$-component of $v$ in $\cgrid{}[(j,b)]$
does not contain any vertical variable in row $(j,b,n)$ outside of $\vars{E}$. 
We just say that $v$ is $E$-separated from row $n$ if $b$ and $j$ are clear from the context.
For $b \in [2,h]$, a set $E\subseteq \cgrid{}$ is a \defining{block-$b$ internal separator}, if $E$ is not a double-crossing separator and for every $k \in [w]$,
there is a $j \in [2]$ such that $x^{j'}_{b-1,c,k}$ is $E$-separated from row $(j,b,c)$ for both $j' \in [2]$.
For $b=1$, a set $E\subseteq \cgrid{}$ is a block-$1$ internal  separator if it is not a double-crossing separator and
if for some side $j$, the variable $x^j_{1,1,k}$ is $E$-separated from row $(j,1,c)$ for every $k\in [w]$. 

Let $E$ be a block-$b$ internal separator of size $w$, with $b>1$.
 Then $E$ contains one equation per column each of which must contain a variable of block $b$. This means that every equation in $E$ must occur either in block $b$ or in the bottom row of block $b+1$. 
Now let $j_k \in [2]$ such that the
$x^{j'}_{b-1,c,k}$ are $E$-separated from row $(j_k,b,c)$.
Let $\cmpnt^{j',k}$ be the $E$-component of the $x^{j'}_{b-1,c,k}$ in $\cgrid{}[(j_k,b)]$.
Set $K := \setcond{\cmpnt^{j',k}}{j' \in [2], k \in [w]}$. Let $\alpha \in \sols{E}$.
The \defining{parity of $(E,\alpha)$} is defined to be $\sum_{\cmpnt \in K} \sum_{v \in \boundary[E]{\cmpnt}} \alpha(v)$. 
Observe, that if neither $e^{1}_{b,1,k}$ nor $e^{2}_{b,1,k}$ is in $E$, then $\cmpnt^{1,k} = \cmpnt^{2,k}$, in which case we denote this component by $\cmpnt^k$. The parity of block-$1$ internal separators is defined analogously except that we now consider the $E$-components of $\cgrid{}[j,1]$ containing an equation of row $(j,1,1)$.

\begin{lemma}
	The parity of an internal separator is well-defined.
\end{lemma}
\begin{proof}
	We show that the components in $K$ used to define the parity do not depend on the chosen sides $j_k$.
	Consider $b>1$, the remaining case is similar. Recall that $E$ contains exactly one equation per column.
	If $x^1_{b-1,c,k},x^2_{b-1,c,k}\in\vars{E}$, their $E$-components in $K$ are the singletons
	$\set{x^1_{b-1,c,k}}$ and $\set{x^2_{b-1,c,k}}$, independently of $j_k$.

	Otherwise, neither bottom variable in column $k$ is assigned, which means that $e^1_{b,1,k}, e^2_{b,1,k} \notin E$. Here we show that $j_k$ is unique.
	Towards a contradiction,
	suppose that both choices of $j_k$ satisfy the separation condition.
	If the unique equation of $E$ in column $k$ lies in block $b$ on side $j$,
	it assigns no vertical variable of block $b$ on side $3-j$.
	Thus the vertical path in column $k$ on side $3-j$ connects the bottom variables to an unassigned top-row variable, a contradiction.
	Thus the equation in column $k$ lies in row $(b+1,1)$; in particular, $b<h$.

	Since $c=3w$ and at most $w-1$ equations of $E$ lie in block $b$,
	there is a row $n\in[2,c-1]$ containing no equation of $E$ on either side.
	Hence, any equation of $E$ in block $b$ on side $j$ would leave a path to an unassigned top-row variable on side $3-j$.
	Hence all equations of $E$ lie in row $(b+1,1)$, making $E$ a block-$(b+1)$ double-crossing separator, contradicting that $E$ is an internal separator.
	Consequently, $j_k$ is unique.
\end{proof}

\begin{example}
Let $b \in [2,h]$, $j \in [2]$ and $E:= \setcond{e^j_{b,w, k}}{k\in[w]}$. Then $E$ is a block-$b$ internal separator of size $w$, since for $j' \in [2]$, $k\in[w]$ we have that $x^{j'}_{b-1,c,k}$ is $E$-separated from row $(j,b,c)$. Then all of these $x^{j'}_{b-1,c,k}$ variables are in the same $E$-component of $\ucgrid[(j,b)]$, namely the component $\cmpnt$ consisting of the bottom $w-1$ rows of $\ucgrid[(j,b)]$. In the language used above $\cmpnt^k = \cmpnt$ for all $k \in [w]$. Then $\boundary[E]{\cmpnt}$ is the set of vertical variables in row $w - 1$ of block $b$. Therefore, for $\alpha \in \sols{E}$ we have that the parity of $(E, \alpha)$ is equal to $\sum_{k \in [w]} \alpha(x^j_{b, w-1,k})$.
\end{example}

A block-$b$ internal separator $E$ of size $w$ can contain equations from both sides of block $b$,
to separate small components at the bottom row of $b$.
Higher up internal separators can only be on one side of block $b$. Moreover, separators `in the middle' of block $b$ cannot contain equations from the bottom row of block $b+1$, so they must only contain equations on one side.
The next lemma tells us that internal separators match our more informal description of a separator.

\begin{lemma}
	\label{lem:separator-determines-parity}
	If $(E,\alpha)$ is a block-$b$ internal separator of size $w$ and parity $p$ with $b>1$,
	then every extension $\beta$ of $\alpha$ to a solution of $\cgrid{}[b]$
	satisfies
	$\sum_{j \in [2], k \in [w]} \beta(x^j_{b-1,c,k}) = p$.
\end{lemma}
\begin{proof}
	Let $\beta$ be such an extension and let $K$ be defined as above.
	Because $\cgrid{}[(j,b)]$ is a Tseitin-system, so is every non-trivial $E$-component in $K$.
	Note that each nontrivial $\cmpnt \in K$ satisfies $\boundary[E]{\cmpnt} \subseteq \dangling{\cmpnt}$,
	all boundary variables in $\boundary[E]{\cmpnt}$ 
	are not in row $(b-1,c)$,
	and all other dangling variables are in row $(b-1,c)$.
	Noting that $\kappa$ is a homogeneous system,
	Lemma~\ref{lem:sol-Tseitin-extend}\ref{itm:tseitin-boundary-parity} implies that $\sum_{v \in \dangling{\cmpnt}}  \beta(v) = 0$,
	and so $\sum_{v \in \vars{\cmpnt}, v \text{ in row} (b-1,c)}  \beta(v) = \sum_{v \in \boundary[E]{\cmpnt}} \alpha(v)$.
	
	For every trivial $\cmpnt = \set{w} \in K$, 
	we have that $w= x^{j}_{b-1,c,k}$ for some $j,k$ and 
	we obviously have $\sum_{v \in \vars{\cmpnt}, v \text{ in row} (b-1,c)} \beta(v) = \beta(w) = \alpha(w) =\sum_{v \in \boundary[E]{\cmpnt}} \alpha(v)$ since $w \in \vars{E}$.
	
	Now note that for every $j,k$, there is exactly one component $\kappa \in K$ 
	such that $x^{j}_{b-1,c,k} \in \vars{\kappa}$.
	Hence we can sum the equations above:
	\begin{align*}
		\sum_{j \in [2], k \in [w]} \beta(x^j_{b-1,c,k}) = 
		\sum_{\kappa \in K} \sum_{
			\substack{
			v \in \vars{\cmpnt}, 
			\\ v \text { in row } (b-1,c)
		}}  
		\beta(v) = \sum_{\kappa \in K} \sum_{v \in \boundary[E]{\cmpnt}} \alpha(v) = p,
	\end{align*}
	where the last step is by the definition of the parity of $(E,\alpha)$.
\end{proof}

A set $E \subseteq \ucgrid{}$ is a \defining{block-$b$ separator}
if $E$ is a block-$b$ internal separator, a block-$b$ double-crossing separator, or a block-$(b+1)$ double-crossing separator. We add this last clause for convenience in some later definitions.
Following our convention,
we also apply these definitions to sets of equations of $\comcwgrid{w}{h}$, by identifying such sets of classes by their union, yielding
subsets of equations in $\cwgrid{w}{h}$.
We make the easy observation, that separators of size $w$ equations span at most $w$ consecutive rows.
\begin{lemma}
	\label{lem:separator-span}
	Let $E \subseteq \ucgrid{}$ of size $w$ be a separator.
	If $E$ is a block-$b$ internal separator,
	then $E$ contains equations from at most $w$ consecutive rows, in block  $b$ or in row $(b+1,1)$.
	If $E$ is a block-$b$ double-crossing separator,
	then $E$ contains one equation per column
	and only in row $(b,1)$.
\end{lemma}
\begin{proof}
	If $E$ is a block-$b$ internal separator that contains equations from more than $w$ consecutive rows in $b$,
	then there is a row $n\in [c]$ such that $E$ contains equations above and below row $n$, but not in row $n$ (since $c = 3w$).
	Thus, for both $j\in[2]$, one equation in row $(j,b,1)$ is in the $E$-component of block $(j,b)$ that contains row $(j,b,n)$ since there are at most $w-1$ equations in between.
	The same reasoning applies to row $(j,b,c)$ and row $(j,n)$.
	Therefore, not all variables in row $(b-1,c)$ are separated from row $c$ in block $(j,b)$ for both $j$.
	This contradicts that $E$ is an internal separator.
	
	If $E$ is a block-$b$ double-crossing separator, then for each $k \in [w]$ the variables $x^1_{b-1,c,k}$ and $x^2_{b-1,c,k}$ are in $\vars{E}$. Therefore, there must be some equation in column $k$. Since, $|E| = w$, there is a unique equation in column $k$ which implies that either $e^1_{b,1,k}$ or $e^2_{b,1,k}$ is in $E$. 
\end{proof}

\subparagraph*{Inconsistency of the System.}
Using these separators, we show that $\cwgrid{w}{h}$ is inconsistent.
More strongly, we show that Falsifier wins the $(w+1)$-equation game on $\cwgrid{w}{h}$.

\begin{lemma}
	\label{lem:crossed-system-Falsifier-wins-equation-game}
	For every $h,w\geq 3$,
	Falsifier wins the $(w+1)$-equation game on $\cwgrid{w}{h}$.
\end{lemma}
\begin{proof}
	Falsifier starts by choosing $E$ to be the equations in row $(1,h,c)$.
	If the obtained position $(E,\alpha)$ is not an odd separator,
	then one of the  unary equations  $f_k$ is vialoated.
	Falsifier chooses it and wins.
	Otherwise, the sum $\sum_{k\in [w]} \alpha(x^1_{h,c,k})$ is one.
	 By summing up the equations in row $(1,h,c)$,
	 also $\sum_{k\in [w]} \alpha(x^1_{h,c-1,k})$ is one.
	Using the $(w+1)$-th equation, Falsifier step by step replaces equation $e^1_{h,c,k}$ by $e^1_{h,c-1,k}$.
	This fixes the assignments to the vertical variables in row $(1,h,c-1)$.
	Summing up the equations in row $(1,h,c-1)$ shows again that the resulting position is again an odd separator.
	In this way, Falsifier moves the separator row by row down maintaining oddness
	until it reaches row $(1,h,1)$.
	The reached position $(E',\alpha')$ is now an odd block-$h$ double-crossing separator. Therefore,
	$\sum_{k\in [w]} \alpha'(x^j_{h-1,c,k})$ is odd for one $j\in [2]$.
	Falsifier step-by-step replaces $e^1_{h,1,k}$ with $e^j_{h-1,c, k}$ using the $(w+1)$-th pebble. This fixes the odd assignment to the vertical variables in row $(h-1,c)$.
	The resulting position is an odd block-$(h-1)$ internal separator.
	Now Falsifier repeats the strategy from block $(1,h)$ for block $(j,h-1)$.
	In this way, Falsifier moves an odd separator down to some block $(j',1)$.
	When all equations are in row $(j',1,1)$, the assignment to the 
	vertical variables in row $(j',0,c)$ is odd and thus violates one unary equation.
	Falsifier chooses this equation an wins.
\end{proof}

\begin{corollary}
	\label{cor:compressed-crossed-system-Falsifier-wins-equation-game}
	For every $w \ge 3$, and every $h$ for which $\compEquiv$ is defined,
	Falsifier wins the $(w+1)$-equation game on the compressed system $\comcwgrid{w}{h}$.
\end{corollary}
\begin{proof}
	Let $h$ be such that $\comcwgrid{w}{h}$ is defined.
	Then Falsifier plays the same strategy as on the uncompressed system given by Lemma~\ref{lem:crossed-system-Falsifier-wins-equation-game}.
	Instead of an equation $e$, Falsifier chooses $e/_\compEquiv$.
	It is easy to see that Falsifier wins on $\comcwgrid{w}{h}$ when Falsifier
	wins on $\cwgrid{w}{h}$ (or possibly earlier).
\end{proof}

Note that at every double-crossing separator Verifier has to choose on which side the odd parity moves.
This choice is the core reason why Duplicator can win the width-$w$ size game with 
$2^{h-O(1)}$ size bound on $\cwgrid{w}{h}$.
The magic of the compression is that the same asymptotic bound holds for the much smaller system $\comcwgrid{w}{h}$. 

\section{The Lower Bound} \label{sec:lower-bound}

In this section, we prove our main result by showing that Duplicator can win the width-$w$ size game on $\comcwgrid{w}{h}$ from position $(\emptyset, \set{0}, 2^{h-O(1)})$. The key ideas are as follows.
Spoiler's optimal strategy is to  form an odd separator in the top block and then push it down row-by-row 
   (as in Falsifier's strategy from Lemma~\ref{lem:crossed-system-Falsifier-wins-equation-game}). Duplicator always plays regular rounds when there is an odd internal separator, unless it is close to a block border. When this occurs, Duplicator plays subtree rounds to remain flexible on which side to send the odd parity in the next block. To make progress, Spoiler must eventually choose a small subtree round (which recall halves the size bound). Since the odd separator has to pass through $h$ blocks, this means that the size bound reaches one (and Duplicator wins) before the odd separator reaches the bottom block. 
   Moreover, Spoiler cannot do any better even after we compress the crossed grid. We next develop the techniques to formalize this.

\subsection{Extending Local Solutions} \label{ss:extend-local-solution}

We first study two types of extensions of local solutions to larger ones.
We fix $w \ge 3$ and $h \ge 3$.
To simplify notation, we set $\ucgrid^1 := \cwgrid{w}{h}$ and $\ucgrid^0 := \homsys{(\cwgrid{w}{h})}$.
Note that $\homsys{(\ucgrid^1/_\compEquiv)} = \ucgrid^0/_{\compEquiv}$.
Now fix $\ucgrid \in \set{\ucgrid^0, \ucgrid^1}$,
so we uniformly consider the homogeneous and the inhomogeneous system.
The notions from the previous section, for example odd/even separators, also apply to the homogeneous case.
We encourage the reader to think about the inhomogeneous system; 
the same arguments also can be used on the homogeneous system (and are often easier).
Before introducing our extension notions we need some definitions.

\begin{definition}[Closure]
Let $E \subseteq \cgrid{}$, $b \in [h]$ and $j\in[2]$.
An $E$-component $\cmpnt$ of $\cgrid{}$ is \defining{$E$-large}
if $\cmpnt$ contains all vertical variables of the same row in some block $b$ on some side $j$.
We call all other $E$-components \defining{$E$-small}.
The \defining{closure} $\cl{E}$ is the union of $E$ and all $E$-small components.
A solution $\alpha \in \sols{E}$ is \defining{closure-consistent} if
$\alpha$ extends to a solution in $\sols{\clo{E}}$.
\end{definition}

\begin{definition}[Right Parity]
	We say that block $b$ is a 
	\defining{border block} if $b \in \set{1,h}$.
	We say that the parity of a border block $b$ is odd if $b=h$
	and $\cgrid{}=\cgrid{}^1$,
	otherwise the parity of block $b$ is even.
	Let $E\subseteq \cgrid{}$ be of size $w$ and $\alpha \in \sols{E}$.
	We say that $(E,\alpha)$ has the \defining{wrong parity} if
either 
\begin{enumerate}
	\item $E$ is a separator in a border block $b$, but not a block-$h$ double-crossing separator, and the parity of $(E,\alpha)$ is not the parity of $b$, or
	\item $E$ is a block-$h$ double-crossing separator and $\sum_{j \in [2], k \in [w]} \alpha(x^j_{h-1,c,k})$ is not the parity of block $h$.
\end{enumerate}
	If $(E,\alpha)$ does not have the wrong parity, then we say $(E,\alpha)$ has the \defining{right parity}.
	When $E$ is clear from the context, we also say that $\alpha$ has the right or wrong parity.
	If $E$ is not of size $w$,
	then $E$ has right parity if for every $F\subseteq E$ of size $w$,
	the pair $(F, \restrict{\alpha}{F})$ has the right parity.
\end{definition}

We also apply these definitions to $\comucgrid$ in the obvious way. This leads to the following notion.

\begin{definition} \label{def:almost_extendable} 
Let $E \subseteq \comucgrid$. 
 An \defining{almost-extension} of a solution $\alpha \in \sols{E}$ is an extension of $\alpha$ to a solution of $\comucgrid \setminus \set{e}$,
 where $e$ is an equation in block $(1,h)$ in an $E$-large component 
 that contains a unary equation in block $(1,h)$.
 We call $\alpha$ \defining{almost-extendable} if there is an almost-extension of~$\alpha$ and $(E,\alpha)$ is of the right parity. 
\end{definition}

Almost-extensions become relevant when not all of the $w$ equations in $E$ are from block $(1,h)$.
In this case, there is a unique $E$-large component in block $(1,h)$, which is well-defined for $\comucgrid$ since block $(1,h)$ is in singleton $\compEquiv$-classes.
In this component, the inconsistency can be moved to every equation in it using Lemma~\ref{lem:sol-Tseitin-extend}\ref{itm:tseitin-move-invonsistency}.
That means, to detect the inconsistency all $w$ equations have to be in block $(1,h)$.
We will now prove that various conditions imply almost-extendability.
Intuitively, if such a local solution is reached in the size game then Spoiler made a bad move and essentially has to restart the game in block $(1,h)$. 

We first focus on the case of an even separator or a non-separator. 
Odd separators are never almost-extendable:
\begin{lemma}
	\label{lem:odd-separator-not-almost-extendable}
	Let $E \subseteq \ucgrid$ be of size $w$ and let $\alpha \in \sols{E}$ such that $(E,\alpha)$ is an odd separator.
	Then $\alpha$ is not almost-extendable.
\end{lemma}
\begin{proof}
Suppose for a contradiction, that
$\beta$ is an almost-extension of $\alpha$. 
Firstly, if $E$ is a block-$1$ internal separator, 
we immediately get a contradiction by Lemma~\ref{lem:sol-Tseitin-extend}.

	If $E$ is a block-$(b+1)$ internal separator for $h-1>b\ge1$,
	then $\sum_{j\in[2],k\in[w]} \beta(x^j_{b,c,k})$ is odd by Lemma~\ref{lem:separator-determines-parity}. 
	This means that $\sum_{k\in[w]} \beta(x^j_{b,c,k})$
	is odd for one $j$ and even for the other. Similarly, if $E$ is a block-$(b+1)$ double-crossing separator then $\sum_{k\in[w]} \beta(x^j_{b,c,k})$
	is odd for at least one $j \in [2]$. 
	Otherwise,  $E$ is a block-$h$ internal separator.
	If it is on side $2$, the same reasoning applies.
	If it is on side $1$, then $\beta$ may not satisfy one equation
	in the unique $E$-large component containing a top unary equation.
	This means that this component is ``above'' the separator,
	so the argument from Lemma~\ref{lem:separator-determines-parity}
	applies to the part of block $(1,h)$ below the separator, and we again obtain
	that $\sum_{k\in[w]} \beta(x^j_{h-1,c,k})$ is odd for one $j$.

	On the other hand, we show that, since $\beta$ is an almost extension, for every $b' \in [1,h-1]$, $j \in [2]$ it holds that  $\sum_{k\in[w]} \beta(x^j_{b',c,k})$ is even. This clearly gives a contradiction; we show the claim by an easy induction.

	In detail, the base case $b' = 1$ follows from the unary equations and Lemma~\ref{lem:block_sol_preserves_parity}. For the induction step, we have that for some $ h-1 > b'\ge 1$, that $\sum_{k\in[w]} \beta(x^j_{b',c,k})$ is even for both $j \in [2]$. Therefore, by Lemma~\ref{lem:parity-crossing} we have that $\sum_{k \in [w]} \beta(x^j_{b'+1,1,k})$ is even for both $j \in [2]$. Therefore, by Lemma~\ref{lem:block_sol_preserves_parity} we have that $\sum_{k \in [w]} \beta(x^j_{b'+1,c,k})$ is even for both $j$.
\end{proof}

So if $\alpha \in \sols{E}$ has an almost extension and $(E,\alpha)$ is a separator, then it is an even separator. In particular, this occurs if $E$ is a separator in a border block. Therefore, if in this scenario $\alpha$ is also almost-extendable, then the parity of the border block must be even. Note that this is always the case in the bottom block and also in the homogenous system, but is not the case in the top block of the inhomogeneous system.

For the analysis the following auxiliary equivalence relations will be helpful.
For $\ell \in \nat$, we define the relation $\approx_\ell$ via
\begin{align*}
	x^j_{b,n,k} \approx_\ell x^j_{b',n,k} &\iff (b' - b) \textit{ is divisible by } \ell\\
	y^j_{b,n,k} \approx_\ell y^j_{b',n,k} &\iff (b' - b) \textit{ is divisible by } \ell
\end{align*}
for all $j \in[2]$, $b,b' \in [h]$, $n \in [c]$, and $k \in [w]$.
That is, all vertical (and horizontal) variables on the same side and in the same column and the same row inside its block are identified whose blocks are equal modulo~$\ell$.
Define for each $k \in [w]$: 
\begin{align*} 
	\sim_k &{}:={} \approx_{p_{k+1}}\\
	\sim   &{}:={} \approx_{1}.
\end{align*} 
Observe that $\sim$ is finer than $\sim_k$ and $\equiv$.
We use the same notation for equivalence classes as for~$\compEquiv$, e.g., $e/_\sim$ or $E/_{\sim_k}$.
We again may identify $E/_\sim$ with the set of equations in the classes in $E/_\sim$ when convenient.
When $E\subseteq{\comucgrid}$, we write $E/_\sim$ for $\setcond{e/_\sim}{e/_\compEquiv \in E}$, which is well-defined since $\sim$ is finer than $\equiv$. We care about the relation ${\sim_k} = {\approx_{p_{k+1}}}$, since $p_{k+1}$ is the common factor of the periods used to compress the vertical variables in columns $k$ and $k+1$.

It will often be convenient to work with assignments which respect some $\approx_{\ell}$. We formalize this in the following definition.

\begin{definition}[Periodic and Cyclic Solutions]

For a set $E \subseteq \cwgrid{w}{h}$, a solution $\alpha \in \sols{E}$ 
is \defining{$p$-periodic in column $k$}, if 
\[
	\alpha(x^j_{b,n,k}) = \alpha(x^j_{b',n,k}) \text{ for all } x^j_{b,n,k},x^j_{b',n,k} \in \vars{E} \text{ such that }
	(b' - b) \textit{ is divisible by } p,\]
\defining{$p$-periodic between column $k$ and $k+1$} if 
\[ \alpha(y^j_{b,n,k+1}) = \alpha(y^j_{b',n,k+1}) \text{ for all } y^j_{b,n,k+1},y^j_{b',n,k+1} \in \vars{E} \text{ such that } (b' - b) \textit{ is divisible by } p,\]
and \defining{$p$-periodic} if for all $k \in [w]$ the solution $\alpha$ is $p$-periodic in column $k$ and $p$-periodic between columns $k$ and $k+1$.

For $b<b' \in [h]$ and $n\in [c-1]$, a solution $\alpha \in \sols{\cgrid{}[(b,n+1),(b',n)]}$ is called \defining{cyclic} if $\alpha(x^j_{b,n,k}) = \alpha(x^j_{b',n,k})$ for all $k \in [w]$.
\end{definition}

The reason that cyclic $\ell$-periodic local solutions are nice is that if they are locally reasonable and can be extending to $\ell$-consecutive blocks, then they are almost-extendable. To do this we take the solution to the consecutive blocks, repeat it and then locally fix things in the top and bottom blocks.

\begin{lemma}
	\label{lem:periodic-solutions}
	Let $E\subseteq \ucgrid$ be a set of at most $w$ equations and $\alpha \in \sols{E}$ be closure-consistent such that $(E, \alpha)$ is of the right parity.
	If there are, for all $k \in [w]$,  positive integers $q_k$ and $r_k$ and, for $\ell := \mathsf{lcm}_{k \in [w]} \set{q_k, r_k}$, the assignment $\alpha$ is $\ell$-periodic and there are some $a\in[2,h-\ell-1]$ and $n\in[c-1]$ such that $\alpha/_{\approx_\ell}$ extends to a cyclic solution  $\beta \in \sols{\cgrid{}[(a,n+1),(a+\ell, n)]}$ such that
	\begin{enumerate}
		\item $\beta$ is $q_k$-periodic in column $k$,
		\item $\beta$ is $r_k$-periodic between columns $k$ and $k+1$,
		\item $q_k$ divides $p_k \cdot p_{k+1}$, and
		\item $r_k$ divides $p_{k+1}$,
	\end{enumerate}
	then $\alpha$ is almost-extendable.
\end{lemma}
\begin{proof}
	Since $(E, \alpha)$ is of the right parity, it suffices to provide an almost-extension of $\alpha$.
	Let $\beta$ be a cyclic solution of $\cgrid{}[(a,n+1),(a+\ell, n)]$ that extends $\alpha/_{\approx_\ell}$ with the required properties.
	We extend $\beta$ to all  variables by setting for all $j\in[2]$, $b\in[h]$, and $m \in [c]$  (as for $(b,m) = (0,c)$ for vertical variables)
	\begin{align*}
		\beta(x^j_{b,m,k}/_\compEquiv) &:= \beta(x^j_{b',m,k})\\
		\beta(y^j_{b,m,k}/_\compEquiv) &:= \beta(y^j_{b',m,k})
	\end{align*}
	for the unique $b'$  such that $\ell$ divides $b'-b$
	and $m \leq n$ implies $a< b' \leq a+ \ell$
	and $m > n$ implies $a\leq b' < a+ \ell$.
	That is, we copy the solution of $\cgrid{}[(a,n+1),(a+\ell, n)]$ in an $\ell$-periodic manner to all variables.
	This definition of $\beta$ is compatible with the compression by Items~3 and~4 that is, the definition of $\beta$ is independent of the representative of the class $x^j_{b,m,k}/_\compEquiv$.

	Because $\beta$ is a cyclic solution, the only possible equations violated by $\beta$ are the unary ones.
	We show how to modify $\beta$ to obtain an almost extension.
	Since $\beta$ is a cyclic solution on $\cgrid{}[(a,n+1),(a+\ell, n)]$,
	the parity on each side in each block is even by Lemma~\ref{lem:parity-crossing},
	which means that $\sum_{k\in[w]} \beta(x^j_{a,n,k}) = \sum_{k\in[w]} \beta(x^j_{a+\ell,n,k})$ is even.
	
	Let $(b,n) \in \set{(1,1),(h,c)}$ and consider the unary equations in block $b$.
	Consider side $j\in[2]$.
	If a unary equation is in an $E$-small component $\cmpnt$, then we may modify $\beta$ on $\vars{\cmpnt}\setminus\vars{E}$ by inserting a solution to $\cmpnt$ extending $\alpha$, which exists since $\alpha$ is closure-consistent. In particular, after doing this all unary equations occurring in small components are violated. Therefore, all violated equations are in the same $E$-large components: 
	 $w$ many equations can split side $j$ of block $b$
	into at most two $E$-large components, one containing vertices from row $1$ and the other one from row $c$.

	If there are two $E$-large components, let $F=E$,
	otherwise let $F$ be the union of $E$ and all equations in row  $(b, c+1-n)$.
	Then $F$ is a separator and all non-satisfied unary constraints are in the $F$-large component, which we again call $\cmpnt$.
	Since $F$ is a separator in block $(b,n$),
	it is via $\sim$ also a separator in $\cgrid{}[(a,n+1),(a+\ell, n)]$.
	So its parity under $\beta$ has to be even because this system is homogeneous (Lemmas~\ref{lem:sol-Tseitin-extend}\ref{itm:tseitin-even-boundary} and~\ref{lem:parity-crossing}).
	This implies that also the parity of the boundary $\boundary[F]{\cmpnt}$ has to be even.
	If $(b,n) = (1,1)$, then $\cmpnt$ itself is homogeneous and the
	assignment of $\beta$ to the boundary $\boundary[F]{\cmpnt}$ extends to a solution of $\cmpnt$ by Lemma~\ref{lem:sol-Tseitin-extend}\ref{itm:tseitin-boundary-parity}.
	This does not change values set by $\alpha$
	since $E \subseteq F$.
	Otherwise, $(b,n,j) = (h,c,1)$ and Lemma~\ref{lem:sol-Tseitin-extend}\ref{itm:tseitin-move-invonsistency} says that it suffices to violate a single equation in $\cmpnt$ (which in particular contains top unary equations). So in both cases, since $\alpha$ has the right parity, we have obtained our required almost extension.
\end{proof}

While almost-extensions describe positions in which Spoiler has made a bad move,
the following notion will describe positions, in which Duplicator has \emph{not} made a bad move.

\begin{definition}[Safe Solutions]
Let $E \subseteq \cgrid{}$.
A $1$-periodic solution $\alpha \in \sols{E}$ is \defining{block-consistent} if 
there is a $b \in [2, h-2]$ and an $n \in [c-1]$ such that $\alpha/_\sim$ extends to a solution of $\cgrid{}[(b,n+1), (b+1, n)]$ and $E/_\sim$ does not contain an equation in row $(b,n+1)$ or $(b+1, n)$.
A  solution $\alpha \in \sols{E}$ is \defining{safe} if
\begin{enumerate}
	\item $\alpha$ is $1$-periodic,
	\item $\alpha$ is closure-consistent,
	\item $(E, \alpha)$ has the right parity, and
	\item $\alpha$ is block-consistent.
\end{enumerate}
\end{definition}

Note that block-consistency of $\alpha$ does not imply closure consistency of $\alpha$ since it does not cover the $E$-components of the top row $(h,c)$ or the bottom row $(1,1)$.
Also note that the choice of $b$ for block-consistency does not matter and that $\cgrid{}[(b,n+1), (b+1, n)]$ contains a member of the $\sim$-class of each non-unary equation and each variable.
For the vertical variables in row~$n$, it actually contains two members (on which the extension to $\cgrid{}[(b,n+1), (b+1, n)]$ may disagree).

\begin{lemma}
	\label{lem:closure-consistent-separator-is-safe}
	Let $E \subseteq \comucgrid{}$ be of size $w$ such that $E/_\sim$ is a separator, let $f \in \comucgrid{}$, and let $\alpha \in \sols{E\cup\set{f}}$.
	If $\alpha$ is $1$-periodic and $\alpha/_\sim$ is closure-consistent,
	then $\alpha$ is block-consistent.	
	In particular, if $(E\cup\set{f},\alpha)$ has the right parity, then $\alpha$ is safe.
\end{lemma}
\begin{proof}
	We need to show that $\alpha$ is block-consistent.
	Because $\alpha/_\sim$ is closure-consistent,
	there is an extension $\alpha'$ of $\alpha/_\sim$ to $\clo{(E \cup \set{f})/_\sim}$.
	By Lemma~\ref{lem:separator-span}, the rows occupied by $E/_{\sim}$ lie in a $w$ consecutive rows, possibly at the end and begin of a block.
	Since $c=3w$, excluding this interval and the row of $f$
	leaves at least $2w-1\ge5$ empty rows in at most three intervals.
	Hence two consecutive rows are empty.
	By Lemma~\ref{lem:separator-span}, the rows occupied by $E/_{\sim}$ lie in a consecutive interval of at most $w$ rows.
	With the additional equation $f$, this interval may contain up to $w+2$ rows, or there are two consecutive free rows between this interval and $f$.
	Since $c = 3w$, there are $2w-2 \geq 4$ rows left. 
	Hence there is an $n \in [c-1]$ such that, for an arbitrary $b \in [2,h-2]$
	the system $\mathcal{B} := \cgrid{}[(b,n+1), (b+1,n)]$ contains no equations of $(E \cup \set{f})/_{\sim}$ in rows $n$ or $n+1$.
	In this proof, we say block $(j,b)$ to mean $\cgrid{}[(j,b)] \cap \mathcal{B}$ and block $(j,b+1)$ to mean $\cgrid{}[(j,b+1)] \cap \mathcal{B}$. Note that since $E/_\sim$ is a separator, there is one equation per column (on some side) in $E$.
	Since $\alpha'$ extends $\alpha/_\sim$, it suffices to show that
	$\restrict{\alpha'}{\mathcal{B}}$ is block-consistent.
	We now identify $\alpha'$ with $\restrict{\alpha'}{\mathcal{B}}$.

	We now show that we can find a solution that extends $\alpha'$ to all
	vertical variables in row $(b,c)$ (and possibly also to some other variables) such
	that this assignment does not violate any equation and for each block $(j',b')$ for $b' \in \set{b,b+1}$ and $j' \in [2]$ a single component remains unassigned.
	
 	First assume this claim and let $b' \in \set{b,b+1}$ and $j' \in [2]$.
	Since all vertical variables in row $(b,c)$ are assigned, each unassigned component $\cmpnt$ of $\mathcal{B}$ is a subsystem of a block $(j',b')$, which in particular is a Tseitin-system.
	Because $(E \cup \set{f})/_\sim$ contains no equation in row $n$ and $n+1$, neither does $\clo{(E \cup \set{f})/_\sim}$ and the vertical variables in row $(b',n)$ on side $j'$ are unassigned dangling variables of $\cmpnt$.
	Thus by Lemma~\ref{lem:sol-Tseitin-extend}\ref{itm:tseitin-boundary-unassigned}, we can extend the existing assignment to $\cmpnt$. The lemma follows.

	It remains to show the claim. We do this via a case distinction.
	\begin{enumerate}
		\item Assume all equations of $E/_{\sim} \cap \mathcal{B}$ are either in block $b$ and occur above row $n+1$ or occur in the first row of block $b+1$. Then all equations in block $b$ must occur on the same side $j$. 
		Otherwise, there is a column $k$ such that no equation contains a variable of column $k$ and side $j$ and a column $k'$ such that no equation contains a variable of column $k'$ and side $3-j$, which contradicts the fact $E/_{\sim}$ is a separator. 
		We consider the Tseitin-subsystem above the separator up to row $(j,b,c)$. If all equations occur in row $(b+1,1)$, we pick $j$ arbitrarily.
		The vertical variables in row $(j,b,c)$ are dangling for this system.
		If $f$ creates $(E\cup\set{f})/_\sim$-small components that overlap with this subsystem,
		these components are already assigned by $\alpha'$.
		If an equation of the subsystem remains unassigned, there is still one dangling variable in row $(j,b,c)$ left and we can extend by Lemma~\ref{lem:sol-Tseitin-extend}\ref{itm:tseitin-boundary-unassigned}.
		
		If $x^{3-j}_{b,c,k}$ is unassigned by $\alpha'$, 
		then there is at most one equation, of which $x^{3-j}_{b,c,k}$ is the only unassigned variable, namely $e^{j'}_{b+1,1,k}$, for some $j' \in [2]$.
		Assign $x^{3-j}_{b,c,k}$ such that this equation is satisfied, and otherwise assign it arbitrarily.
		Thus all vertical variables in row $(b,c)$ are assigned without violating an equation.

		Now all components that are still unassigned have a dangling variable in row $(b,n)$ or $(b+1,n)$ and is a Tseitin-systems.
		Lemma~\ref{lem:sol-Tseitin-extend}\ref{itm:tseitin-boundary-unassigned}
		provides the required extensions.

		\item Otherwise,  $E/_\sim$ is an internal-separator and all equations are in block $b+1$ up to row $n$.
		Then for each $k \in [w]$,
		there is a minimal $j_k \in [2]$ such that $E/_\sim$ separates
		both $x^{j'}_{b,c,k}$ from row $(j_k, b+1, c)$.
		Let $\kappa^{j',k}$ be the $E/_{\sim}$ component of the $x^{j'}_{b,c,k}$ in block $(j_k,b+1)$. We show that we can extend $\alpha'$ to each $\kappa^{j',k}$. If $\kappa^{j',k}$ is trivial, then some $e^j_{b+1,1,k}$ is in $E/_\sim$ and so $\kappa^{j',k}$ is already assigned. Otherwise, $\kappa^{1,k} = \kappa^{2,k}$ and we denote this component by $\kappa^k$. 
		Since there are no equations in row $n$ and $n+1$,
		this component is between row $1$ and row $n-1$.
		We show that we can extend $\alpha'$ to each $\kappa^k$.
		This ensures that for every $k \in [w]$ there is some $j \in [2]$ so that the variables of $e^j_{b+1,1,k}$ are assigned, and thus all variables in row $(b,c)$ are assigned.
		
		So let $k \in [w]$. If $\kappa^k$ is an $(E\cup\set{f})/_\sim$-small-component in $\ucgrid$ and not just in 
		block $(j_k, b+1)$
		then it is already assigned. 
		
		Assume $f$ splits $\kappa^k$ into multiple $(E\cup\set{f})/_\sim$-components in block $(j_k, b+1)$.
		Then each of the new components is either a small $(E\cup\set{f})/_\sim$-component in $\ucgrid$ or contains an equation in row $(b+1,1)$.
		In the former case, the component is already assigned by $\alpha'$.
		In the latter case, the component contains unassigned dangling variables in row $(b,c)$. Thus, it can be assigned using Lemma~\ref{lem:sol-Tseitin-extend}\ref{itm:tseitin-boundary-unassigned}.
		
		Otherwise, there has to be a $k'$ such that $\kappa^k = \kappa^{k'}$
		and $f\neq e^{3-j_k}_{b+1,1,k'}$.
		(If such a $k'$ does not exist, then $\kappa^k$ is an $(E\cup\set{f})/_\sim$-component since this is the only case of $f$ such that $\alpha'$ would assign both $x^{j}_{b,c,k'}$ for $j\in[2]$).
		Thus,  $x^{j}_{b,c,k'}$ for one $j\in[2]$ is not assigned by $\alpha'$.
		Since this variable is a dangling variable of $\cmpnt^k$, Lemma~\ref{lem:sol-Tseitin-extend}\ref{itm:tseitin-boundary-unassigned} can be used to extend $\alpha'$ to $\cmpnt^k$. \qedhere
	\end{enumerate}
\end{proof}

 \begin{lemma}
	\label{lem:block_consistent_change parity}
	Let $E\subseteq\cgrid{}$ be of size at most $w$ and let $\alpha \in \sols{E}$ be block-consistent
	witnessed by some $b\in[2,h-2]$, $n\in [c-1]$, and a suitable extension of $\alpha/_\sim$.

	If $E/_\sim$ is
	 neither a double-crossing separator nor a block-(b+1) internal-separator all of whose equations lie in block $(b+1)$,
	then, for each $p \in \set{0,1}$, the block-consistency of $\alpha$ can also be witnessed by $b$, $n$, and an extension $\beta'\in\sols{\cgrid{}[(b,n+1),(b+1,n)]}$ of $\alpha/_{\sim}$ such that
	$\sum_{k\in[w]} \beta'(x^j_{b+1,n,k}) = p$ for both $j$.
 \end{lemma}

 \begin{proof}
	Let $p \in \set{0,1}$ and $\beta\in\sols{\cgrid{}[(b,n+1),(b+1,n)]}$ witness the block-consistency of $\alpha$.
	If $\beta$ satisfies the claim, we are done.

	Otherwise, define \[p_{b',j} := \begin{cases}
			\sum_{ k\in[w]} \beta (x^{j}_{b,c,k})  & \text{if } b' =b\\
			\sum_{ k\in[w]} \beta (x^{j}_{b+1,1,k})  & \text{if } b' =b+1.
		\end{cases}\] 
	We know that $p_{b+1,1}=p_{b+1,2} \neq p$.
 
	We claim that there are $k \in [w]$ and $j \in [2]$ such that $x^j_{b,c,k}\notin\vars{E/_\sim}$ and $x^j_{b,c,k}$ is not $E/_{\sim}$-separated from row $n$ in any of the blocks $(j,b)$, $(1,b+1)$, $(2,b+1)$.

	First, assume that $E/_{\sim}$ is a block-$b$ internal separator containing an equation in row $(b+1,1)$.
	Then there is a column $k$ where $E/_{\sim}$ contains an equation $e^{3-j}_{b,c,k}$ from row $c$.
	The claim follows for $k$ and $j$.
	Second, assume that $E/_{\sim}$ is an internal separator only containing equations from rows $n+1$ to $c$. Then all equations in $E/_{\sim}$ must lie on the same side and the claim clearly follows.
		
		Finally, suppose $E/_{\sim}$ is not a separator. Since $E/_{\sim}$ is not an internal separator, there exists a $k \in [w]$ such that $x^1_{b,c,k}$ and $x^2_{b,c,k}$ are not $E/_{\sim}$-separated from row $n$ in the blocks $(1,b+1)$ and $(2,b+1)$. Therefore, if there exists some $j \in [2]$ such that $x^j_{b,c,k}$ is not $E/_{\sim}$-separated from row $n$ of block  $b$ the claim follows. Otherwise, there must be an equation in $E/_{\sim}$ on side $j$ and column $k$ for both $j \in [2]$. Since, $|E| \leq w$, this implies there is some column not containing any equation from $E$ and so the claim follows. We now use it to construct $\beta'$.

		We flip $x^j_{b,c,k}$ and, in each of the blocks $(j,b)$, $(1,b+1)$, $(2,b+1)$, flip the variables shared by consecutive equations along a path of equations not in $E/_\sim$ from the equation containing $x^j_{b,c,k}$ to an equation containing a vertical variable in row $n$ outside $\vars{E/_\sim}$, and flip that variable. Every equation in $\cgrid{}[(b,n+1),(b+1,n)]$ then sees an even number of flips, so we obtain a solution $\beta'$ extending $\alpha/_\sim$, and the parity of row $n$ changes on both sides of block $b+1$.
 \end{proof}

The next lemma is a key building block of the lower bound: it says that a safe assignment to a set of equations of size $w$ is always either almost-extendable or an odd separator modulo $\sim$.

\begin{lemma}
	\label{lem:safe-nont-odd-almost-extends}
	Let $E \subseteq \cgrid{}$ be of size at most $w$
	and $\alpha \in \sols{E}$ be safe.
	If $(E/_\sim,\alpha/_\sim)$ is not an odd separator, then
	\begin{enumerate}
		\item $\alpha/_\sim$ 
		extends to a cyclic solution of $\cgrid{}[b]$ for every $b\in [2,h-1]$, and
		\item $\alpha$ is almost-extendable.
	\end{enumerate}
\end{lemma}
\begin{proof}
	Let $\beta$ be an extension of $\alpha/_{\sim}$ to $\cgrid{}[(b,n+1), (b+1, n)]$ for some $n\in [c-1]$ witnessing that $\alpha$ is safe.

	\begin{claim}
		There is an extension $\beta'$ of $\alpha$ to $\cgrid{}[(b,n+1), (b+1, n)]$ such that
		for every $j \in [2]$ and $b' \in \set{b, b+1}$ we have
		 that $\sum_{k\in[w]} \beta' (x^j_{b',n,k})$ is even.
	\end{claim}
	\begin{claimproof}
		
		Again define \[p_{b',j} := \begin{cases}
			\sum_{ k\in[w]} \beta (x^{j}_{b,c,k})  & \text{if } b' =b\\
			\sum_{ k\in[w]} \beta (x^{j}_{b+1,1,k})  & \text{if } b' =b+1.
		\end{cases}\] 
		
		If every $p_{b',j}$ is even, then we can set $\beta' = \beta$, since $\sum_{k \in [w]} \beta(x^j_{b',n,k})$ is even for $b' \in \set{b,b+1}$ and $j \in [2]$ by Lemma~\ref{lem:sol-Tseitin-extend}\ref{itm:tseitin-boundary-parity}. In detail, for $b' = b$ we consider the subsystem $\ucgrid[(j,b,n+1), (j,b,c)]$ where the dangling variables are the vertical in row $n$ and row $c$. By Lemma~\ref{lem:sol-Tseitin-extend}\ref{itm:tseitin-boundary-parity} the sum of these dangling variables under $\beta$ is even, so since $p_{b,j}$ is even, $\sum_{k \in [w]} \beta(x^j_{b,n,k}) = 0$. The case where $b' = b+1$ is similar using the system $\ucgrid[(j,b+1,1), (j,b+1,n)]$.

		We now show how to achieve that the $p_{b',j'}$ are even.
		Lemma~\ref{lem:parity-crossing} shows that
		$p_{b+1,1} = p_{b,1} + p_{b,2} = p_{b+1, 2}$. 
		First assume that $E/_\sim$ is 
		double-crossing.
		Then $p_{b,j}$ is even for both $j$.
		Thus, $p_{b+1,j}$ is also even for both $j$ and so we are done setting $\beta' = \beta$.

		So assume that $E/_{\sim}$ is not a double-crossing separator 
		but an even internal separator such that 
		$E/_{\sim}$ contains no equations from row $n+1$ to $c$
		(so every equation is in or above row $(b+1,1)$  in $\cgrid{}[(b,n+1), (b+1, n)]$).
		Since $(E/_{\sim}, \alpha/_{\sim})$ is an even separator, it follows that $p_{b,1}+p_{b,2}$ is even by Lemma~\ref{lem:separator-determines-parity}
		and so are $p_{b+1,1}$ and $p_{b+1,2}$.
		If $p_{b,1}$ and $p_{b,2}$ are even, we are done.
		Otherwise, $p_{b,1}$ and $p_{b,2}$ are both odd.
		Because $E/_\sim$ is not a double-crossing separator,
		there has to be a $k$ such that $e^j_{b+1,1,k}/_\sim \notin E/_\sim$ for both $j$.
		Hence, we can alter the assignment to $x^j_{b,c,k}$ for both $j$, so that $p_{b,1}$ and $p_{b,2}$ become even.
		This satisfies all equations in block $b+1$ in $\cgrid{}[(b,n+1), (b+1, n)]$. 
		For block $b$, we find a satisfying extension by Lemma~\ref{lem:sol-Tseitin-extend}\ref{itm:tseitin-boundary-unassigned}.

		It remains to consider the case that 
		$E/_\sim$ is either not a separator or is an even internal separator 
		that contains an equation between rows $n+1$ and $c$.
		First, suppose that $E/_{\sim}$ is a separator and contains an equation from rows $1$ to $n-1$.
		Then since $E/_{\sim}$ is a separator, does not contain any equations from rows $n$ and $n+1$, and contains an equation in rows $n+1$ to $c$, it follows that every equation of $E/_{\sim}$ lies in rows $n+1$ to $c$ or in row $1$.
		In both cases, we can assume by Lemma~\ref{lem:block_consistent_change parity} that $p_{b+1,1}$ and $p_{b+1,2}$ are even.
		If $p_{b,1}$ and $p_{b,2}$ are even, we are done.
		Otherwise, $p_{b,1}$ and $p_{b,2}$ are odd:
		\begin{itemize}
			\item If $E/_{\sim}$ is a block-$b$ internal separator, then every equation of $E/_{\sim}$ in block $b$ occurs on the same side $j$. Moreover, $(E/_{\sim}, \alpha/_{\sim})$ is an even separator. All equations of block $(j,b)$ above the separator lie in rows $n+2$ to $c$ where $\beta$ is a solution. By applying Lemma~\ref{lem:sol-Tseitin-extend}\ref{itm:tseitin-boundary-parity} to these equations we obtain that $p_{b,j}$ is equal to the parity of $(E/_{\sim}, \alpha/_{\sim})$, i.e. is even.
			Therefore, so is $p_{b,3-j} = p_{b+1,1} - p_{b,j}$ and we are done.
			\item Suppose there is a $k$ such that $x^j_{b,c,k}$ is not $E/_{\sim}$-separated from row $n$ in block $(j,b)$ for both $j$. Then we flip the assignment to these two variables.  Since $x^{j}_{b,c,k}$ is in the $E/_\sim$-large component in block $(j,b)$, we can extend this change to a solution by Lemma~\ref{lem:sol-Tseitin-extend}\ref{itm:tseitin-boundary-unassigned}. This makes both $p_{b,1}$ and $p_{b,2}$ even.
			\item Otherwise, $E/_{\sim}$ is not a separator and for each $k \in [w]$ there is a $j \in [2]$ such that $x^j_{b,c,k}$ is $E/_{\sim}$-separated from row $n$ in block $(j,b)$. Therefore, $E$ contains exactly one equation per column. Moreover, all equations in $E/_{\sim} \cap \mathcal{B}$ must occur in either in row one of block $b+1$ or above row $n+1$ of block $b$. Since $E/_{\sim}$ is not a separator, for each $j \in [2]$, there is a $k \in [w]$ such that $x^{j'}_{b-1,c,k}$ is not $E/_{\sim}$-separated from row $(j,b,c)$ for some $j' \in [2]$. In particular, for each $j \in [2]$ there is a $k_j$ such that $x^j_{b,c,k_j}$ is in the $E/_{\sim}$-large component. This means that for $j \in [2]$ neither $e^1_{b+1,1,k_j}$ nor $e^2_{b+1,1,k_j}$ is in $E/_{\sim}$. Therefore, we flip the values to $x^1_{b,c,k_1}, x^2_{b,c,k_2}$ as well as every $x^{j'}_{b+1,1,k_j}$ for $j,j' \in [2]$. This makes $p_{b,1}$ and $p_{b,2}$ even, while $p_{b+1,1}$ and $p_{b+1,2}$ are flipped twice and so remain even.
			We can extend this change to a solution by Lemma~\ref{lem:sol-Tseitin-extend}\ref{itm:tseitin-boundary-unassigned}. \qedhere
		\end{itemize}
	\end{claimproof}

	Using the extension $\beta'$, we can 
	assign the vertical variables in row $(b,n)$ and $(b+1,n)$ in a cyclic
	manner (respecting even parity per side):
	since there are no equations in rows $(b, n+1)$ and $(b+1,n)$,
	we can reassign the variables of the equations in row $(j,b,n+1)$
	and $(j,b+1,n)$:
	Fix the assignment to the vertical variables in row $(b,n+1)$ and $(b+1,n-1)$
	according to $\beta'$. This has to be even on both sides.
	Then fix a cyclic assignment on the vertical variables in rows $(b, n)$ and $(b+1,n)$, that is also even on both sides.
	By Lemma~\ref{lem:sol-Tseitin-extend}\ref{itm:tseitin-boundary-parity} we can assign the horizontal variables of row $(b,n+1)$ and $(b+1,n)$ to satisfy all equations.
	So the assumptions of Lemma~\ref{lem:periodic-solutions} are satisfied
	implying that $\alpha$ is almost-extendable.
	
	Because $\beta'$ is cyclic and $1$-periodic, it can be copied to a cyclic solution of $\cgrid{}[b]$; since we considered $\alpha/_\sim$,
	this is true for every $b\in [2,h-1]$.
\end{proof}

 We have a position $(E, \set{\alpha},r)$ in the size game \defining{almost-extendable} if $\alpha$ is almost-extendable. Then Duplicator can play so that from an almost-extendable position we stay almost-extendable until Spoiler chooses a separator at the top of the grid. This is facilitated by the next two lemmas.

\begin{lemma}
	\label{lem:move-up-fix-parity}
	Let $E\subseteq \comucgrid$ be of size at most $w$ such that $E$ is not a separator in block $h$, let $\alpha \in \sols{E}$ be closure-consistent and have the right parity, and let $e \in \comucgrid$ be such that $E \cup \set{e}$ is a separator in block $h$.
	Then there is an extension $\beta \in \sols{E\cup \set{e}}$ of $\alpha$ to $e$ such that $\beta$ is closure-consistent and has the right parity.
\end{lemma}
\begin{proof}
	Let $\alpha'$ be an extension of $\alpha$ to $\cl{E}$.
	Since $E \cup \set{e}$ is a separator in block $h$ we have that for all $k \in [w]$ there exists a $j_k \in [2]$ such that both $x^j_{h-1,c,k}$ are $(E \cup \set{e})$-separated from row $(j_k,h,c)$ in block $(j_k,h)$.
	Moreover,  one such $x^j_{h-1,c,k}$ is not $E$-separated from row $c$ on both sides, since $E$ is not a separator in block $h$.
	This implies that $e$ is in block $h$, say on side $\ell$.
	Let $\kappa$ be the $E$-component of $x^j_{h-1,c,k}$ in block $(\ell,h)$.
	This component $\kappa$ is split by $e$ into $(E \cup\set{e})$ components
	$\kappa_1, \dots, \kappa_m$ because $(E \cup\set{e})$ is a separator,
	but $\kappa$ contains equations in row $c$ and $1$.
	
	Each $\kappa_i$ contains an unassigned variable $v_i\in\vars{e}$ that occurs in no other $\kappa_{i'}$.
	Let $\kappa_1$ be the component of $x^j_{h-1,c,k}$, and let $\kappa_2$ be a component containing a vertical variable of row $(\ell,h,c)$ outside $\vars{E\cup\set{e}}$, which exists since $E\cup\set{e}$ separates $x^j_{h-1,c,k}$ from row $(\ell,h,c)$.
	We assign the unassigned variables of $e$ other than the $v_i$ arbitrarily, and then
	\begin{itemize}
		\item $v_1$ so that $v_1$ so that some block-$h$ separator $E'\subseteq E\cup\set{e}$ of size $w$ has the right parity under $\beta$; this is possible since $v_1$ occurs exactly once in the sum defining the parity: in $\boundary{\kappa_1}$, or, if $\kappa_1$ is trivial, $e=e^\ell_{h,1,k}$ and $v_1=x^j_{h-1,c,k}$;
		\item $v_i$ for $i\ge3$ so that the assignment to $\boundary{\kappa_i}$ extends to a solution of $\kappa_i$, which is possible by Lemma~\ref{lem:sol-Tseitin-extend}\ref{itm:tseitin-boundary-parity} since $v_i$ occurs exactly once in $\boundary{\kappa_i}$;
		\item $v_2$ so that $e$ is satisfied.
	\end{itemize}
	The $E$-small components extend by $\alpha'$, and $\kappa_1$ is not small unless it is trivial, since $x^j_{h-1,c,k}$ also occurs in block $h-1$.
	It remains to consider $\kappa_2$ if it is small and contains no dangling variable.
	Then $\ell=1$ and the separating equations lie in rows $(1,h,c-1)$ and $(1,h,c)$, so $\kappa_2$ consists of equations of row $(1,h,c)$ and unary equations $f_k$; we can extend to a solution of it by Lemma~\ref{lem:sol-Tseitin-extend}\ref{itm:tseitin-boundary-parity}, since 
	$(E\cup\set{e},\beta)$ has the right parity. All block-$h$ separators $E'\subseteq E\cup\set{e}$ of size $w$ have the same parity under $\beta$: each equals $\sum_{j,k}\gamma(x^j_{h-1,c,k})$ for any extension $\gamma$ of $\beta$ satisfying all equations below the separators, and such a $\gamma$ exists by the choice of the $v_i$ and since $\alpha'$ solves the $E$-small components. Hence $\beta$ has the right parity.
	Hence, $\beta$ is closure-consistent.
\end{proof}

\begin{lemma}
	\label{lem:extend-almost-extension}
	Let $F\subseteq \comucgrid$ of  size at most $w$ and $\alpha \in \sols{F}$ be almost-extendable.
	Then for every $f \in \comucgrid$ there is an extension $\beta \in \sols{F \cup \set{f}}$ of $\alpha$ such that
	for every $f$-neighbor $E$ of $F$:
	\begin{enumerate}
		\item the solution $\restrict{\beta}{E}$ is almost-extendable or
		\item $(E, \restrict{\beta}{E})$ is a block-$h$ separator of the right parity
		and $\restrict{\beta}{E}$ is safe.
	\end{enumerate}
\end{lemma}

\begin{proof}
	Let $\alpha'$ be an almost extension of $\alpha$.
	If $F$ is a block-$h$ separator,
	then $(F,\alpha)$ is even because odd separators are not almost-extendable
	by Lemma~\ref{lem:odd-separator-not-almost-extendable}.
	In particular, $\sum_{j \in [2], k \in [w]} \alpha'(x^j_{h-1,c,k})$ is even.
	Because $(F,\alpha)$ has the right parity,
	 $\comucgrid=\comucgrid^0$ is homogeneous. Thus, we can extend $\alpha$ to a true solution of $\comucgrid$ by reassigning $\alpha'$ on the $F$-large component above the separator in block $(1,h)$ using Lemma~\ref{lem:sol-Tseitin-extend}\ref{itm:tseitin-boundary-parity}. 
	This defines an extension to $f$, which is clearly almost-extendable. Since every restriction of an almost-extendable map is almost-extendable we have our required extension.
	
	Next suppose that $F$ is not a block-$h$ separator.
	If $F\cup \set{f}$ is also not a block-$h$ separator,
	we change $\alpha'$ on the $F$-large component to violate an equation different from $f$ using Lemma~\ref{lem:sol-Tseitin-extend}\ref{itm:tseitin-move-invonsistency}  yielding an almost-extension $\alpha''$ of $\alpha$. We define $\beta:= \restrict{\alpha''}{F \cup \set{f}}$, so $\beta$ is also almost-extendable which completes this case.
	
	The only remaining case occurs when $F$ is not a block-$h$ separator but $F\cup \set{f}$ is.
	So $f$ is in block $h$. Assume that $f$ is on side $j$.
	Then $f$ has to be in the unique $F$-large component in block $(j,h)$, otherwise adding $f$ cannot turn $F$ into a separator.
	We choose $\beta$ as given by Lemma~\ref{lem:move-up-fix-parity}.
	Then $\beta$ has the right parity.
	Now let $E$ be an $f$-neighbor of $F$ and $\beta' := \restrict{\beta}{E}$. If $E=F$, the claim is trivial since $\beta'=\alpha$.
	If $E\neq F$ and $E$ is a block-$h$ separator,
	then $(E,\beta')$ has the right parity.
	Also, in this case all equations are in block $h$ in at most $w$ consecutive rows, in particular not in both row $1$ and row $c$. So $\beta'$ is $1$-periodic.
	Moreover, the closure consistency of $\beta$ guaranteed by Lemma~\ref{lem:move-up-fix-parity} implies that $\beta'/_\sim$ extends to $\cl{E/_\sim}$.
	Therefore, $(E,\beta')$ is safe by Lemma~\ref{lem:closure-consistent-separator-is-safe}.
	
	If $E$ is not a block-$h$ separator, we can reassign $\alpha'$ in the $F$-large component in block $(1,h)$ to agree with $\beta$ on $f$:
	The equation $f$ may split this component into multiple ones.
	Small $E$-components can be assigned because $\beta$ is closure-consistent
	(and are in particular composed of $(F\cup\set{f})$ small components).
	That is, on all variables of $f$ which are boundary variables of small $E$-components,
	the current assignment agrees with $\beta$.
	Only a single large $E$-component remains because $E$ is not a separator.
	Thus, all remaining variables of $f$ are dangling variables of the same component, thus we can assign the component violating some other equation
	by Lemma~\ref{lem:sol-Tseitin-extend}\ref{itm:tseitin-move-invonsistency}.
	 Therefore, $\beta'$ is almost-extendable which completes the proof.
\end{proof}

For the next lemma we need the following notion. We say that $F \subseteq \ucgrid$ is \defining{block-$b$ almost double-crossing} if
\begin{enumerate}
	\item  $F$ is a block-$b$ internal separator for $b\in [2,h]$,
	\item there is a $k\in[w]$ and a $j\in [2]$ such that $\opgridequation{j}{b}{2}{k} \in F$, and
	\item  for every $k' \in [w]\setminus\set{k}$ there exists a $j'\in [2]$ such that $\opgridequation{j'}{b}{1}{k'} \in F$.
\end{enumerate}
This means that replacing $\opgridequation{j}{b}{2}{k}$ with $\opgridequation{j}{b}{1}{k}$ turns $F$ into a block-$b$ double-crossing separator.

 The next lemma will allow Duplicator to maintain safety or reach an almost-extendable assignment.
 The exception in the below statement never comes up in the lower bound since Duplicator never allows odd separators to get so close to the bottom. 

\begin{lemma} \label{lem:safe_to_safe}
	Let $E \subseteq \comucgrid{}$ be of size $w$ and $\alpha \in \sols{E}$ be safe.
	Then 
	\begin{enumerate}
		\item for every $f \in \comucgrid{}$, there is an extension $\beta \in \sols{E\cup \set{f}}$ of $\alpha$ such that
		for every $f$-neighbor $F$ of $E$,  the restriction $\restrict{\beta}{F}$ is safe or almost-extendable; or
		\item $(E,\alpha)$ is a block-$2$ odd almost double-crossing separator. 
	\end{enumerate}
\end{lemma}
\begin{proof}
	We make a case distinction.
	
	\subparagraph{Non-odd separator.}

	If $(E/_\sim, \alpha/_\sim)$ is not an odd separator,
	then $\alpha$ is almost-extendable by Lemma~\ref{lem:safe-nont-odd-almost-extends}.
	By Lemma~\ref{lem:extend-almost-extension} there exists an extension $\beta \in \sols{E\cup\set{f}}$ such that for every $f$-neighbor $F$ of $E$ we have that $\restrict{\beta}{F}$ is either almost-extendable or safe.  In the rest of the cases, we assume that $(E/_\sim, \alpha/_\sim)$ is an odd separator. Moreover, by the same argument, we may also assume that $\alpha$ is not almost-extendable.

	\subparagraph{Odd separator that can move into block $h$.}
	Suppose $E \cup \set{f}$ is a separator in block $h$ but that $E$ is not a separator in block $h$. Then $E$ has to contain at least $w-1$ equations from block $h$ and so since $E/_{\sim}$ is a separator, it follows that $E/_{\sim}$ has to be a block-$(h-1)$ internal separator with $w-1$ equations lying in row $(h,1)$ (otherwise $\alpha$ would be almost extendable).
	By Lemma~\ref{lem:move-up-fix-parity},
	there is an extension $\beta$ of $\alpha$ to $f$
	such that $(E \cup \set{f}, \beta)$ is a block-$h$ separator of the right parity and $\beta$ is closure-consistent.
	Since the equations of $E$ lie in rows $(h-1,c)$ and $(h,1)$,
	all equations of $E \cup \set{f}$ are in at most three consecutive rows (between rows $(h-1,c)$ and $(h,2)$),
	which means that $\beta$ is also $1$-periodic.
	In this case, closure-consistency of $\beta$ also implies extendability to $\clo{(E \cup \set{f})/_\sim}$ since there are no small components at row $(h,c)$.
	By Lemma~\ref{lem:closure-consistent-separator-is-safe}, $\beta$ is also block-consistent. Thus $\beta$ is safe.

	\subparagraph{Odd separator that can move to block $1$.}
	Let $E \cup \set{f}$ be a block-$1$-separator.
	Since $\alpha$ is of the right parity and $(E/_{\sim},\alpha/_{\sim})$ is an odd separator we have that $E$ is not a block-$1$-separator. Similarly to in the previous case, it follows that $E$ contains $w-1$ equations in block $1$ or row $(2,1)$ and the remaining equation of $E$ lies between rows $(1,c)$ to $(2,2)$ (otherwise, $\alpha$ would be almost extendable). 
	Since $E$ is not a block-$1$ separator but $E/_\sim$ is a separator,
	$E \cup \set{f}$ contains a block-$2$ double-crossing separator or a block-$1$ internal separator with equations in rows $(j,1,c)$ and $(2,1)$,
	and in both cases $E$ is a block-$2$ odd almost double-crossing separator.

	\subparagraph{Odd separator in a border block.}
	Assume that all equations of $E$ are contained in the same border block $b$. 
	Then $b = h$, since a separator in block $1$ of the right parity is even.
	Moreover, since $\alpha$ has the right parity, the parity of $(E,\alpha)$ viewed as a block-$h$ separator
	(for a double-crossing separator, we mean $\sum_{j \in [2], k\in[w]} \alpha(x^j_{h-1,c,k})$) is the parity of block $h$.
	We make another case distinction.
	If $f$ is in a row above the separator (that is, in an $E$-component containing a top unary equation),
	we consider an extension of $\alpha$ up to row $(h,c)$ by Lemma~\ref{lem:sol-Tseitin-extend}\ref{itm:tseitin-boundary-parity} (applied for each side independently) and assign the variables of $f$ according to this extension.
	Thus, the assignment is closure-consistent and has the right parity.
	It is clearly $1$-periodic.
	By Lemma~\ref{lem:closure-consistent-separator-is-safe}, it is block-consistent and thus safe, and so is every restriction of it.

	If $f$ is below the separator,
	we consider the solution that witnesses that $\alpha$ is block-consistent
	and assign the variables according to it via $\sim$.
	The resulting assignment is clearly $1$-periodic, closure-consistent,
	and block-consistent by Lemma~\ref{lem:closure-consistent-separator-is-safe}.
	It also does not change the parity of a separator in block $h$.

		\subparagraph{Odd separator in a non-border block.}

	Finally, assume that the previous cases do not apply.	
	Let $b \in [2,h-1]$, $n \in [c-1]$, and $\beta \in \sols{\cgrid{}[(b,n+1), (b+1, n)]}$
	witness that $\alpha$ is safe. By definition, $\beta$ extends $\alpha/_\sim$.
	First assume that $f$ is not in row $(h,c)$ and $(1,1)$.
	Define an extension $\alpha'$ of $\alpha$ to $f$ by the assignment of $\beta$
	to $f/_\sim$.
	It suffices to show that $\alpha'$ is safe.
	Because the previous cases do not apply, $\alpha'$ has the right parity.
	We show that $\alpha'$ is closure-consistent.
	Because $f$ is not in row $(h,c)$ or $(1,1)$, every $(E\cup \set{f})$-small component that is not $E$-small is modulo $\sim$ also a $(E\cup \set{f})/_\sim$-small component of $\cgrid{}[(b,n+1), (b+1, n)]$. Thus, $\alpha'$ is closure-consistent.
	Moreover, $\alpha'$ is block-consistent by
	 Lemma~\ref{lem:closure-consistent-separator-is-safe}.
	
	The case that $f$ is in row $(h,c)$ or $(1,1)$ remains.
	In these cases we show that there is an extension $\beta$ of $\alpha$ such that $\restrict{\beta}{F}$ is almost extendable for every $f$-neighbor $F \neq E$ of $E$.
	Because $E$ is not a separator in a border block,
	all of its equations in the same block as $f$ are at least $w$ rows apart from $f$
	(otherwise $\alpha$ is almost-extendable and we conclude by Lemma~\ref{lem:extend-almost-extension}).
	In particular, $f$ is in the $E$-large component in its block and none of its
	variables are already assigned. Assign $f$ arbitrarily not violating a unary equation.
	This defines an extension $\beta$ of $\alpha$ to $f$.
	
	Now let $F$ be an $f$-neighbor of $E$.
	If $F=E$, the claim is trivial since $\restrict{\beta}{F}=\alpha$.
	Otherwise, $F \setminus \set{f} \subseteq E$ consists of $w-1$ equations, so $(F \setminus \set{f})/_\sim$ is not a separator.
	The assignment $\restrict{\alpha}{F\setminus \set{f}}$ is safe since it is a restriction of a safe assignment.
	Thus, by Lemma~\ref{lem:safe-nont-odd-almost-extends},
	there is an almost-extension $\beta'$ of $\restrict{\alpha}{F\setminus \set{f}}$.
	We then reassign $\beta'$ in the $E$-large component of the block of $f$
	to agree with $\beta$ on $f$ using Lemma~\ref{lem:sol-Tseitin-extend}\ref{itm:tseitin-boundary-parity}.
	Thus, $\restrict{\beta}{F}$ is almost-extendable.
\end{proof}

\subsection{Pairwise-Separators and Critical Separators} \label{SS:pairwise_critical}

This section considers sets of equations $E$ such that $E/_\sim$ is a separator but $E/_\compEquiv$ is not.
If, for some assignment, the separator $E/_\sim$ is odd, the results from the previous section cannot be used to show almost-extendability.
To extend it anyway, we need to use the exact definition of the compression more carefully.
Fortunately, we can do this by adapting results from~\cite{DBLP:journals/jacm/GroheLNS25, RezendeFJN025} to the crossed grid. Note that some of the results we need from \cite{RezendeFJN025} are only present/proved in the full version \cite{RezendeFJN024Arxiv}, so we also cite this full version when appropriate.

To avoid corner cases, this section will focus on sets of non-unary equations. This will be not restrictive later on.
Instead of assigning the variable $z$ of a unary equation,
we will assign the variables of a non-unary equation which include $z$.
We call a set $E\subseteq\cgrid{}$ or $E\subseteq\comucgrid{}$ \defining{non-unary} if it only contains non-unary equations.

\begin{definition}[Pairwise Separator]
	A set of variables $W \subseteq \vars{\ucgrid}$ is a \defining{pairwise separator}
	if, for every $k \in [w]$, there is a side $j\in[2]$, a block $b \in[h]$, and a row $n \in [c]$ such that
	$W/_{\sim_k}$ contains the variables $x^j_{b,n,k}$ and 
	$x^j_{b,n,k+1}$.
	 A set of equations $E \subseteq \ucgrid$ is a pairwise separator if $\vars{E}$ is a pairwise separator.
\end{definition}

\begin{lemma}
	\label{lem:not-pairwise-separator-almost-extendable}
	Let $F\subseteq \comucgrid$ be a non-pairwise separating set of at most $w$ non-unary equations.
	If $F/_\sim$ is a separator,
	then every safe $\alpha \in \sols{F}$ is almost-extendable.
\end{lemma}

\begin{proof}
	Because $F/_\sim$ is a separator, $F$ contains for each column $k$ exactly one equation in column $k$ (on some side).
	Let $k \in [w]$ be a column for which $F$ is not pairwise separating (so between columns $k$ and $k+1$).
	Let $e := e^j_{b,n,k}/_{\approx_{p_kp_{k+1}}}$ be the $\approx_{p_kp_{k+1}}$-class of the unique equation in column $k$ in $F$.
	That is, we additionally to $\compEquiv$ also identify the bottom and top block modulo $p_kp_{k+1}$.
	Note that we can pick our representative $e^j_{b,n,k}$ of $e$ so that $2 < b < h-p_{k+1}-2$.

	Set $f := e^j_{b,n,k-1}/_\compEquiv$ and $E := (F\setminus \set{e})\cup\set{f}$. 
	For $E$, there is one column, namely $k$, without an equation in $E$. So $\cl{E/_{\sim}}$ is not a separator.
	
	Since $\alpha$ is safe it is block-consistent. Let $\gamma$ be a witness to this and set $\alpha' :=\restrict{\gamma}{E}$, which is clearly $1$-periodic.
	As witnessed by $\gamma$, the map $\alpha'/_\sim$ is also closure-consistent.
	Since $F$ is non-pairwise separating, $F$ and thus $E$ is not a separator in a border block. Thus $(E,\alpha')$ has trivially the right parity. Since $\alpha'$ is block consistent (witnessed by $\gamma)$ 
	we obtain that $\alpha'$ is safe and hence there is cyclic extension of $\alpha'/_\sim$ to the block $\cgrid{}[b]$ by Lemma~\ref{lem:safe-nont-odd-almost-extends}.
	By cyclically repeating this extension,
	we obtain a solution $\beta$ of $\mathcal{B} := \cgrid{}[b-1,b+p_{k+1}-1]$.
	
	We observe that $\beta$ may disagree with $\alpha/_\sim$ on $\vars{e}$
	but agrees with $\alpha/_\sim$ on $y^j_{b,n,k}$, since $y^j_{b,n,k}$ is in $\vars{F} \cap \vars{E}$.
	We now show how $\beta$ can be modified to obtain a cyclic solution of $\mathcal{B}$ that extends $\alpha/_{\sim_k}$.
	
	\begin{claim}
		There is a set $W\subseteq \vars{\mathcal{B}}$ of variables in columns $k$ and $k+1$ which does not intersect  the first or last row of $\mathcal{B}$
		such that:
		\begin{itemize}
			\item $W/_\compEquiv$ is disjoint from $\vars{E}$ and $\vars{e} \setminus \set{y^j_{b,n,k}}\subseteq W$
			\item Every solution $\gamma \in \sols{e}$ such that $\gamma(y^j_{b,n,k}) = 0$ extends to a solution of $\mathcal{B}$ that is zero on $\vars{\mathcal{B}} \setminus W$.
		\end{itemize}
	\end{claim}
	\begin{claimproof}
	We make a case distinction on $n$.
	
	\begin{itemize}
		\item Case $n=c$.
		Because $F$ is not pairwise separating for columns $k$ and $k+1$, $F$ (and thus $E$) cannot contain any 
		of the equations $G/_\compEquiv$ for 
		\[G:= \set*{e^j_{b,c-1,k+1}, e^j_{b,c,k+1}, e^j_{b+1,1,k+1}, e^{3-j}_{b+1,1,k}}\]
		because otherwise $\vars{F}$ contains
		$x^j_{b,c-1,k}$ and $x^j_{b,c-1,k+1}$ (in case of $e^j_{b,c-1,k+1}/_\compEquiv$) or
		 $x^j_{b,c,k}$ and $x^j_{b,c,k+1}$ (in the other cases).
		 This means that for the following set of variables $W$,
		 the set $W/_\compEquiv$ is disjoint from $\vars{E}$:
		 \begin{align*}
		 	W &:= \setcond*{x^j_{b,n',k'}}{k' \in \set{k,k+1}, n' \in \set{c-1,c}} \cup {}\\
		 	&\hiddenEqq  \set*{y^j_{b,c-1,k+1}, y^j_{b,c,k+1},y^1_{b+1,1,k+1},y^2_{b+1,1,k+1}}
		 \end{align*}
		 
		 Because  $2 < b < h-p_{k+1}-2$, these variables are surely not in the top or bottom row of $\mathcal{B}$.
		 
		 Since $e$ is of arity $4$ and we fix one variable, 
		 there are $2^2$ many solutions $\gamma \in \sols{e}$ that are zero on $y^j_{b,n,k}$.
		 For readability, we identity an assignment with the set of variables set to one.
		 Let $x = y^j_{b,c,k+1}$ and $Z = \set{x^j_{b,c,k}, x^j_{b,c-1,k}}$
		 be the remaining variables of $e$.
		 Then the set $\Gamma = \setcond{\set{x,z}}{z \in Z} \subseteq \sols{e}$ 
		 is linearly independent and therefore spans the set of all  solutions $\gamma \in \sols{e}$ such that $y^j_{b,c,k} \notin \gamma$. 
		 For each solution $\gamma \in \Gamma$, we provide a set $X\subseteq W \setminus \vars{e}$ such that
		  $\gamma \cup X$ is the desired solution. 
		 \begin{align*}
		 	x^j_{b,c,k}:&& X &:= \set{x^j_{b,c,k+1}, y^1_{b+1,1,k+1}, y^2_{b+1,1,k+1}}\\
		 	x^j_{b,c-1,k}:&& X&:= \set{ x^j_{b,c-1,k+1}, y^j_{b,c-1, k+1}}
		 \end{align*}
		 The remaining cases follow the same approach and we only define the involved sets.
		\item Case $n=1$.
		\[G:= \set*{e^j_{b-1,c,k+1}, e^{3-j}_{b-1,c,k+1}, e^j_{b,1,k+1},  e^{3-j}_{b+1,1,k},  e^j_{b+1,2,k+1}}\]
		\begin{align*}
			W &:= \setcond*{x^{j'}_{b,n',k'}}{k' \in \set{k,k+1}, (j',b,n) \in \set{(1,b-1,c),(2,b-1,c),(j,b,1)}} \cup {}\\
			&\hiddenEqq \set*{y^1_{b-1,c,k+1},y^2_{b-1,c,k+1}, y^j_{b,1,k+1},y^j_{b,2,k+1}}
		\end{align*}
		$e$ is of arity $5$, pick $x=y^j_{b,1,k+1}$, and set $Z=\set{x^j_{b,1,k}, x^1_{b-1,c,k}, x^2_{b-1,c,k+1}}$.
		\begin{align*}
			x^j_{b,1,k}:&& X &:= \set{x^j_{b,1,k+1}, y^j_{b,2,k+1}}\\
			x^{j'}_{b-1,c,k}:&& X&:= \set{ x^{j'}_{b-1,c,k+1}, y^{j'}_{b-1,c, k+1}, y^{3-j}_{b,1, k+1}} \text{ for both } j' \in [2]
		\end{align*}
		\item Case $n \in [2,c-1]$.
		\[G:= \setcond*{e^j_{b,n',k+1}}{n' \in [n-1,n+1]}\]
		\begin{align*}
			W &:= \setcond*{x^{j}_{b,n',k'}}{k' \in \set{k,k+1}, n' \in [n-1,n+1]} \cup {}\\
			&\hiddenEqq  \setcond*{y^{j}_{b,n',k+1}}{n' \in [n-1,n+1]}
		\end{align*}
		$e$ is of arity $4$, pick $x=y^j_{b,n,k+1}$, and set $Z=\set{x^j_{b,n,k}, x^j_{b,n-1,k}}$.
		\begin{align*}
			x^j_{b,n,k}:&& X &:= \set{x^j_{b,n,k+1}, y^j_{b,n+1,k+1}}\\
			x^j_{b,n-1,k}:&& X&:= \set{ x^j_{b,n,k+1}, y^j_{b,n, k+1}} \qedhere
		\end{align*}
	\end{itemize}
	\end{claimproof}
	
	 By the previous claim, there is a solution $\beta'$ to $\mathcal{B}$
	that is only nonzero on $W$ and extends $\restrict{\alpha}{e} - \restrict{\beta}{e}$.
	Because $\mathcal{B}$ is homogeneous,
	$\beta+\beta'$ is also a solution of $\mathcal{B}$.
	The solution extends $\alpha/_{\sim{p_k}}$ on $e$ by construction and on $F\setminus \set{e}$ because $W$ is disjoint from $\vars{E}$.
	It is cyclic, because $\beta$ is cyclic and $\beta'$ does not modify the variables in the first or last row of $\mathcal{B}$.
	It is $p_{k+1}$-periodic between columns $k$ and $k+1$,
	$p_{k+1}$-periodic in column $k$ and $k+1$,
	and $1$-periodic in and between all other columns.
	Also, $\beta + \beta'$ extends $\alpha/_{\sim_{k}} = \alpha/_{\approx_{p_{k+1}}}$. 
	Thus, applying Lemma~\ref{lem:periodic-solutions} to $\alpha$, we conclude that $\alpha$ is almost-extendable.
\end{proof}

Recall that $\compEquiv^+$ also identified variables in the top and bottom block. 
We aso exploit the notation $e/_{\compEquiv^+}$ and $E/_{\compEquiv^+}$ as defined before for the other equivalences.
The relation $\compEquiv^+$ is used only to locate virtual representatives; the equation sets and assignments in the game remain compressed by $\compEquiv$.
Since $\sim$ also identified variables in the top and bottom block,
we have ${\compEquiv^+} \subseteq {\sim}$.
Since $\compEquiv^+$ is defined using modular arithmetic,
for the next step in the analysis it will be convenient to allow separators to wrap around from block $h$ to block $1$.
Formally, we define this as follows:
\begin{definition}[Wrapping Separator]
	Define $e^j_{b,n,k} + z := e^j_{1+( (b-1+z) \bmod h),n,k}$ for all $b\in[h]$, $n\in[c]$, $k\in[w]$, $z\in \ZZ$.
	A set $S\subseteq \ucgrid$ of non-unary equations is a \defining{wrapping separator}
	if $S+z:=\setcond{e+z}{e \in S}$ is a separator for some $z \in \ZZ$.
	The set $S$ is \defining{truly wrapping} if $S\cap \cgrid{}[h]\neq \emptyset$ and $S\cap\cgrid{}[1] \neq \emptyset$.
\end{definition}
Intuitively, $S+z$ shifts the equations in $S$ by $z$ blocks, while wrapping around, so ``block'' $0$ is block $h$.
Note that $S$ is a wrapping separator if and only if $S$ is a separator or $S+1$ is a separator, which exactly concerns the case of a separator that wraps around from block $h$ to block $1$.

\begin{definition}[Virtual Separator]
	A wrapping separator $S\subseteq \ucgrid$  is a \defining{virtual separator} for a 
	set of equations $E\subseteq \comucgrid$ if 
	\begin{enumerate}
		\item for every $k\in[w]$,
		for which $E$ contains only one class 
		$e/_\compEquiv$
		in column $k$, the separator $S$ contains one equation in $e/_{\compEquiv^+}$ and no other in that column, and
		\item for all other $k$, the separator $S$ does not contain more equations 
		in column $k$ than $E$ does.
	\end{enumerate}
\end{definition}
In particular, the representative in a column containing a single equation class need only agree with that class modulo $\compEquiv^+$.

\begin{definition}[Critical Set]
	A set of equations $E\subseteq \comucgrid$ is \defining{critical} if $E$ is a pairwise separator and there exists a virtual separator for $E$.
\end{definition}

Both notions of a virtual separator and a critical set originate from~\cite{RezendeFJN025} and we adapt them to the crossed setting.
Intuitively, critical sets are sets of equations,
which, if used as an odd separator, are not almost-extendable.
We show that in the size game the virtual separators for critical sets cannot ``jump''
in the grid.
They have to be moved row by row.
We remark that we will use virtual separators mostly only for sets $E$ of size $w$,
in which only Point~1 of the definition applies.

To lift results from the uncrossed setting to the crossed one,
define for a set $F \subseteq \comucgrid$ of non-unary equations and $j\in [2]$, the set
\[F_j := \setcond*{e^j_{b,n,k}/_\compEquiv}{e^{j'}_{b,n,k}/_\compEquiv \in F}\]
which moves all equations in $F$ to side $j$.
We see that the set of side-$j$-variables in $\vars{F}$ is a subset of the ones of $\vars{F_j}$.

We aim to lift results of~\cite{RezendeFJN025} to $F_j$.
The compression in~\cite{RezendeFJN025} is obtained from a so-called graph compression, but also results in an $\FF_2$-system.
Let $G_j$ be the graph defined as follows:
all non-unary side-$j$-equations become vertices of $G_j$ and are  adjacent if they share a variable. This is precisely the cylindrical grid in~\cite{RezendeFJN025}, which is compressed.
The compression in~\cite{RezendeFJN025} defines equivalence classes on edges and vertices of $G_j$, corresponding to our classes of variables and equations on side $j$, respectively (variables fixed by unary equations are not represented in the graph).

The set $F_j$ can be identified with a set of vertex equivalence classes.
The most important difference is the different notions of separators.
For us, we distinguish between internal separators and double-crossing separators.
In~\cite{RezendeFJN025}, a separator is defined for a graph.
A ``graph-separator'' of $G_j$ is an equation set (so a vertex set of $G_j$),
that blocks all paths from the first to the last row of $G_j$.\footnote{The authors of~\cite{RezendeFJN025} view the grid in a left-to-right manner,
	so our columns corresponds to rows in~\cite{RezendeFJN025}.}
This means, that ``graph-separators'' use the same notion as block separators, but not only within one block (plus the row directly above).
Our notion of double crossing separators is more restrictive, since all equations of a $w$-sized double-crossing separator have to be in the same row at the block borders.
Hence, for $F_j$, our notion of a separator is more restrictive.
For correspondence of our wrapping separators 
we have to allow wrapping around in the cylicdrical graph, which makes it a toroidal grid (see also~\cite{DBLP:journals/jacm/GroheLNS25}).
``Toroidal graph separators'' block cycles in the toridal grid, that contain vertices from every row.

We now relate the central notions of pairwise separators and critical sets to the corresponding ones in~\cite{RezendeFJN025}.
\begin{itemize}
	\item If $F_j$ is a pairwise side separator, then for every $k \in [w]$   the set $\vars{F_j/_{\sim_k}}$ contains the variables $x^j_{b,n,k}$ and $x^j_{b,n,k+1}$ for some $b$ and $n$.
	This implies that $F_j/_{\sim_k}$ contains the
	equations $e^j_{b,n',k}$ and $e^j_{b,n'',k+1}$ for some $n\leq n',n'' \leq n+1$ (or also $e^j_{b+1,1,k'}$ and $e^j_{b,c,k''}$ such that $\set{k',k''} = \set{k, k+1}$ if $n = c$).
	This is precisely the notion of a two-separating vertex set in~\cite{RezendeFJN025}.
	Note that the converse direction is also true,
	thus our notion of a pairwise side separator is equivalent to two-separating for $F_j$ (on side $j$).
	\item If $F_j$ is critical, then $F_j$ is a pairwise separator (which we have already seen corresponds to two-separating) and there exists a virtual separator $S$ for $F_j$.
	Moving every equation of $S$ to side $j$ gives a virtual graph-separator for $F_j$, with representatives matched modulo $\compEquiv^+$.
	This corresponds to a virtual cordon in~\cite{RezendeFJN024Arxiv} (which we here call ``virtual graph-separators'' or ``virtual size-$j$ graph-separators''). This is the same notion as our virtual separators, apart from the different notion of a separator.
	Since our notion of separators is more restrictive, the converse direction  from virtual graph-separators to virtual separators is false.
\end{itemize}

\begin{lemma}
	\label{lem:critical-virtual-separator-conincide}
	Let $F \subseteq \comucgrid$ be critical, non-unary, and of size at most $w+1$.
	Then all virtual separators for $F$ coincide on all columns of which $F$ contains only one equation.
\end{lemma}
\begin{proof}
	Let $I \subseteq[w]$ be the columns of which $F$ contains at most one equation.
	By~\cite[Proposition~4.10]{RezendeFJN024Arxiv}, the virtual side-$j$ graph-separators of $F_j$ coincide on the columns in $I$.
	Let $S\subseteq\ucgrid$ be a virtual separator for $F$, and let $S_j$ be obtained by moving every equation of $S$ to side $j$.
	Then $S_j$ is a virtual separator for $F_j$,
	and in particular $S_j$ is a virtual side-$j$ graph-separator.
	Because all virtual side-$j$ graph-separators of $F_j$ coincide on $I$,
	all virtual separators $S$ of $F$ have to coincide on $I$.
\end{proof}

\begin{corollary} \label{cor:critical-then-unique-v-separator}
	Let $F \subseteq \comucgrid$ be critical, non-unary, and of size $w$. Then there exists a unique virtual separator for $F$.
\end{corollary}
\begin{proof}
	This is an immediate consequence of Lemma~\ref{lem:critical-virtual-separator-conincide}, because $F$ contains exactly one equation per column (on some side).
\end{proof}

The unique virtual separator of a critical and non-unary set $F$ of size $w$ is denoted by~$S_F$.

\begin{corollary} \label{cor:sep_no_jump}
	Let $E,F \subseteq \comucgrid$ be critical, non-unary, and of size $w$ such that $F\setminus E=\set{\eclass{e^j_{b,n,k}}}$.
	Then $|S_E \cap S_F| \ge w-1$. In particular, $S_E$ and $S_F$ contain the same equation in all columns $k' \neq k$.
\end{corollary}
\begin{proof}
	This is a consequence of Lemma~\ref{lem:critical-virtual-separator-conincide} applied to $E\cup F$. Equality need not hold: the two distinct actual classes in the changed column may have the same $\compEquiv^+$-image, in which case $S_E=S_F$.
\end{proof}

Finally, the next lemma will be used in the lower bound to show that it is a mistake for Spoiler to ever choose a non-critical set of equations.

\begin{lemma}
	\label{lem:critical-to-critical}
	Let $E\subseteq \comucgrid$ be non-unary and of size at most $w+1$, and let $F\subseteq E$.
	If $E$ is critical and $F$ is a pairwise separator, then either $F$ is critical or $F/_\sim$ is not a separator.
\end{lemma}

\begin{proof}
	Let $E$ be critical and $F$ be pairwise separating.
	If $F=E$, the conclusion is immediate. Otherwise, pairwise separation requires an equation in every column, so $|F|=w$ and $F$ contains exactly one equation per column.
	Clearly, $E_j$ is critical and $F_j$ is pairwise separating for both $j\in[2]$.
	
	As seen earlier, $E_j$ is critical also in the sense of~\cite{RezendeFJN024Arxiv} for graph-separators.
	By~\cite[Lemma~4.11]{RezendeFJN024Arxiv}, $F_j$ is critical in the sense of~\cite{RezendeFJN024Arxiv}.
	So there is a virtual graph-separator $S$  for $F_j$.
	Because $|F_j| = w$, $S$ contains one equation of each class in $F_j/_{\compEquiv^+}$.
	This implies that $S/_{\compEquiv^+} = F_j/_{\compEquiv^+}$. Since ${\compEquiv^+}\subseteq{\sim}$, also $S/_\sim = F_j/_\sim$.
	
	First assume that $S$ is not a wrapping separator in our notion.
	Because $S$ is a toroidal graph-separator, the equations in $S$ are within $w$ many consecutive rows.
	Because $c=3w$, this also means that $S/_\sim$ is not a separator in our notion because modulo $\sim$ we do not need to shift the separator.
	Because $S/_\sim = F_j/_\sim$, also $F_j/_\sim$ is not a separator.
	So $F/_\sim$ is not a separator, either.

	Second assume that $S$ is actually a wrapping separator in our stronger notion.
	Let 
	\[S' := \setcond*{e^{j'}_{b,n,k}}{e^j_{b,n,k}\in S, e^{j'}_{b,n,k}/_{\compEquiv^+}\in F/_{\compEquiv^+}}.\]
	Intuitively, $S'$ is obtained from $S$ by moving an equation $e\in S$ to side $3-j$, if the class of $e$ was moved to side $j$ in $F_j$; otherwise do not move $e$.
	Thus the equations of $S'$ match the classes in $F$ modulo $\compEquiv^+$.
	If $S'$ is a wrapping separator,
	then $S'$ is a virtual separator for $F$.
	Thus, $F$ is critical because $F$ is pairwise separating.
	
	Otherwise, $S'$ is not a wrapping separator.
	Because $S$ is a wrapping separator, all equations in $S$ and $S'$
	are within $w$ consecutive rows (in a wrapping manner, so row $(1,1)$ and $(h,c)$ count as consecutive).
	Thus, $S'/_\sim$ is not a separator, either.
	Since $S'/_{\compEquiv^+} = F/_{\compEquiv^+}$ and hence $S'/_\sim = F/_\sim$,
	this implies that $F/_\sim$ is not a separator, either.
\end{proof}

\begin{lemma}
\label{lem:non-critical-subset-almost-extendable}
Let $F\subseteq\comucgrid$ be critical, non-unary, and of size $w$ such that $S_F\not\subseteq\cgrid{}[2]$, and let $\alpha\in\sols{F}$ be safe.
For every non-unary $f\in\comucgrid\setminus F$, there is an extension $\beta\in\sols{F\cup\set{f}}$ of $\alpha$ such that for every $E\subseteq F\cup\set{f}$ of size at most $w$,
\begin{enumerate}
\item $\restrict{\beta}{E}$ is almost-extendable, or
\item $\restrict{\beta}{E}$ is safe, $E$ is critical, and $(E/_\sim,\restrict{\beta}{E}/_\sim)$ is an odd separator.
\end{enumerate}
\end{lemma}
\begin{proof}
Uniqueness of $S_F$ excludes a separator in block $2$, so the exceptional case of Lemma~\ref{lem:safe_to_safe} does not apply.
Choose an extension $\beta$ by that lemma, whose restrictions are safe or almost-extendable.
Let $E\subseteq F\cup\set{f}$ of size at most $w$
such that the restriction $\beta':=\restrict{\beta}{E}$ is safe but not almost-extendable.
By Lemma~\ref{lem:safe-nont-odd-almost-extends}, $(E/_\sim,\beta'/_\sim)$ is an odd separator.
Moreover, $E$ is pairwise separating: otherwise Lemma~\ref{lem:not-pairwise-separator-almost-extendable} gives almost-extendability since $\beta'$ is safe.
Since $F\cup\set{f}$ is critical, Lemma~\ref{lem:critical-to-critical} now implies that $E$ is critical, proving Assertion~2.
\end{proof}

In the final part of this section, we consider 
virtual separators
which wrap around from block $h$ into block $1$ in a modulo-like manner.
While this is possible and results in a virtual separator entering block $1$ quickly, we show that in this case we can always choose an almost extendable assignment before the separator becomes `dangerous'. The key property we maintain to ensure this is the following.

\begin{definition}
	For a non-unary $E \subseteq \comucgrid{}$, a safe assignment $\alpha \in \sols{E}$ is called \defining{upwards even} if block-consistency can be witnessed by $b\in[2,h-2]$,$n\in[c-1]$, and an extension $\beta \in\sols{\cgrid{}[(b,n+1),(b+1,n)]}$ of $\alpha/_\sim$ such that for both $j \in[2]$
	\[ \sum_{k\in[w]} \beta(x^j_{b+1,n,k}) = 0.\]
	The assignment $\alpha$ is \defining{wrap-safe}
	if $E$ is of size $w$ and critical, $E\cap \cgrid{}[1]=\emptyset$, $S_E \subseteq \cgrid{}[h] \cup \cgrid{}[(1,1)]$, $S_E\cap \cgrid{}[(1,1)] \neq \emptyset$,
	and $\alpha$ is safe and upwards even.
\end{definition}

Upward-evenness refers to the upper boundary of the witness.
The notion becomes important for separators $E$  which 'cross' between blocks, in which case an odd separator $(E, \alpha)$ 
can still be upwards even.
Intuitively, an upwards-even separator cannot be moved upwards without getting a true even internal separator and  becoming almost-extendable.

This gives the following strategy for virtual separator moving from block $h$ to block $1$:
ensure that, when moving through the wrapping block border  between block $h$ and $1$,
the crossing separator becomes upwards even, and than that moving it up further in block $1$ makes it almost extendable.
This approach only works, because for a set $E \subseteq \comucgrid{}$ of size $w$ with its unique virtual separator $S_E$ in block $h$,
there is no equation modulo $\compEquiv^+$ in block $1$ in $E$ and thus the unary equations at the bottom of the grid are not important.
When an equation from block $1$ is added, we can use existing lemmas to show almost extendability.
If the new equation is only modulo $\compEquiv^+$ in block $1$, we maintain a safe solution following the upwards even approach.
This is captured by the notion of being wrap-safe.

\begin{lemma}
\label{lem:virtual-top-no-bottom}
Let $E\subseteq\comucgrid$ be critical, non-unary, and of size $w$ such that $S_E\subseteq\cgrid{}[h]$.
Then $E/_{\compEquiv^+} \cap \cgrid{}[1]/_{\compEquiv^+} = \emptyset$.
In particular, $E \cap \cgrid{}[1] =\emptyset$.
\end{lemma}
\begin{proof}
	Since $E$ is of size $w$, the virtual separator $S_E$
	satisfies $(S_E)/_{\compEquiv^+} = E/_{\compEquiv^+}$.
	In column $k$, equivalence to an equation in block $1$ would require $p_kp_{k+1}$ to divide $h-1$.
	This is impossible since $p_kp_{k+1}>1$ divides $h$.
\end{proof}

\begin{lemma}
\label{lem:first-actual-bottom-equation}
Let $F\subseteq\comucgrid$ be non-unary and of size $w$ such that $F\cap\cgrid{}[1]=\emptyset$, and let $\alpha\in\sols{F}$ be safe.
If $f$ is an equation in row $1$ or $2$ of block $1$, then there is an extension $\beta\in\sols{F\cup\set{f}}$ of $\alpha$ such that $\restrict{\beta}{E'}$ is almost-extendable for every $E'\subseteq F\cup\set{f}$ of size at most $w$ with $E'\neq F$.
\end{lemma}

\begin{proof}
Extend $\alpha$ to $\beta$ by assigning all variables of $f$ to zero.
These variables are singleton classes and none occurs in $F$: equations outside block $1$ can only use variables in row $c$ of block $1$.
Let  $E' \subseteq F\cup \set{f}$ of size at most $w$ with $E' \neq F$.
Set $F' := E'\setminus \set{f}$.
Because $\alpha$ is safe, so is $\restrict{\alpha}{F'}$.
Because $F'$ has size at most $w-1$, $F'/_\sim$ is not a separator
and by Lemma~\ref{lem:safe-nont-odd-almost-extends}
there is an almost-extension $\gamma$ of $\restrict{\alpha}{F'}$.
Because $F$ contains no equation in block $1$,
Lemma~\ref{lem:sol-Tseitin-extend}\ref{itm:tseitin-boundary-parity} 
provides us with a solution of the side of $f$ in block $1$
that agrees with the all-zero assignment to $f$
and the assignment of $\gamma$ to the variables in row $(1,c)$ (since $\gamma$ satisfies this Tseitin-system).
Combining $\gamma$ with this solution yields an almost extension of $\restrict{\beta}{E'}$.
The restriction has the right parity: it contains at most one equation in block $1$ and fewer than $w$ equations in block $h$. 
\end{proof}

\begin{lemma}
\label{lem:wrap-prepare-even-upwards}
Let $F\subseteq\comucgrid$ be critical, non-unary, and of size $w$ such that $S_F\subseteq\cgrid{}[h]$, and let $f\in\comucgrid\setminus F$ be $\compEquiv^+$-equivalent to an equation in row $(1,1)$ and thus be non-unary.
For every safe $\alpha\in\sols{F}$ there is an extension $\beta\in\sols{F\cup\set{f}}$ of $\alpha$ such that
for every $E\subseteq F\cup\set{f}$ of size at most $w$ with $E\neq F$,
\begin{enumerate}
	\item $\restrict{\beta}{E}$ is almost-extendable, or
	\item $\restrict{\beta}{E}$ is wrap-safe.
\end{enumerate}
\end{lemma}
\begin{proof}
Let $\alpha\in\sols{F}$ be safe.
First suppose that no critical set $\hat{E}\subseteq F\cup\set{f}$ of size $w$ distinct from $F$ exists.
Since $S_F\subseteq\cgrid{}[h]$, Lemma~\ref{lem:non-critical-subset-almost-extendable} gives Assertion~1 for every $E\neq F$.

Otherwise, fix such a critical set $\hat{E}$.
By Corollary~\ref{cor:sep_no_jump}, $S_{\hat{E}}$ and $S_F$ share $w-1$ equations.
Together with the assumption on $f$, this forces the additional equation of $S_{\hat{E}}$ into row $(1,1)$ and the equations of $S_F$ into consecutive rows at the top of block $h$, including row $c$.
By Lemma~\ref{lem:virtual-top-no-bottom}, $F$ contains no equation in block $1$.
If $f$ is in block $1$, Lemma~\ref{lem:first-actual-bottom-equation} gives Assertion~1.

Otherwise, $f$ is in neither border block and its variables are not constrained by unary equations.
Choose a block-consistency witness for $\alpha$ with $b\in[2,h-2]$ and cut $n=2$, extending or restricting across the empty rows below the separator as necessary.
Since $F/_\sim$ is internal above the lower cut, Lemma~\ref{lem:block_consistent_change parity} gives a witness $\gamma$ with even upper parity.
Assign $f$ according to $\gamma$ via $\sim$, obtaining $\beta$.

Let $E\subseteq F\cup\set{f}$ have size at most $w$ with $E\neq F$, and put $\beta':=\restrict{\beta}{E}$.
It is $1$-periodic, and $\gamma$ witnesses block-consistency and upwards-evenness.
Closure-consistency follows from that of $\alpha$ for small components in the border blocks and from $\gamma$ for the others.
If $E=\hat{E}$, its unique virtual separator is truly wrapping, excluding an actual border-block separator and hence giving the right parity.
Thus $\beta'$ is safe and upwards even; since $E\cap\cgrid{}[1]=\emptyset$, it is wrap-safe.
Otherwise, $E$ omits a column, so it has the right parity and $\beta'$ is safe.
Lemma~\ref{lem:safe-nont-odd-almost-extends} then gives Assertion~1.
\end{proof}

\begin{lemma}
\label{lem:wrap-maintain-even-upwards}
Let $F\subseteq\comucgrid$ be non-unary and of size $w$, let $\alpha\in\sols{F}$ be wrap-safe, and let $f\in\comucgrid\setminus F$ be non-unary.
Then there is an extension $\beta\in \sols{F\cup\set{f}}$ of $\alpha$ such that
for every $E\subseteq F\cup\set{f}$ of size at most $w$ with $E\neq F$,
\begin{enumerate}
	\item $\restrict{\beta}{E}$ is almost-extendable, or
	\item $\restrict{\beta}{E}$ is wrap-safe, or
	\item $\restrict{\beta}{E}$ is safe, $E$ is critical, $S_{E} \subseteq \cgrid{}[h]$, and $(E/_\sim,\restrict{\beta}{E}/_\sim)$ is an odd separator. Moreover, if $F\cap\cgrid{}[h]\neq\emptyset$, then $E\subseteq\cgrid{}[h]$.
\end{enumerate}
\end{lemma}
\begin{proof}
First suppose that no critical set $\hat{E}\subseteq F\cup\set{f}$ of size $w$ distinct from $F$ exists.
Wrap-safety implies $S_F\not\subseteq\cgrid{}[2]$, so Lemma~\ref{lem:non-critical-subset-almost-extendable} gives Assertion~1 for every $E\neq F$.

Otherwise, fix a critical set $\hat{E}\subseteq F\cup\set{f}$ of size $w$ with $\hat{E}\neq F$; then $\hat{E}\setminus F=\set{f}$.
By Corollary~\ref{cor:sep_no_jump}, $S_{\hat{E}}$ and $S_F$ share $w-1$ equations.
Since separators span at most $w$ rows,
the new equation lies in rows $w+1$ to $c$ in block $h$ or in rows $(1,1)$ or $(1,2)$, and $S_{\hat{E}}$ either remains truly wrapping or contained entirely in block $h$ or $1$.
Throughout, consider $E\subseteq F\cup\set{f}$ of size at most $w$ with $E\neq F$, and write $\beta':=\restrict{\beta}{E}$.
If $f$ is in block $1$, Lemma~\ref{lem:first-actual-bottom-equation} shows almost-extendability for Assertion~1.

Suppose $f$ is in block $h$.
Choose a closure-consistent extension $\beta$ with the right parity by Lemma~\ref{lem:move-up-fix-parity} if a top-block separator is formed, and otherwise by the usual Tseitin extension within block $h$.
The latter respects the unary constraints since $F$ does not prescribe the lower boundary and contains fewer than $w$ equations in this block.
If $E$ is an actual separator in block $h$, then $E=\hat{E}$ and Lemma~\ref{lem:closure-consistent-separator-is-safe} gives safety; its parity determines Assertion~1 or~3.
Otherwise, the restriction to $E\setminus\set{f}$ is almost-extendable because it is safe and has size less than $w$.
Restore the assignment to $f$ within block $h$ as in the proof of Lemma~\ref{lem:extend-almost-extension}: closure-consistency handles the small components, and the large component accommodates the possible violated equation, preserving the fixed boundary variables.
This gives Assertion~1.

Now suppose $f$ is outside the border blocks.
Choose a witness $\gamma$ for upwards-evenness of $\alpha$ with $n=3$ as done in the former proof.
Since $c= 3w$, rows $(b,n+1)$ and $(b+1,n)$ do not contain equations of $(F\cup\set{f})/_\sim$.
Assign $f$ according to $\gamma$ via $\sim$.
As in the preceding proof, $\beta'$ is $1$-periodic and closure-consistent, and $\gamma$ witnesses block-consistency and upwards-evenness.
The set $E$ is not a separator in a border-block: block $1$ contains no equations, and a top-block separator would already lie in $F$, contradicting uniqueness of $S_F$.
Thus $\beta'$ is safe.

If $E\neq \hat{E}$, it omits a column, giving Assertion~1 by Lemma~\ref{lem:safe-nont-odd-almost-extends}.
It remains to consider $E=\hat{E}$. We distinguish three cases.
\begin{enumerate}
\item If $S_{\hat{E}}\subseteq\cgrid{}[h]\cup\cgrid{}[(1,1)]$ and $S_{\hat{E}}\cap\cgrid{}[(1,1)]\neq\emptyset$, then $\beta'$ is wrap-safe, giving Assertion~2.
This includes the case that $S_{\hat{E}}$ lies entirely in row $(1,1)$.
\item Otherwise, if $S_{\hat{E}}\subseteq\cgrid{}[1]$, it contains an equation in row $(1,2)$.
This internal separator has the even upper parity of $\gamma$, giving Assertion~1 by Lemma~\ref{lem:safe-nont-odd-almost-extends}.
\item Otherwise, $S_{\hat{E}}\subseteq\cgrid{}[h]$.
Even parity likewise gives Assertion~1, while odd parity gives Assertion~3 if $F\cap\cgrid{}[h]=\emptyset$.
If $F\cap\cgrid{}[h]\neq\emptyset$, choose $e\in F\cap\cgrid{}[h]$.
It remains in $\hat{E}$, since the removed equation is modulo $\compEquiv^+$ in row $(1,1)$.
The restriction to $\hat{E}\setminus\set{e}$ is safe and hence almost-extendable.
Restore $e$ by the same block $h$ argument above; since $f\notin\cgrid{}[h]$, $\hat{E}$ is not separator in block $1$, so Assertion~1 applies. \qedhere
\end{enumerate}
\end{proof}

\subsection{Lower Bound Proof}
\label{sec:lb-proof}

It is crucial for the lower bound to measure the minimum \defining{separator height} of a separator of a set of equations $F \subseteq \comucgrid$ of size $w$ such that $F$ is critical. We can naturally measure this by the unique virtual separator $S_F$ for $F$ (recall Corollary~\ref{cor:critical-then-unique-v-separator}). Then we define $\sheight{F}$ to be the minimum $b \in [h]$ such that $S_F$ contains an equation from block $b$. In our lower bound proof, it is crucial that this height cannot decrease too quickly. We formalize this via the following lemma, which essentially follows from Corollary~\ref{cor:sep_no_jump}.

\begin{lemma} \label{lem:sep_no_jump}
Let $E, F \subseteq \comucgrid$ be sets of equations of size $w$ such that $E, F$ are critical and $|E \setminus F| = 1$.
Assume that neither $S_E$ nor $S_F$ is truly wrapping.
Then $|\sheight{E} - \sheight{F}| \le 1$. Moreover,
\begin{enumerate}
\item if $F/_{\sim}$ is double-crossing then $\sheight{F} \ge \sheight{E}$. \label{itm:double_crossing_height}
\item if $F/_{\sim}$ is not double-crossing and $\sheight{F} \in [3,h-1]$, then either  \label{itm:not_double_crossing_height}
\begin{enumerate}
\item $\sheight{F} \le \sheight{E}$, or 
\item $F/_{\sim}$ is almost double-crossing and for every safe $\alpha \in \sols{F}$, there exists a $\beta \in \exts{\sols{E \cup F}}{\set{\alpha}}$ such that $\restrict{\beta}{E'}$ is almost-extendable for every $E' \subseteq E \cup F$ of size at most $w$ such that $E' \neq F$. 
\end{enumerate} 
\end{enumerate} 
\end{lemma}

\begin{proof}
Let $F \setminus E = \set{e^j_{b,n,k}/_\compEquiv}$.
By Corollary~\ref{cor:sep_no_jump}, $S_E$ and $S_F$ contain the same equation in all columns $k' \neq k$.
Since neither separator is truly wrapping, each is contained in its minimum block and possibly the next block.
A shared equation therefore gives $\sheight{E}\leq\sheight{F}+1$ and, symmetrically, $\sheight{F}\leq\sheight{E}+1$.
Thus $|\sheight{E}-\sheight{F}|\leq 1$; throughout the remaining proof, block adjacency is likewise non-wrapping.

Now suppose that $F/_{\sim}$ is double-crossing. Then every equation in $F$ is of the form $\eclass{e^j_{b,1,k}}$, i.e.\! every equation must be in row one. Therefore, every equation in $S_F$ is of the form $e^j_{b,1,k}$ and since $S_F$ is a separator it follows that there must be some $b \in [h]$ such that every equation in $S_F$ lies in the first row of this block. By Corollary~\ref{cor:sep_no_jump}, all equations in $S_E$ apart from one lie in the first row of block $b$. Since $S_E$ is a separator, it follows that every equation in $S_E$ lies either in block $b$ or block $b-1$. In particular, $\sheight{F} \ge \sheight{E}$.

Next suppose that $F/_{\sim}$ is not double-crossing and that $b:=\sheight{E} < \sheight{F}$. Then to complete the proof it is enough to show that condition (b) holds. We first show that $F/_{\sim}$ is almost double-crossing. 

Since $|\sheight{E} - \sheight{F}| \le 1$ we have that $\sheight{F} = b+1$. Moreover, by Corollary~\ref{cor:sep_no_jump}, $S_E$ contains exactly one equation from block $b$. Further, since $S_E$ is a separator it follows that every other equation in $S_E$ lies in the first row of block $b+1$. As $S_F$ is also a separator and is not double-crossing it follows that there exists $j_{k'} \in [2]$ for each $k' \in [w]$ such that 
 \[
S_F= \setcond{\opgridequation{j_{k'}}{b+1}{1}{k'}}{k' \in [w] \setminus \set{k}} \cup \set{{\opgridequation{j_k}{b+1}{2}{k}}}.
 \]
 Then since $S_E$ is a separator and contains an equation from block $b$ it follows that $S_F \setminus S_E = \set{\opgridequation{j_k}{b+1}{2}{k}}$. Moreover, the unique equation $S_E \setminus S_F$ occurs in block $b$, row $c$, and column $k$. Next observe that every equation in $E$ (resp. $F$) is $\compEquiv^+$-equivalent to an equation in $S_E$ (resp. $S_F$). Since $\compEquiv^+\subseteq\sim$, we have $E/_{\sim}=S_E/_{\sim}$ and $F/_{\sim}=S_F/_{\sim}$. In particular, $F/_{\sim}$ is almost double-crossing.

Now let $\alpha \in \sols{F}$ be safe. It remains to show that there exists $\beta \in \exts{\sols{E \cup F}}{\set{\alpha}}$ such that $\restrict{\beta}{E'}$ is almost-extendable for every $E' \subseteq E \cup F$ of size at most $w$ such that $E' \neq F$.  

For this first observe that since $\sheight{F} \ge 3$, $F$ is not a block-$2$-separator (else it would be its own virtual separator). Therefore, we may apply Lemma~\ref{lem:safe_to_safe} to obtain some $\beta \in \exts{\sols{E \cup F}}{\set{\alpha}}$ such that for all $E' \subseteq E \cup F$ of size at most $w$, we have that $\restrict{\beta}{E'}$ is safe or almost-extendable. Suppose $E' \neq E$ and $E' \neq F$. Then $E'/_{\sim}$ is not a separator, since it contains two equations from column $k$. If $\restrict{\beta}{E'}$ is safe it is also almost-extendable by Lemma~\ref{lem:safe-nont-odd-almost-extends}. Therefore, for all such $E'$ we have that $\restrict{\beta}{E'}$ is almost-extendable. Moreover, again by Lemma~\ref{lem:safe-nont-odd-almost-extends}, if $(E/_{\sim},\restrict{\beta}{E}/_{\sim})$ is an even separator then it is also almost-extendable, in which case $\beta$ is the required map.

So it remains to consider the case where $(E/_{\sim},\restrict{\beta}{E}/_{\sim})$ is an odd separator. Let $e$ be the unique equation in $E \setminus F$. Define $\beta' \in \sols{E \cup F}$ to be the 1-periodic map such that $\beta'$ agrees with $\beta$ everywhere except for on the vertical variables of $e$. Note that since $\vars{e} \cap \vars{F} = \emptyset$, we have that $\beta'$ extends $\alpha$ and indeed $\beta' \in \sols{E \cup F}$. Moreover, $(E/_{\sim}, \restrict{\beta'}{E}/_{\sim})$ is an even separator, because $\beta'$ disagrees with $\beta$ exactly on the vertical variables of $e$. So it remains to show that $\beta'$ is safe, as then we can use Lemma~\ref{lem:safe-nont-odd-almost-extends} to infer almost-extendability.

For this we check the conditions. By definition $\beta'$ is 1-periodic. Also, since $\cl{E \cup F} = E \cup F$, it follows that $\beta'$ is closure-consistent.
To check the right parity, 
 recall that $\sheight{E}=\sheight{F}-1\in[2,h-2]$ and $\sheight{F}\in[3,h-1]$. Therefore, neither $E$ nor $F$ is a separator in a border block: if either was then by the uniqueness of virtual separators so would either $S_E$ or $S_F$, which would contradict the aforementioned separator heights and the fact that $S_E$ is not double crossing.
Moreover, every size-$w$ separator contained in $E\cup F$ is either $E$ or $F$, since it must contain one equation per column. Therefore, by Lemma~\ref{lem:closure-consistent-separator-is-safe} we obtain that $\beta'$ is safe, which completes the proof.
\end{proof}

We are now ready to prove our size game lower bound; recall that we are fixing $w \ge 3$.

\begin{lemma}
	\label{lem:lower-bound}
	Duplicator has a winning strategy for the width-$w$ size game on $\ucgrid^1/_{\compEquiv}$ in position $(\emptyset, \{0\}, 2^{h-3})$, where in the first round Duplicator chooses a regular round.
\end{lemma}

\begin{proof}

To state the Duplicator strategy, we first define different types of positions occurring in the game. Observe that, every position $P$ reached in this game is either a position in the width-$w$ size game on $\ucgrid^1/_{\compEquiv}$ or $\ucgrid^0/_{\compEquiv}$. We say a position $P = (F, W, r)$ on $\comucgrid \in \set{\ucgrid^0/_{\compEquiv},\ucgrid^1/_{\compEquiv}}$ is:
\begin{itemize}
	\item \defining{homogeneous} if $\comucgrid = \ucgrid^0/_{\compEquiv}$,
	\item \defining{non-homogeneous} if $\comucgrid = \ucgrid^1/_{\compEquiv}$,
	\item \defining{safe} if every $\alpha \in W$ is safe,
	\item \defining{odd separator} if every $\beta \in W$ is $1$-periodic,
	$\beta/_{\sim} \in \sols{F/_{\sim}}$, and $(F/_\sim, \beta/_\sim)$ is an odd separator.
\end{itemize}
For convenience, whenever $F$ has some property we also say that $P$ has that property. For example, we say that $P$ is pairwise separating if $F$ is pairwise separating.

Duplicator is able to win the game by maintaining the following invariant.
Let $P = (F, W, r)$ be a position reached on $\comucgrid \in \set{\ucgrid^0/_{\compEquiv},\ucgrid^1/_{\compEquiv}}$ in the width-$w$ size game with $r >1$. Let $V$ be the linear space of $W$. Then $P$ satisfies at least one of the following properties.
 \begin{enumerate} [({I}a)]
 	\item $P$ is \defining{almost-extendable}. This means that $W = \set{\alpha}$ for some almost-extendable $\alpha$.
 	\label{itm:inv-almost-extendable}
 	\item $P$ is \defining{wrap-safe}. This means that $W = \set{\alpha}$ for some wrap-safe $\alpha$.
 	\label{itm:inv-wrap-safe}
 	\item $P$ is a \defining{non-critical odd separator}. This means that $P$ is a safe, non-critical, pairwise separating odd separator.
 	\label{itm:inv-non-critical-odd}
 	\item $P$ is \defining{high}. This means that $P$ is a safe, critical odd separator such that
 	$r \le 2^{\sheight{F}-2}$ and if $P$ is double-crossing then $r \le 2^{\sheight{F}-3}$.
 	\label{itm:inv-high}
 		\item $P$ is \defining{switchable}. This means that $P$ is a safe critical odd separator which is double-crossing such that 
 		\label{itm:inv-swtichable}
 		\begin{enumerate}[(a)]
 				\item
 				 $r \le 2^{\sheight{F}-2}$ and, \label{con:switch_high}
 			\item there is some $b \in [h]$, $k \in [w]$, $\alpha_1, \alpha_2 \in W$  
			 such that
 			\[ \alpha_1/_{\sim}(v) = \alpha_2/_{\sim}(v) \text{ iff } v \not \in \{\eclassw{ \opvvariable{1}{b}{c}{k}}, \eclassw{\opvvariable{2}{b}{c}{k}} \}. 
 			\]
			In particular, we may assume w.l.o.g. that
			$(F/_{\sim}, \alpha_j/_{\sim})$ has even side-$j$-parity for both $j \in [2]$.
			  \label{con:switch_switcher} 
 			\end{enumerate}
 \end{enumerate}

We first consider the initial position and show that Duplicator can reach a position satisfying the invariant.
Recall, that from the initial position, Spoiler picks a non-empty set $E$ of at most $w$ equations and Duplicator picks $\alpha \in \sols{E}$. Duplicator constructs $\alpha$ by starting with the empty map and extending it to the equations of $E$ one at a time while maintaining the invariant. Note that the empty map is almost-extendable, as witnessed by the extension which assigns zero to every variable. 
Now suppose that we have $\alpha \in \sols{E'}$ for some $E' \subsetneq E$ such that $\alpha$ is almost-extendable. Let $e \in E \setminus E'$. Then by Lemma~\ref{lem:extend-almost-extension} we can extend $\alpha$ to $e$ and either maintain that $\alpha$ is almost-extendable or obtain $\alpha$ such that $(E' \cup \set{e}, \alpha)$ is a block-$h$ separator with odd parity and that $\alpha$ is safe. 
Moreover, in the latter case $|E' \cup \set{e}| = w$ which implies that $E' \cup \set{e} = E$. Therefore, at the end of this process we obtain $\alpha \in \sols{E}$ such that either $\alpha$ is almost-extendable or $(E,\alpha)$ is a block-$h$ odd separator and $\alpha$ is safe. We reach the position $P = (E, \set{\alpha},2^{h-3})$. Then if $\alpha$ is almost-extendable then $P$ is almost-extendable and we satisfy \invref{itm:inv-almost-extendable}. Otherwise, $P$ is a safe, odd separator. 
Moreover, $E$ is a block-$h$ separator so $E$ is critical and $\sheight{E} = h$. Therefore, $P$ is high \invref{itm:inv-high}. So in either case $P$ satisfies the invariant. Moreover, we will see in the case distinction below that from $P$ Duplicator next chooses a regular round.

We now prove by induction on $r$ that if a position $P$ satisfies the invariant, then Duplicator wins the game from $P$.
Since from our initial position we can reach a position satisfying the invariant and from this position Duplicator chooses a regular round, this implies the lemma.
For the base case $r=1$, Duplicator wins. This is because $P$ is always either almost-extendable or safe and so Spoiler does not win during the round (recall that Spoiler wins during a round if there is some $\alpha \in W$ violating an equation of $\comucgrid$, which is impossible if $\alpha$ is almost-extendable or safe).

 We prove that if the position $P = (F, W, r)$ reached at the start of some round satisfies one of \invref{itm:inv-almost-extendable}-\invref{itm:inv-swtichable}, then so does the position $Q$ reached at the end of the round. To do this, we distinguish five cases. Again observe that since $P$ is always either almost-extendable or safe, Spoiler does not win at the start of the round.
 
 For the proof the following definitions will be useful. Recall, that both a subtree or a regular round begins with Spoiler choosing $f \in \comucgrid$. We write $W^{+}:= W^{+f}$ and $V^{+}$
 for the linear space of $W^{+}$; explicitly $V^{+} = \exts{\sols{\homsys{(F \cup \set{f})}}}{V}$, where recall $V$ is the linear space of $W$.  Before carrying out our case distinction we prove one easy but useful claim; it intuitively says that if we satisfy the invariant then there cannot be an odd separator near the bottom of the grid.

 \begin{claim} \label{claim:no_low_sep}
	If $ P = (F,W, r)$ with $r>1$ satisfies one of \invref{itm:inv-non-critical-odd}-\invref{itm:inv-swtichable}, then $F$ is not a block-$1$ separator or a block-$2$ internal separator. Moreover, if $P$ satisfies \invref{itm:inv-high} or \invref{itm:inv-swtichable} then $\sheight{F} \ge 3$.
 \end{claim}

 \begin{claimproof}
If $P$ satisfies \invref{itm:inv-non-critical-odd}, then $F$ is not critical so in particular $F$ is not a separator. If $P$ satisfies \invref{itm:inv-high} or \invref{itm:inv-swtichable} then $F$ is critical and $2 \le r \le 2^{\sheight{F} - 2}$, which implies that $\sheight{F} \ge 3$. If $F$ were a block-$1$ separator or a block-$2$ internal separator, then $F$ is its own unique virtual separator and so $\sheight{F} \le 2$, a contradiction.
 \end{claimproof}
We now begin our case distinction.

\subparagraph*{Case 1: $P$ is almost-extendable~\invref{itm:inv-almost-extendable}.}
In this case, Duplicator chooses a regular round in the size game. The round begins by Spoiler choosing $f \in \comucgrid$. Then since $\alpha$ is almost-extendable by Lemma~\ref{lem:extend-almost-extension}, there is a $\beta \in \sols{F \cup \set{f}}$ extending $\alpha$ such that for all $f$-neighbors $E$ of $F$ either
\begin{itemize}
	\item $\restrict{\beta}{E}$ is almost-extendable, or
	\item $(E, \restrict{\beta}{E})$ is a block-$h$ separator of the right parity and $\restrict{\beta}{E}$ is safe.
\end{itemize}

 In particular, $\beta \in W^{+}$ and so Duplicator chooses this $\beta$. Spoiler chooses some $f$-neighbor $E$ of $F$ and we reach the position $Q=(E, \set{\alpha'}, r-1)$, where $\alpha' := \restrict{\beta}{E}$. We show that $Q$ satisfies the invariant. 

If $\alpha'$ is almost-extendable then $Q$ is almost-extendable \invref{itm:inv-almost-extendable}. Otherwise, by the choice of $\beta$ we have that $(E, \alpha')$ is a block-$h$ separator of the right parity and $\alpha'$ is safe. If $(E, \alpha')$ is not an odd separator, then by Lemma~\ref{lem:safe-nont-odd-almost-extends} $\alpha'$, and hence $Q$, is again almost-extendable \invref{itm:inv-almost-extendable}. Otherwise, $(E, \alpha')$ is an odd block-$h$ separator. Therefore, $E$ is critical and $\sheight{E} = h$: the actual separator in block $h$ is a virtual separator for $E$, and uniqueness forces it to be $S_E$. It follows that $r-1 \le 2^{h-3} = 2^{\sheight{E}-3}$. Moreover, since $\alpha'$ is safe, $Q$ is safe. Therefore, $Q$ is high \invref{itm:inv-high}.

\subparagraph{Case 2: $P$ is wrap-safe~\invref{itm:inv-wrap-safe}.}
Duplicator chooses a regular round which begins by Spoiler choosing $f$.
By Lemma~\ref{lem:wrap-maintain-even-upwards},
there is an extension $\beta \in \sols{F\cup \set{f}}$ extending the single assignment $\alpha$ in $P$ such that for  all $f$-neighbors $E\neq F$ of $F$
\begin{itemize}
	\item $\restrict{\beta}{E}$ is almost-extendable, or
	\item $\restrict{\beta}{E}$ is wrap-safe, or 
	\item $\restrict{\beta}{E}$ is safe, $E$ is critical, $S_{E} \subseteq \cgrid{}[h]$, and $(E/_\sim,\restrict{\beta}{E}/_\sim)$ is an odd separator.
\end{itemize}
Duplicator chooses this $\beta$ and Spoiler an $f$-neighbor $E$ of $F$.
If $E=F$, the position clearly stays wrap safe.
Otherwise,
in the first two cases, clearly Invariants~\invref{itm:inv-almost-extendable} or~\invref{itm:inv-wrap-safe} are satisfied.
In the third case, the resulting position $Q=(E,\set{\restrict{\beta}{E}}, r-1)$ is a safe and critical odd separator
with $\sheight{E} = h$. This implies that $r-1 \leq 2^{\sheight{E}-3}$, so $Q$ is high \invref{itm:inv-high}.

\subparagraph*{Case 3: $P$ is a non-critical odd separator~\invref{itm:inv-non-critical-odd}.}

Again Duplicator chooses a regular round and Spoiler chooses $f = \eclass{e^j_{b,n,k}} \in \comucgrid$. As $\alpha$ is safe we can apply Lemma~\ref{lem:safe_to_safe} and Claim~\ref{claim:no_low_sep} to find a $\beta \in W^{+}$ such that for every $f$-neighbor $E$ of $F$ we have that $\restrict{\beta}{E}$ is either safe or almost-extendable. 

Duplicator chooses this $\beta$ and Spoiler chooses an $f$-neighbor $E$ of $F$. We reach the position $Q=(E, \set{\alpha'}, r-1)$, where $\alpha' := \restrict{\beta}{E}$. If $\alpha'$ is almost-extendable then $Q$ is almost-extendable \invref{itm:inv-almost-extendable}. 

Otherwise, by the choice of $\beta$ we have that $\alpha'$ is safe. Moreover, $(E/_{\sim}, \alpha'/_{\sim})$ is an odd separator, since otherwise by Lemma~\ref{lem:safe-nont-odd-almost-extends} we would have that $\alpha'$ is almost-extendable. Also $E$ is pairwise separating, as otherwise by Lemma~\ref{lem:not-pairwise-separator-almost-extendable}, we would again have that $\alpha'$ is almost-extendable. Suppose for a contradiction that $E$ is critical. Then also $E \cup F$ is critical. So by Lemma~\ref{lem:critical-to-critical} since $F$ is pairwise separating, either $F$ is critical or $F/_{\sim}$ is not a separator. But $F$ has neither of these properties, a contradiction. Therefore, $E$ is not critical and so $Q$ satisfies \invref{itm:inv-non-critical-odd}.

 \subparagraph*{Case 4: $P$ is high~\invref{itm:inv-high}}

 Here we need to make a further case distinction, intuitively based on whether  the next position could be switchable. Formally we say that $P$ is \defining{almost switchable} if $P$ is high, $r > 2^{\sheight{F}-3}$, and $F/_{\sim}$ is almost double-crossing, i.e. there exists $b \in [2,h]$, $k \in [w]$ and $j_{k'}$ for $k' \in [w]$ such that 
 \[
 F/_{\sim} = \setcond{\eclassw{\opgridequation{j_{k'}}{b}{1}{k'}}}{k' \in [w] \setminus \set{k}} \cup \set{\eclassw{\opgridequation{j_k}{b}{2}{k}}}.
 \]
The idea is that if $\eclassw{\opgridequation{j_k}{b}{2}{k}}$ is moved `one position down' to $\eclassw{\opgridequation{j_k}{b}{1}{k}}$ we obtain a double-crossing separator. If this does occur to maintain the invariant Duplicator has to ensure a switchable position is reached. Therefore, if $P$ is almost switchable then Duplicator plays a subtree round, otherwise Duplicator plays a regular round, similarly to in Case 3. We now formally carry out this case distinction.

\subparagraph*{Case 4a: $P$ is high and not almost switchable}

This is very similar to Cases~2 and~3. Again Duplicator chooses a regular round which begins by Spoiler choosing $f = \eclass{e^j_{b,n,k}} \in \comucgrid$.

If there is a critical $f$-neighbor $\hat{E}$ of $F$
such that $S_{\hat{E}}$ contains an equation in row $(1,1)$,
then the invariant ensures $\sheight{F}\geq 3$.
This implies  $S_{F} \subseteq \cgrid{}[h]$ because separators span consecutive rows, so $S_{F}$ is not in block $1$.
By Lemma~\ref{lem:wrap-prepare-even-upwards} and since $\alpha$ is safe,
there is an extension $\beta$ of $\alpha$ to a solution of $f$ such that for every $f$-neighbor $E \neq F$,
the restriction $\restrict{\beta}{E}$ is almost-extendable or wrap-safe.
Duplicator chooses this $\beta$ and Spoiler an $f$-neighbor $E$. If $E = F$ we reach position
  $(F,\set{\alpha},r-1)$ satisfying Invariant~\invref{itm:inv-high}. Otherwise, we reach a position which is
either almost extendable or wrap safe, satisfying
Invariants~\invref{itm:inv-almost-extendable} or~\invref{itm:inv-wrap-safe}.

Otherwise, every critical $f$-neighbor $E$ of $F$ (including $F$ itself)
satisfies $S_E\cap \cgrid{}[(1,1)]=\emptyset$.
Hence, $S_E$ is not truly wrapping.
By Lemma~\ref{lem:non-critical-subset-almost-extendable},
there exists a $\beta \in W^{+}$
such that for every $f$-neighbor $E$ of $F$ 
the restriction $\restrict{\beta}{E}$ is almost-extendable or $E$ is critical, $\restrict{\beta}{E}$ is safe, and $(E/_{\sim}, (\restrict{\beta}{E})/_{\sim})$ is an odd separator. 

Duplicator chooses this $\beta$ and then Spoiler chooses an $f$-neighbor $E$ of $F$. We reach the position $Q = (E, \set{\alpha'}, r-1)$, where $\alpha' := \restrict{\beta}{E}$.
If $\alpha'$ is almost-extendable then $Q$ is almost-extendable satisfying Invariant~\invref{itm:inv-almost-extendable}.

It remains to consider the case where $Q$ is a critical odd separator. Then by considering $\sheight{E}$, we show that $Q$ is high. For this we use that $P$ is high so we have $r \le 2^{\sheight{F} - 2}$. This leads to the following case distinction.
\begin{itemize}
	\item If $r \le 2^{\sheight{F} -3}$ then:
	\begin{itemize}
		\item if $E/_{\sim}$ is not double-crossing, then we just use that $|\sheight{E} - \sheight{F}| \le 1$ by Lemma~\ref{lem:sep_no_jump}. Therefore, $r-1 \le 2^{{\sheight{F} - 3}} \le 2^{\sheight{E}-2}$ and $Q$ is high \invref{itm:inv-high}. 
		\item Otherwise, if $E/_{\sim}$ is double-crossing, then $\sheight{E} \ge \sheight{F}$ by Lemma~\ref{lem:sep_no_jump}. Therefore, $r-1 \le 2^{\sheight{E}-3}$ and so $Q$ is again high \invref{itm:inv-high}.
	\end{itemize}
	\item Suppose $r > 2^{\sheight{F}-3}$. 
	 Then, $F/_{\sim}$ is not almost double-crossing since $P$ is not almost switchable and $F/_{\sim}$ is not double-crossing since $P$ is high. Since $r \le 2^{h-3}$ it follows that $\sheight{F} \neq h$. Combining this with Claim~\ref{claim:no_low_sep} we obtain that $\sheight{F} \in [3,h-1]$. We make a further case distinction based on whether $E/_{\sim}$ is double-crossing.
	\begin{itemize}
		\item Suppose $E/_{\sim}$ is not double-crossing. Then, since $F/_{\sim}$ is not double-crossing by Lemma~\ref{lem:sep_no_jump} we obtain that $\sheight{F} \le \sheight{E}$ and so $r-1 \le 2^{\sheight{E}-2}$. Therefore, $Q$ is high \invref{itm:inv-high}.
		\item Suppose $E/_{\sim}$ is double-crossing. Then $S_E$ is a double-crossing separator and $S_F$ is not an almost double-crossing separator. Since $S_E$ is double-crossing and $S_F$ is not, they are distinct. Thus Corollary~\ref{cor:sep_no_jump} gives $|S_E \cap S_F| = w-1$. Therefore, since $S_F$ is not double-crossing it must contain an equation in block $\sheight{E} - 1$, i.e. $\sheight{E} = \sheight{F} +1$. Therefore, $r-1 \le 2^{\sheight{E}-3}$ and so $Q$ is high.
	\end{itemize}
	
\end{itemize}

\subparagraph*{Case 4b: $P$ is high and almost switchable.}

Recall that in this case there exists $b \in [2,h]$, $k \in [w]$ and $j_{k'}$ for $k' \in [w]$ such that 
 \[
 F/_{\sim} = \setcond*{\eclassw{\opgridequation{j_{k'}}{b}{1}{k'}}}{k' \in [w] \setminus \set{k}} \cup \set*{\eclassw{\opgridequation{j_k}{b}{2}{k}}}.
 \]
Moreover, $r >2^{\sheight{F}- 3}$. Since $r \le 2^{h-3}$ it follows that $\sheight{F} \neq h$. Combining this with Claim~\ref{claim:no_low_sep} we obtain that $\sheight{F} \in [3,h-1]$. Here Duplicator chooses a subtree round and Spoiler chooses some $f \in \comucgrid$. 

The most interesting case occurs when $f/_{\sim} = \eclassw{\opgridequation{j}{b}{1}{k}}$, for some $j \in [2]$. Let $\gamma^{+} \in \sols{\homsys{F \cup \set{f}}}$ be the map which takes the value zero everywhere except on variables $\sim$-equivalent to $\opvvariable{1}{b-1}{c}{k}$ or  $\opvvariable{2}{b-1}{c}{k}$, where it takes the value one. Duplicator picks $\beta \in W^{+}$ via Lemma~\ref{lem:non-critical-subset-almost-extendable}, i.e. $\beta$ is such that for every $f$-neighbor $E$ of $F$ we have that $\restrict{\beta}{E}$ is almost-extendable or
$E$ is critical, $\restrict{\beta}{E}$ is safe, and $(E/_\sim,\restrict{\beta}{E}/_\sim)$ is an odd separator. Duplicator then sets $\beta' := \beta + \gamma^{+}$.

Observe that $\beta'$  is safe. Since $\sheight{F} \in [3,h-1]$ we have that $F$ is not a separator in a border block. Therefore, $(F \cup \set{f}, \beta')$ has the right parity. Also, since $\cl{F \cup \set{f}} = F \cup \set{f}$, it follows that $\beta'$ is closure-consistent. Therefore, by Lemma~\ref{lem:closure-consistent-separator-is-safe} we obtain that $\beta'$ is indeed safe.

Duplicator sets $W^{+} := \set{\beta,\beta'}$. In the game, Spoiler next picks an $f$-neighbor $E$ of $F$, we set $W^{-} := \restrict{W^{+}}{E} = \set{\restrict{\beta}{E},\restrict{\beta'}{E}}$, and Spoiler decides whether to play a big subtree round or a small subtree round.

Suppose Spoiler decides to play a big subtree round. Then if there is some almost-extendable $\alpha' \in W^{-}$, then Duplicator chooses $W' := \set{\alpha'}$ and we reach the almost-extendable position $Q = (E, \set{\alpha'}, r-1)$.

So suppose that no such $\alpha'$ exists. In particular, $\restrict{\beta}{E}$ is not almost-extendable. Then by the choice of $\beta$, we know that $E$ is critical, $\restrict{\beta}{E}$ is safe,
and $(E/_{\sim}, \restrict{\beta}{E}/_{\sim})$ is an odd separator.
Since $E/_{\sim}$ is a separator and $E/_{\sim} \subseteq F/_{\sim} \cup \set{\eclassw{\opgridequation{j}{b}{1}{k}}}$, it follows that either $E = F$ or $E/_{\sim}$ is double-crossing.
If $E = F$, then as $\restrict{\beta}{F} = \alpha = \restrict{\beta'}{F}$, taking $W' := \set{\alpha}$ we reach the position $Q= (F, \set{\alpha}, r-1)$. Since $P$ is high so is $Q$ \invref{itm:inv-high}.

Otherwise, $E/_{\sim}$ is double-crossing. Then Duplicator sets $W' := W^{-}$, so that we reach the position $Q = (E, W^{-}, r-1) = (E, \set{\restrict{\beta}{E}, \restrict{\beta'}{E}},r-1)$. Then $Q$ is switchable \invref{itm:inv-swtichable}. In detail, by Corollary~\ref{cor:sep_no_jump} it is easy to see that $\sheight{E} = \sheight{F}$. So since $P$ is high we have $r -1 < r \le 2^{\sheight{F}- 2} = 2^{\sheight{E}-2}$. Moreover, note that $(E/_{\sim}, \restrict{\beta}{E}/_{\sim})$ and $(E/_{\sim},\restrict{\beta'}{E}/_{\sim})$ have opposite side-$j$-parities for both $j \in [2]$.
So to show condition (b) in \invref{itm:inv-swtichable}, it is enough to show that $(E/_{\sim},\restrict{\beta'}{E})$ has odd side-$j$-parity for exactly one value of $j$. This in turn holds because $(F/_{\sim}, \alpha/_{\sim})$ is an odd internal separator, $\beta'$ is a safe extension of $\alpha$ and by Lemma~\ref{lem:separator-determines-parity}.

So suppose that Spoiler picks a small subtree round. Writing $V^{-}$ for the linear space of $W^{-}$, Spoiler picks $\alpha' \in V^{-}$ and we reach the position $Q= (\homsys{E}, \set{\alpha'}, \lfloor r/2 \rfloor)$. Then $\alpha' \in \set{0_E, \gamma}$, where $\gamma:= \restrict{\gamma^{+}}{E}$. If $\alpha' = 0_E$, then $Q$ is clearly almost-extendable \invref{itm:inv-almost-extendable}. So suppose $\alpha' = \gamma$. If $Q$ is almost-extendable we again satisfy \invref{itm:inv-almost-extendable} and are done. So suppose $Q$ is not almost-extendable. By the same argument that $\beta'$ is safe we also have that $\gamma^{+}$ is safe. Therefore, since $\gamma$ is not almost-extendable,
the arguments as in the proof of Lemma~\ref{lem:non-critical-subset-almost-extendable} $(E/_{\sim}, \alpha'/_{\sim})$ is a critical and safe  odd separator. Moreover, $\lfloor r/2 \rfloor \le 2^{\sheight{F}-3} = 2^{\sheight{E}-3}$, so $Q$ is high \invref{itm:inv-high}.

It remains to consider the case where $f/_{\sim} \neq \eclassw{\opgridequation{j}{b}{1}{k}}$ for any $j \in [2]$. Then intuitively Duplicator will simulate a regular round in which they play as in the previous case. To do this, Duplicator set $ W^{+} := \set{\beta}$ for $\beta$ chosen as follows. If there is some $\beta$ an extension of $\alpha$ such that $\restrict{\beta}{E}$ is almost extendable for every $f$-neighbor $E \neq F$, then Duplicator chooses it. 
Otherwise, Duplicator chooses $\beta$  via Lemma~\ref{lem:non-critical-subset-almost-extendable} exactly as previous cases. In this case Lemma~\ref{lem:sep_no_jump}\eqref{itm:not_double_crossing_height} gives $\sheight{F} \le \sheight{E}$ for every critical $f$-neighbor $E$ of $F$.  If Spoiler chooses a small subtree round, then we reach the position $Q= (\homsys{E}, \set{0_E}, \lfloor r/2 \rfloor)$ which is clearly almost-extendable. If Spoiler chooses a big subtree round then we reach the position $Q = (E, \set{\alpha'}, r-1)$, where $\alpha' := \restrict{\beta}{E}$. Either $Q$ is almost-extendable, or $E = F$ and $Q$ is high, or (in the second case) $E$ is a critical safe odd separator that is not double-crossing with $r - 1 \le 2^{\sheight{F}-2} \le 2^{\sheight{E}-2}$, so $Q$ is high.

\subparagraph*{Case 5: $P$ is switchable \invref{itm:inv-swtichable}.}

Since $P$ is double-crossing it follows that there is some $b \in [2,h]$ and a sequence $(j_k)_{k \in [w]} \in \{1,2\}^w$ such that
\[
F/_{\sim} = \setcond*{\eclassw{\opgridequation{j_k}{b}{1}{k}}}{k \in [w]}.
\]
Moreover, since $P$ is switchable for $j \in [2]$ there exists $\alpha_j \in W$  satisfying the properties from \invref{itm:inv-swtichable}(b). 		
Duplicator picks a subtree round which begins with Spoiler choosing an equation $f \in \comucgrid$. Duplicator picks the set $\set{\beta_1, \beta_2} \subseteq W^{+}$ as follows. Let $k'$ be the column given by \invref{itm:inv-swtichable}(b), and let $\gamma^{+} \in \sols{\homsys{F \cup \set{f}}}$ be the map which takes the value one on all variables $\sim$-equivalent to $\opvvariable{1}{b-1}{c}{k'}$ or $\opvvariable{2}{b-1}{c}{k'}$, and, if $f/_\sim = \eclassw{e^{j'}_{b-1,c,k'}}$ for some $j' \in [2]$, also on all variables $\sim$-equivalent to $\opvvariable{j'}{b-1}{c-1}{k'}$, and zero everywhere else.
 Duplicator picks $\beta_1$ by extending $\alpha_1$ via Lemma~\ref{lem:non-critical-subset-almost-extendable} and sets $\beta_2 := \beta_1 + \gamma^+$. Then $\beta_2$ extends $\alpha_2$.

Next Spoiler chooses an $f$-neighbor $E$ of $F$, we set $W^{-} := \set{\alpha'_1, \alpha'_2}$, where $\alpha'_j := \restrict{\beta_j}{E}$, and Spoiler decides whether to play a big or a small subtree round. 
The case where Spoiler chooses a small subtree round can be handled analogously to in Case (4); the key point is that since the round number halves the best that Spoiler can do is reach a high position.

In the big subtree round,
if some $\alpha'_j$ is almost-extendable, Duplicator chooses $W'=\{\alpha'_j\}$ and we proceed as before.
Otherwise, by the choice of the extensions, both assignments are safe and form a critical odd separator.
We first consider the special case where $E/_{\sim} = F/_{\sim}$. Then Duplicator sets $W' := W^{-}$, so we reach the position $Q = (E, W^{-}, r-1)$. Note that since $E/_{\sim} = F/_{\sim}$ we have that $\alpha'_j/_{\sim} = \alpha_j/_{\sim}$ for $j \in [2]$. Moreover, by Lemma~\ref{lem:sep_no_jump} we have $\sheight{F} = \sheight{E}$. Therefore, since $P$ is switchable, $Q$ is switchable.

So suppose $E/_{\sim}$ is a separator and that $E/_{\sim} \neq F/_{\sim}$. It follows that 
\[ 
f/_{\sim} \in \set*{ \eclassw{e^1_{b-1,c,k}}, \eclassw{e^2_{b-1,c,k}},
 \eclassw{e^{3-j_k}_{b,1,k}},
 \eclassw{e^{j_k}_{b,2,k}},  \eclassw{e^{3-j_k}_{b,2,k}}}.
 \]
 Intuitively, the separator can move down, stay at the same height, or move up. We deal with each of these in turn; the most interesting case is when the separator moves down, because this is when Duplicator can remove an odd separator in a big subtree round which is the key to the lower bound.

 So consider this case, i.e. suppose $f/_{\sim} = \eclassw{e^j_{b-1,c,k}}$, for some $j \in [2]$.
 Then $\setcond{\opvvariable{j}{b-1}{c}{k'}}{k' \in [w]} \subseteq \vars{E/_{\sim}}$,
 since $E/_{\sim}$ is a separator. Then in the big subtree  Duplicator sets $W' := \set{\alpha'_j}$. Then the game reaches the position $Q = (E, \set{\alpha'_j},r-1)$. Since $\alpha_j$ has even side-$j$-parity so does $\alpha'_j$. Therefore, $(E/_{\sim},\alpha'_j/_{\sim})$ is an even separator and so $Q$ is almost-extendable by Lemma~\ref{lem:safe-nont-odd-almost-extends}. 

Next consider the case $f/_{\sim} = \eclassw{e^{3-j_k}_{b,1,k}}$. Then Duplicator sets $W' := W^{-}$ and we reach the position $Q =(E, W^{-},r-1)$.
 Then $\sheight{E} = \sheight{F}$ by Corollary~\ref{cor:sep_no_jump}. Therefore, by the construction of the $\alpha'_j$ $Q$ is clearly switchable.

 Finally, consider the case where $f/_{\sim} = \eclassw{e^{j_k}_{b,2,k}}$ or $f/_{\sim} = \eclassw{e^{3-j_k}_{b,2,k}}$. Then Duplicator chooses the set $W' := \set{\alpha'_1}$, so we reach a high position $Q =(E, \set{\alpha'_1},r-1)$ since $\sheight{E} \ge \sheight{F}$ by Corollary~\ref{cor:sep_no_jump}.

Now we have shown that in all cases the game reaches a position satisfying the invariant. Since $r$ decreases in all cases,
Duplicator has a winning strategy by the inductive hypothesis.
\end{proof}

\begin{theorem}[Restatement of Theorem~\ref{thm:main-conf}]
	For every $w \ge 3$ and every sufficiently large $n$,
	there are two graphs $G_1$ and $G_2$ of order $\Theta(n)$,
	such that 
	\begin{enumerate}
		\item there is a graph $H$ of treewidth at most $w$ such that $\Hom{H}{G_1} \neq \Hom{H}{G_2}$, and
		\item for every graph $H$ of treewidth at most $w$ such that $\Hom{H}{G_1} \neq \Hom{H}{G_2}$, it holds that $|H| = 2^{\Omega(n^{w/2})}$.
	\end{enumerate}
\end{theorem}
\begin{proof}
	Fix $w$ and $n$.
	Let $t := \sqrt{n}$. Let $h$ be defined from $t$, as in the definition of the compression, and consider $\comcwgrid{w}{h}$.
	By Theorem~\ref{thm:indistingusihability-game-equation-game-summary},
	there is an $\FF_2$-system $S$ such that
	\begin{enumerate}[a)]
		\item $|\CFI{S}| \in O(|\CFI{\comcwgrid{w}{h}}|)$,
		\item the arity of $S$ is the arity of $\comcwgrid{w}{h}$ plus one,
		\item for all trees $T$ with root $r$ holds that if Prover wins the $(w+1)$-indistinguishability game on $\comcwgrid{w}{h}$ in position $(T,r)$,
		then all graphs $H$ of treewidth $w$ with a smooth  tree-decomposition $D$ such that $T_D = T$ satisfy $\bIso{(H,D)}{\CFI{\homsys{S}}} = \bIso{(H,D)}{\CFI{S}}$, and
		\item if Falsifier wins the $(w+1)$-equation game on $\comcwgrid{w}{h}$, then
		there is a graph $H$ of treewidth at most $w$ such that $\Hom{H}{\CFI{\homsys{S}}} \neq \Hom{H}{\CFI{S}}$.
	\end{enumerate}
	Set $G_1 := \CFI{\homsys{S}}$ and $G_2 := \CFI{S}$. First consider the order of the CFI graphs.
	By Lemma~\ref{lem:compressed-crossed-size}, $|\comcwgrid{w}{h}| = \Theta(w^2 \cdot t^2) = \Theta(n)$
	and by  Lemma~\ref{lem:cfi-size}, $|\CFI{\comcwgrid{w}{h}}|= \Theta(n)$.
	Thus $|\CFI{\homsys{S}}| = |\CFI{S}| = \Theta(n) $.

	Falsifier wins the $(w+1)$-equation game on $\comcwgrid{w}{h}$
	by Corollary~\ref{cor:compressed-crossed-system-Falsifier-wins-equation-game},
	so the precondition for Item~d) is satisfied.
	This proves Item~1 of the theorem.

	We show Item~2.
	By Lemma~\ref{lem:lower-bound},
	Duplicator wins the width-$w$ size game in position $(\emptyset, \set{0}, 2^{h-3})$ on $\comcwgrid{w}{h}$, where Duplicator has to start with a regular round.
	This implies by Lemma~\ref{lem:size-game-to-indistinguishability-game-initial}
	that Prover wins the $(w+1)$-indistinguishability game on $\comcwgrid{w}{h}$ in position $(T,r)$ for every tree $T$ with root $r$ of order at most $2^{h-3}$.
	By Item~c), this implies that 
	all graphs $H$ of treewidth $w$ with a smooth  tree-decomposition $D$ such that $|T_D|\leq 2^{h-3}$ satisfy $\bIso{(H,D)}{\CFI{\homsys{S}}} = \bIso{(H,D)}{\CFI{S}}$.
	This implies by Lemma~\ref{lem:biso-indistinguishability-implies-indistinguishability}
	that all graphs $H$ of treewidth $w$ and order at most $2^{h-3}$
	satisfy $\Hom{H}{\CFI{\homsys{S}}} = \Hom{H}{\CFI{S}}$.
	Moreover, $h \in \Theta(t^w) = \Theta(n^{w/2})$. Therefore, $2^{h-3} = 2^{\Omega(n^{w/2})}$ and the theorem follows.
\end{proof}

\section{Conclusion}

In this work, we have proved an almost tight $2^{\Omega(n^{k/2})}$ lower bound on the size of a graph witnessing treewidth-$k$-distinguishability. This is an exponential improvement over the previously best $\Omega(n^{k/2})$ lower bounds and contributes to a more fine-grained understanding of homomorphism counting from graphs of bounded treewidth.

To obtain this result we had to overcome two key challenges. Firstly, the lack of available techniques for proving such a lower bound. For this we introduced  a new combinatorial game, which allows one to prove equal homomorphism counts via local reasoning.
Secondly, we had to learn how to play this game on our instances.
This was difficult compared to existing bounds, not just because the game itself is more complicated, but also because the analysis of the equation systems is more involved.

This game can be useful to prove other lower bounds in different settings. 
Moreover, we believe that the techniques and insights developed in this work will be useful in other related areas, in particular, for obtaining improved robust lower bounds on the iteration number of $k$-WL. This would have consequences in other areas such as proof complexity and circuit complexity (cf. \cite{RezendeFJN025, DBLP:conf/mfcs/BerkholzLV25}). 

\bibliographystyle{plainurl}
\bibliography{hom_counts_lower_bound}

\end{document}